\documentclass[a4,twocolappendix]{emulateapj}

\usepackage{aas_macros}
\usepackage{graphicx}
\usepackage{natbib}
\usepackage[usenames]{color}
\usepackage{subfigure}
\usepackage{multirow}
\usepackage{amsmath}
\usepackage{xspace}
\usepackage{flafter}
\usepackage[breaklinks,colorlinks,citecolor=blue]{hyperref}

\def\setsymbol#1#2{\expandafter\def\csname #1\endcsname{#2}}
\def\getsymbol#1{\csname #1\endcsname}

\def\Planck{\textit{Planck}}

\def\all2013resultspapers{\nocite{planck2013-p01, planck2013-p02, planck2013-p02a, planck2013-p02d, planck2013-p02b, planck2013-p03, planck2013-p03c, planck2013-p03f, planck2013-p03d, planck2013-p03e, planck2013-p01a, planck2013-p06, planck2013-p03a, planck2013-pip88, planck2013-p08, planck2013-p11, planck2013-p12, planck2013-p13, planck2013-p14, planck2013-p15, planck2013-p05b, planck2013-p17, planck2013-p09, planck2013-p09a, planck2013-p20, planck2013-p19, planck2013-pipaberration, planck2013-p05, planck2013-p05a, planck2013-pip56, planck2013-p06b, planck2013-p01a}}

\newbox\tablebox    \newdimen\tablewidth
\def\leaderfil{\leaders\hbox to 5pt{\hss.\hss}\hfil}
\def\endPlancktable{\tablewidth=\columnwidth 
    $$\hss\copy\tablebox\hss$$
    \vskip-\lastskip\vskip -2pt}

\def\tablenote#1 #2\par{\begingroup \parindent=0.8em
    \abovedisplayshortskip=0pt\belowdisplayshortskip=0pt
    \noindent
    $$\hss\vbox{\hsize\tablewidth \hangindent=\parindent \hangafter=1 \noindent
    \hbox to \parindent{$^#1$\hss}\strut#2\strut\par}\hss$$
    \endgroup}
\def\doubleline{\vskip 3pt\hrule \vskip 1.5pt \hrule \vskip 5pt}

\def\L2{\ifmmode L_2\else $L_2$\fi}

\def\DeltaT{\ifmmode \Delta T\else $\Delta T$\fi}
\def\deltat{\ifmmode \Delta t\else $\Delta t$\fi}
\def\fknee{\ifmmode f_{\rm knee}\else $f_{\rm knee}$\fi}
\def\Fmax{\ifmmode F_{\rm max}\else $F_{\rm max}$\fi}
\def\solar{\ifmmode{\rm M}_{\mathord\odot}\else${\rm M}_{\mathord\odot}$\fi}
\def\Msolar{\ifmmode{\rm M}_{\mathord\odot}\else${\rm M}_{\mathord\odot}$\fi}
\def\Lsolar{\ifmmode{\rm L}_{\mathord\odot}\else${\rm L}_{\mathord\odot}$\fi}
\def\inv{\ifmmode^{-1}\else$^{-1}$\fi}
\def\mo{\ifmmode^{-1}\else$^{-1}$\fi}
\def\sup#1{\ifmmode ^{\rm #1}\else $^{\rm #1}$\fi}
\def\expo#1{\ifmmode \times 10^{#1}\else $\times 10^{#1}$\fi}
\def\,{\thinspace}
\def\lsim{\mathrel{\raise .4ex\hbox{\rlap{$<$}\lower 1.2ex\hbox{$\sim$}}}}
\def\gsim{\mathrel{\raise .4ex\hbox{\rlap{$>$}\lower 1.2ex\hbox{$\sim$}}}}

\def\simprop{\mathrel{\raise .4ex\hbox{\rlap{$\propto$}\lower 1.2ex\hbox{$\sim$}}}}
\def\deg{\ifmmode^\circ\else$^\circ$\fi}
\def\pdeg{\ifmmode $\setbox0=\hbox{$^{\circ}$}\rlap{\hskip.11\wd0 .}$^{\circ}
          \else \setbox0=\hbox{$^{\circ}$}\rlap{\hskip.11\wd0 .}$^{\circ}$\fi}
\def\arcs{\ifmmode {^{\scriptstyle\prime\prime}}
          \else $^{\scriptstyle\prime\prime}$\fi}
\def\arcm{\ifmmode {^{\scriptstyle\prime}}
          \else $^{\scriptstyle\prime}$\fi}
\newdimen\sa  \newdimen\sb
\def\parcs{\sa=.07em \sb=.03em
     \ifmmode \hbox{\rlap{.}}^{\scriptstyle\prime\kern -\sb\prime}\hbox{\kern -\sa}
     \else \rlap{.}$^{\scriptstyle\prime\kern -\sb\prime}$\kern -\sa\fi}
\def\parcm{\sa=.08em \sb=.03em
     \ifmmode \hbox{\rlap{.}\kern\sa}^{\scriptstyle\prime}\hbox{\kern-\sb}
     \else \rlap{.}\kern\sa$^{\scriptstyle\prime}$\kern-\sb\fi}
\def\ra[#1 #2 #3.#4]{#1\sup{h}#2\sup{m}#3\sup{s}\llap.#4}
\def\dec[#1 #2 #3.#4]{#1\deg#2\arcm#3\arcs\llap.#4}
\def\deco[#1 #2 #3]{#1\deg#2\arcm#3\arcs}
\def\rra[#1 #2]{#1\sup{h}#2\sup{m}}

\def\dots{\relax\ifmmode \ldots\else $\ldots$\fi}
\def\WHzsr{\ifmmode $W\,Hz\mo\,sr\mo$\else W\,Hz\mo\,sr\mo\fi}
\def\mHz{\ifmmode $\,mHz$\else \,mHz\fi}
\def\GHz{\ifmmode $\,GHz$\else \,GHz\fi}
\def\mKs{\ifmmode $\,mK\,s$^{1/2}\else \,mK\,s$^{1/2}$\fi}
\def\muKs{\ifmmode \,\mu$K\,s$^{1/2}\else \,$\mu$K\,s$^{1/2}$\fi}
\def\muKRJs{\ifmmode \,\mu$K$_{\rm RJ}$\,s$^{1/2}\else \,$\mu$K$_{\rm RJ}$\,s$^{1/2}$\fi}
\def\muKHz{\ifmmode \,\mu$K\,Hz$^{-1/2}\else \,$\mu$K\,Hz$^{-1/2}$\fi}
\def\MJysr{\ifmmode \,$MJy\,sr\mo$\else \,MJy\,sr\mo\fi}
\def\MJysrmK{\ifmmode \,$MJy\,sr\mo$\,mK$_{\rm CMB}\mo\else \,MJy\,sr\mo\,mK$_{\rm CMB}\mo$\fi}
\def\microns{\ifmmode \,\mu$m$\else \,$\mu$m\fi}

\def\muK{\ifmmode \,\mu$K$\else \,$\mu$\hbox{K}\fi}
\def\microK{\ifmmode \,\mu$K$\else \,$\mu$\hbox{K}\fi}
\def\muW{\ifmmode \,\mu$W$\else \,$\mu$\hbox{W}\fi}
\def\kms{\ifmmode $\,km\,s$^{-1}\else \,km\,s$^{-1}$\fi}
\def\kmsMpc{\ifmmode $\,\kms\,Mpc\mo$\else \,\kms\,Mpc\mo\fi}
\providecommand{\sorthelp}[1]{}

\newcommand{\WMAP}{\emph{WMAP}}
\newcommand{\QUIET}{QUIET}

\shorttitle{QUIET measurements of the Galactic plane at 43 and 95\,GHz}
\shortauthors{QUIET Collaboration}

\begin{document}

\title{The Q/U Imaging Experiment: \\Polarization measurements of the Galactic plane at 43 and 95\,GHz}

\author{
  QUIET Collaboration---
 T.~M.~Ruud\altaffilmark{1,*},
 U.~Fuskeland\altaffilmark{1},
 I.~K.~Wehus\altaffilmark{2},
 M.~Vidal\altaffilmark{3},
 D.~Araujo\altaffilmark{4},
 C.~Bischoff\altaffilmark{5,6},
 I.~Buder\altaffilmark{5,6},
 Y.~Chinone\altaffilmark{7,8},
 K.~Cleary\altaffilmark{9},
 R.~N.~Dumoulin\altaffilmark{4},
 A.~Kusaka\altaffilmark{10,11},
 R.~Monsalve\altaffilmark{12},
 S.~K.~N\ae ss\altaffilmark{13,1},
 L.~B.~Newburgh\altaffilmark{14},
 R.~A.~Reeves\altaffilmark{15},
 J.~T.~L.~Zwart\altaffilmark{4,16,17},
 L.~Bronfman\altaffilmark{18},
 R. D. Davies\altaffilmark{3},
 R.~Davis\altaffilmark{3},
 C.~Dickinson\altaffilmark{3},
 H.~K.~Eriksen\altaffilmark{1},
 T.~Gaier\altaffilmark{2},
 J.~O.~Gundersen\altaffilmark{19},
 M.~Hasegawa\altaffilmark{7},
 M.~Hazumi\altaffilmark{7},
 K.~M.~Huffenberger\altaffilmark{20},
 M.~E.~Jones\altaffilmark{13},
 C.~R.~Lawrence\altaffilmark{2},
 E.~M.~Leitch\altaffilmark{2},
 M.~Limon\altaffilmark{4},
 A.~D.~Miller\altaffilmark{4},
 T.~J.~Pearson\altaffilmark{9},
 L.~Piccirillo\altaffilmark{3},
 S.~J.~E.~Radford\altaffilmark{9},
 A.~C.~S.~Readhead\altaffilmark{9},
 D.~Samtleben\altaffilmark{21,22},
 M.~Seiffert\altaffilmark{2},
 M.~C.~Shepherd\altaffilmark{9},
 S.~T.~Staggs\altaffilmark{11},
 O.~Tajima\altaffilmark{5,7},
 K.~L.~Thompson\altaffilmark{23}
}

\vspace{+0.2in}

\altaffiltext{1}{Institute of Theoretical Astrophysics, University of Oslo, P.O. Box 1029 Blindern, N-0315 Oslo, Norway}
\altaffiltext{2}{Jet Propulsion Laboratory, California Institute of Technology, 4800 Oak Grove Drive, Pasadena, CA, USA 91109}
\altaffiltext{3}{Jodrell Bank Centre for Astrophysics, Alan Turing Building, School of Physics and Astronomy, The University of Manchester, Oxford Road, Manchester M13 9PL, UK}
\altaffiltext{4}{Department of Physics and Columbia Astrophysics Laboratory, Columbia University, New York, NY 10027, USA}
\altaffiltext{5}{Kavli Institute for Cosmological Physics, Department of Physics, Enrico Fermi Institute, The University of Chicago, Chicago, IL 60637, USA}
\altaffiltext{6}{Harvard-Smithsonian Center for Astrophysics, 60 Garden Street MS 42, Cambridge, MA 02138, USA}
\altaffiltext{7}{High Energy Accelerator Research Organization (KEK), 1-1 Oho, Tsukuba, Ibaraki 305-0801, Japan}
\altaffiltext{8}{Department of Physics, University of California, Berkeley, CA 94720, USA}
\altaffiltext{9}{Cahill Center for Astronomy and Astrophysics, California Institute of Technology, 1200 E. California Blvd M/C 249-17, Pasadena, CA 91125, USA}
\altaffiltext{10}{Physics Division, Lawrence Berkeley National Laboratory, 1 Cyclotron Road, Berkeley, CA 94720, USA}
\altaffiltext{11}{Joseph Henry Laboratories of Physics, Jadwin Hall, Princeton University, Princeton, NJ 08544, USA}
\altaffiltext{12}{School of Earth and Space Exploration, Arizona State University, 781 E. Terrace Road, Tempe, AZ 85287, USA}
\altaffiltext{13}{Department of Astrophysics, University of Oxford, Keble Road, Oxford OX1 3RH, UK}
\altaffiltext{14}{Dunlap Institute, University of Toronto, 50 St. George St., Toronto, ON M5S 3H4}
\altaffiltext{15}{CePIA, Departamento de Astronom\'ia, Universidad de Concepci\'on, Casilla 160-C, Concepci\'on, Chile}
\altaffiltext{16}{Physics Department, University of the Western Cape, Private Bag X17, Bellville 7535, South Africa}
\altaffiltext{17}{Astrophysics, Cosmology \& Gravity Centre, Department of Astronomy, University of Cape Town, Private Bag X3, Rondebosch 7701, South Africa}
\altaffiltext{18}{Departamento de Astronom\'ia, Universidad de Chile, Casilla 36-D, Santiago, Chile}
\altaffiltext{19}{Department of Physics, University of Miami, 1320 Campo Sano Drive, Coral Gables, FL 33146, USA}
\altaffiltext{20}{Florida State University, Physics, Tallahassee, FL 32306, USA}
\altaffiltext{21}{Huygens-Kamerlingh Onnes Laboratorium, Universiteit Leiden, The Netherlands}
\altaffiltext{22}{Nikhef, Science Park,  Amsterdam, The Netherlands}
\altaffiltext{23}{Stanford University and Kavli Institute for Particle Astrophysics and Cosmology, Stanford, CA 94305 USA}
\altaffiltext{*}{Corresponding author: \url{t.m.ruud@astro.uio.no}}

\slugcomment{
Submitted to ApJ---This paper should be cited as ``QUIET (2015)''
}
\journalinfo{Submitted to ApJ---Draft version \today}

\begin{abstract}
  We present polarization observations of two Galactic plane fields
  centered on Galactic coordinates $(l,b)=(0\deg,0\deg)$ and
  $(329\deg,0\deg)$ at Q- (43\,GHz) and W-band (95\,GHz), covering
  between 301 and 539 square degrees depending on frequency and
  field. These measurements were made with the \QUIET\ instrument
  between 2008 October and 2010 December, and include a total of 1263
  hours of observations. The resulting maps represent the deepest
  large-area Galactic polarization observations published to date at
  the relevant frequencies with instrumental rms noise varying
  between 1.8 and 2.8$\,\mu\textrm{K}\,$deg, \mbox{2.3--6} times
  deeper than corresponding \WMAP\ and \Planck\ maps. The angular
  resolution is $27\parcm3$ and $12\parcm8$ FWHM at Q- and W-band,
  respectively.  We find excellent agreement between the \QUIET\ and
  \WMAP\ maps over the entire fields, and no compelling evidence for significant residual
  instrumental systematic errors in either experiment, whereas the
  \Planck\ 44\,GHz map deviates from these in a manner consistent with reported systematic uncertainties for this
  channel. We combine \QUIET\ and \WMAP\ data
  to compute inverse-variance-weighted average maps,
  effectively retaining small angular scales from \QUIET\ and large
  angular scales from \WMAP.  From these combined maps, we derive
  constraints on several important astrophysical quantities, including
  a robust detection of polarized
  synchrotron spectral index steepening of $\approx0.2$ off the plane,
  as well as the Faraday rotation measure toward the Galactic center
  (RM$\,=-4000\pm200\,\textrm{rad}\,\textrm{m}^{-2}$), all of which
  are consistent with previously published results. Both the raw
  \QUIET\ and the co-added \QUIET+\WMAP\ maps are made publicly
  available together with all necessary ancillary information.
\end{abstract}


\maketitle

\nocite{Chiang:2010}

\section{Introduction}
\label{sec:intro}

The field of cosmic microwave background (CMB) cosmology has undergone
an important transition during the last two years. Until 2013, the
primary limitation of most CMB experiments, whether targeting
temperature or polarization fluctuations, was instrumental
noise. Contamination from astrophysical foregrounds and systematic errors
was generally small compared to the instrumental noise level or
intrinsic cosmic variance of the observations, and only minor
corrections for either were required to produce robust cosmological
results. Three examples among many are 
CBI \citep{CBI2004,CBI2007}, BOOMERanG \citep{Boomerang2006},
and \WMAP\ \citep{wmap9}. In this
noise-dominated regime, the CMB field as a whole made rapid progress
for more than two decades, with each new experiment improving
cosmological parameter constraints in accordance with its respective
noise level. Eventually, this process led to the current highly
successful $\Lambda$CDM 'concordance' cosmological model, which today
describes virtually all currently available cosmological observations
with only six free parameters
\citep{wmap9,planck2013-p11,planck2014-a15}.

\begin{table*}[tb] 
\begingroup 
\newdimen\tblskip \tblskip=5pt
\caption{Summary and comparison of field characteristics$^{\rm a}$\label{tab:patch_summary}}

\vskip -2mm
\footnotesize 
\setbox\tablebox=\vbox{ %
\newdimen\digitwidth 
\setbox0=\hbox{\rm 0}
\digitwidth=\wd0
\catcode`*=\active
\def*{\kern\digitwidth}
\newdimen\signwidth
\setbox0=\hbox{+}
\signwidth=\wd0
\catcode`!=\active
\def!{\kern\signwidth}
\newdimen\decimalwidth
\setbox0=\hbox{.}
\decimalwidth=\wd0
\catcode`@=\active
\def@{\kern\signwidth}
\halign{ \hbox to 1.5in{#\leaderfil}\tabskip=3em&
  \hfil#\hfil\tabskip=0.5em&
  \hfil#\hfil\tabskip=3em&
  \hfil#\hfil\tabskip=0.5em&
  \hfil#\hfil\tabskip=0em\cr
\noalign{\doubleline}
\omit& \multispan2 \hfil Q-band \hfil& \multispan2 \hfil W-band\hfil\cr
\noalign{\vskip -3pt}
\omit&\multispan2\hrulefill&\multispan2\hrulefill\cr
\noalign{\vskip 2pt}
\omit \hfil Feature\hfil& G-1& G-2& G-1& G-2\cr
\noalign{\vskip 3pt\hrule\vskip 5pt}
Field center, $(l,b)$&  $(329\deg,0\deg)$& $(0\deg,0\deg)$& $(329\deg,0\deg)$& $(0\deg,0\deg)$\cr
\phantom{Field} sky area&  483$\,$deg$^2$& 301$\,$deg$^2$&$\,$573$\,$deg$^2$&$\,$539 deg$^2$\cr
\phantom{Field} $N_{\textrm{pix}}$ ($N_{\textrm{side}}=512$)&  36\,831& 22\,983& 43\,668& 41\,090\cr
\noalign{\vskip 4pt}
\multispan5FWHM angular resolution\hfil\cr
\noalign{\vskip 2pt}
\hglue 1em\QUIET&\multispan{2}\hfil$27\parcm3$\hfil&\multispan{2}\hfil$12\parcm8$\hfil\cr
\hglue 1em\WMAP& \multispan{2}\hfil$30\parcm6$\hfil&\multispan{2}\hfil$13\parcm2$\hfil\cr
\hglue 1em\Planck&\multispan{2}\hfil$27\parcm0$\hfil&\multispan{2}\hfil*$\cdots$\hfil\cr
\noalign{\vskip 4pt}
\multispan5Effective frequency, $\nu_{\textrm{eff}}$\hfil\cr
\noalign{\vskip 2pt}
\hglue 1em\QUIET&\multispan{2}\hfil43.1$\,$GHz\hfil&\multispan{2}\hfil94.5$\,$GHz\hfil\cr
\hglue 1em\WMAP& \multispan{2}\hfil40.5$\,$GHz\hfil&\multispan{2}\hfil94.2$\,$GHz\hfil\cr
\hglue 1em\Planck&\multispan{2}\hfil44.1$\,$GHz\hfil&\multispan{2}\hfil*$\cdots$\hfil\cr
\noalign{\vskip 4pt}
\multispan5Noise $Q$/$U$ rms per $7'$ pixel\hfil\cr
\noalign{\vskip 2pt}
\hglue 1em\QUIET&  17\muK& 24\muK& 15\muK& *21\muK\cr
\hglue 1em\WMAP&   58\muK& 64\muK& 96\muK& 108\muK\cr
\hglue 1em\Planck& 52\muK& 55\muK& *$\cdots$& *$\cdots$\cr
\noalign{\vskip 4pt}
\multispan5\vtop{\hsize=1.95in \strut Linear regression ($y = ax + b$) slope, $a$ (\S~\ref{sec:comparison})\strut\par}\hfil\cr
\noalign{\vskip 2pt}
\hglue 1em$x=$\QUIET; $y=$\WMAP,             $Q$& $1.06\pm0.04$& $1.05\pm0.04$& *$\cdots$& *$\cdots$\cr
\hglue 1em\phantom{$x=$\QUIET; $y=$\WMAP,}   $U$& $1.11\pm0.17$& $1.00\pm0.03$& *$\cdots$& *$\cdots$\cr
\hglue 1em$x=$\QUIET; $y=$\Planck,           $Q$& $1.33\pm0.30$& $0.95\pm0.11$& *$\cdots$& *$\cdots$\cr
\hglue 1em\phantom{$x=$\QUIET; $y=$\Planck,} $U$& $0.86\pm0.12$& $1.00\pm0.04$& *$\cdots$& *$\cdots$\cr
\hglue 1em$x=$\WMAP; $y=$\Planck,            $Q$& $1.19\pm0.27$& $0.90\pm0.14$& *$\cdots$& *$\cdots$\cr
\hglue 1em\phantom{$x=$\WMAP; $y=$\Planck,}  $U$& $0.78\pm0.14$& $1.00\pm0.05$& *$\cdots$& *$\cdots$\cr
\noalign{\vskip 4pt}
\multispan5\vtop{\hsize=1.95in \strut Noise-weighted mean and standard deviation of deck split null map (\S\,\ref{sub:quietmaps})\strut\par}\hfil\cr
\noalign{\vskip 2pt}
\hglue 1emStokes $Q$&           !$0.22\pm1.22\muK$& $-0.32\pm1.08\muK$& $!0.09\pm1.03\muK$& $-0.17 \pm 1.01\muK$\cr
\hglue 1em\phantom{Stokes} $U$& $-0.03\pm1.16\muK$& !$0.10\pm1.05\muK$& $-0.12\pm1.04\muK$& !$0.20\pm1.01\muK$\cr
\noalign{\vskip 4pt}
\multispan5\vtop{\hsize=1.95in \strut \QUIET\ systematic uncertainties \citep{quiet_instrument}\strut\par}\hfil\cr
\noalign{\vskip 2pt}
\hglue 1emAbsolute responsivity& \multispan2\hfil{6\%}\hfil& \multispan2\hfil{8\%}\hfil\cr
\hglue 1emAbsolute detector angle&    \multispan2\hfil$1\pdeg7$\hfil& \multispan2\hfil$0\pdeg5$\hfil\cr
\noalign{\vskip 3pt\hrule}
}}
\endPlancktable 
\endgroup
\vspace*{1mm}\hspace*{17mm}\textit{a} Note that \Planck\ has not yet released W-band polarization maps, and the corresponding table entries are\par
\hspace*{17mm}\phantom{\textit{a}} therefore empty.\par
\end{table*}

This situation changed dramatically with the
\Planck\ release in 2013 March \citep{planck2013-p01}, and later with
the BICEP2 release the following year \citep{bicep2014}. The exquisite
instrumental sensitivity of \Planck\ resulted in a CMB temperature
likelihood that is, for the first time, limited by confusion from
astrophysical foregrounds rather than instrumental noise
\citep{planck2013-p08,planck2014-a13}. Likewise, BICEP2 was the first
CMB B-mode polarization experiment to become foreground-limited in
polarization \citep{pb2015}.

To continue rapid progress towards a more refined cosmological model,
in particular with respect to large-scale polarization, reionization,
and inflation \citep[e.g.,][and references therein]{liddle:2000}, a
thorough understanding of relevant astrophysical foregrounds is
paramount. 
Great progress has already been made on this
\citep[e.g.,][]{Finkbeiner1999,Costa2008,wmap9,Ichiki2014},
and in early 2015 the \Planck\ collaboration presented the most
detailed full-sky model for the frequency range between 30 and
353\,GHz to date, including both polarized synchrotron and
thermal-dust emission over the full sky
\citep{planck2014-a12}. According to this model, the frequency minimum
for polarized foregrounds on degree angular scales occurs between 70
and 80\,GHz, varying only weakly with multipole range, probably
depending somewhat on sky location.

In order to improve on this foreground model, better measurements are
required with respect to both depth and frequency coverage.  In
addition, control of instrumental systematic errors is of course
critical. As described in
\citet{planck2014-a01,planck2014-a03,planck2014-a09,planck2014-a12},
there are still outstanding issues with the most recent
\Planck\ polarization observations, both below and above the
foreground frequency minimum at $70\,$GHz.  Cross-checks and
comparisons with external data sets, including \WMAP, can be helpful
in identifying such issues.  Other data sets anticipated in the very
near future that should be useful in the effort to map out the
foregrounds include
S-PASS (2.3\,GHz; \citealp{carretti2009}), C-BASS (5\,GHz;
\citealp{king2010,king2014}) and QUIJOTE (10--40\,GHz;
\citealp{rubino-martin2012}), all observing at low frequencies.

In this paper, we present data that fit naturally into this 
larger astrophysical foreground program: measurements at 43 and 95\,GHz of two
fields in the Galactic plane taken by the QUIET instrument
\citep{quiet_instrument} between 2008 October and 2010 December.
\QUIET\ was a pathfinder experiment designed to improve limits
on B-mode polarization and demonstrate the low level of systematic
error achievable through the combination of careful
monolithic-microwave-integrated-circuit (MMIC) receiver module design,
instrument design, and survey strategy.  The instrument employed
detector arrays comprising 19 Q-band (43\,GHz) and 90 W-band
(95\,GHz) detector modules, observing from the Atacama Desert in
Chile.  The experiment reported the cleanest microwave polarization
spectra with respect to instrumental systematic errors at the time. The sum
of all instrumental systematic errors was constrained to correspond to a
tensor-to-scalar ratio of $r\lesssim 0.01$ \citep{quiet_instrument}. This result was
only barely surpassed by the very recent and vastly more sensitive
BICEP2 observations, which reported an equivalent limit on
instrumental systematic errors of $r\lesssim 0.006$
\citep{bicep2_instrument}. Cosmological CMB $E$ and $B$ angular power
spectra were reported in \citet{quiet2011,quiet2012}, while
constraints on polarized point sources were reported by
\citet{quiet_ptsrc}.

The rest of this paper is organized as follows. In \S~\ref{sec:method}
we review the \QUIET\ data selection and processing pipeline as
applied to the Galactic plane analysis, emphasizing those steps that
are different compared to the original CMB-oriented analysis. We
discuss the Q-band maps derived for the Galactic center field in
\S~\ref{sec:maps}, while equivalent discussions and figures for the
remaining observations are deferred to Appendix~\ref{sec:support}. In
\S~\ref{sec:physics} we derive constraints on important astrophysical
quantities such as the spectral index of synchrotron emission and the
Faraday rotation measure toward the Galactic center, both of which are
critical for performing robust astrophysical component separation. We
summarize and conclude in \S~\ref{sec:conclusions}. All final data
products (sky maps, mask, noise covariance matrices and beam profiles)
are available on the LAMBDA
website\footnote{http://lambda.gsfc.nasa.gov}. Following both
\WMAP\ and \Planck, we adopt the HEALPix \citep{Gorski:2004by}
convention for polarization, which differs from the IAU convention in
the sign of the Stokes $U$ parameter. All maps are provided in
Galactic coordinates.

\section{Observations and data processing}
\label{sec:method}

\begin{figure}[t]
\centering
\includegraphics[width=\linewidth]{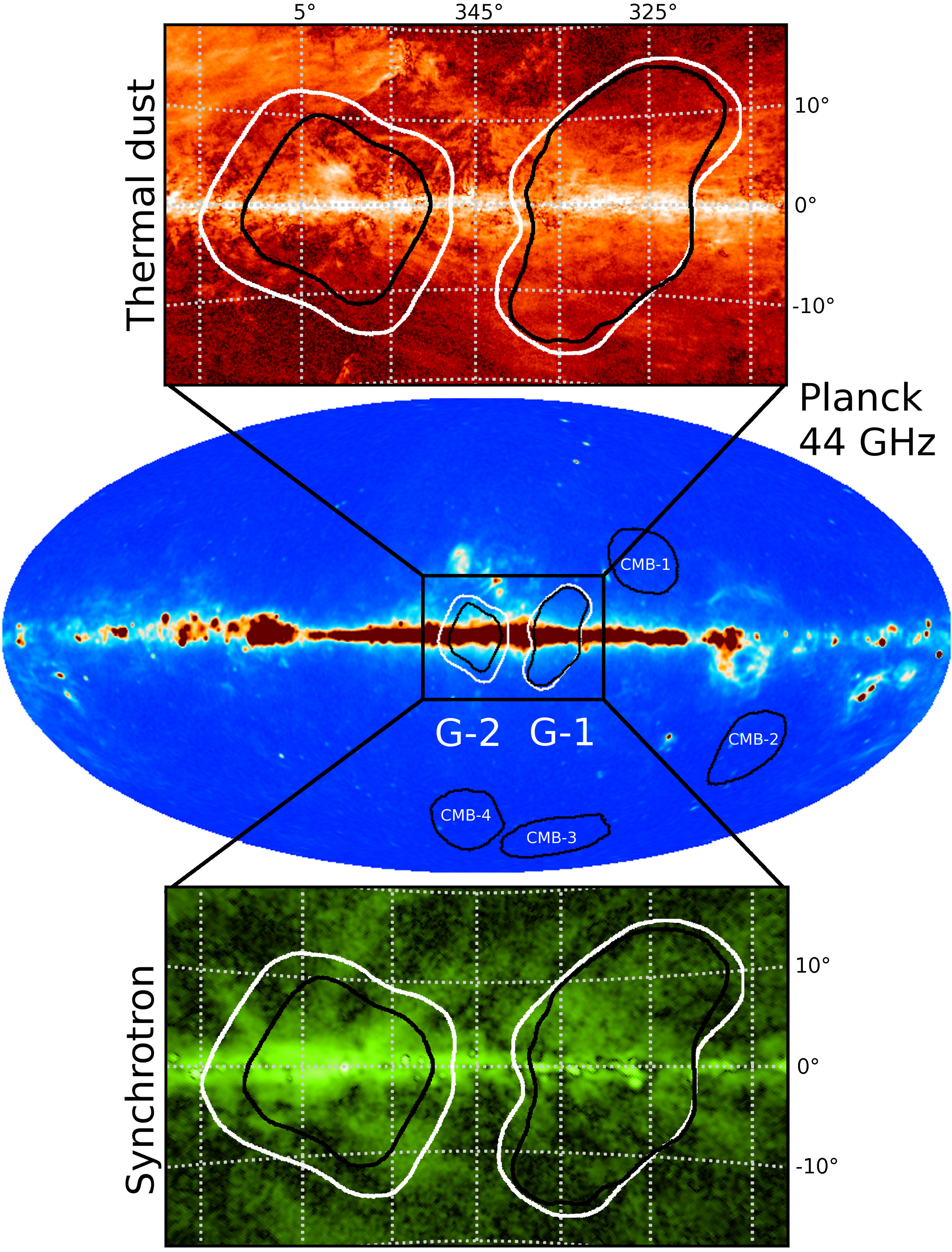}
\vspace{-0.1in}
\caption{Positions of the QUIET fields superimposed on
  \Planck\ foreground maps. The white and black outlines of the Galactic
  fields show the masks applied to the W-band (95\,GHz) and Q-band
  (43\,GHz) data, respectively. The central map is the
  \Planck\ 44\,GHz temperature map, smoothed with a $1\deg$ FWHM
  Gaussian beam, with intensity range from -0.2 mK to 1 mK. The upper
  and lower panels compare the \QUIET\ fields to the
  \Planck\ polarized thermal dust and synchrotron foreground maps. Grid
  cell width is $10\deg$.}
\label{fig:patches}
\end{figure}

The basic data selection and processing steps were described in detail
in \citet{quiet2011,quiet2012}. We briefly review the
main steps in the following, describing in greater detail a few notable differences between the
previous and the updated analysis. The most important of these is
co-addition with a second experiment (typically \WMAP), which is
essential in order to robustly measure angular scales comparable to
the size of the observed field. 
While CMB power spectrum or cosmological parameter estimation can be done without them,
these modes are
essential for deriving astrophysical spectral parameters, such
as the synchrotron spectral index or thermal dust temperature.  A
second difference is in the data-selection criteria, which are slightly less
stringent in this Galactic analysis than in the original CMB
analysis because the signal-to-noise ratio is higher for the Galactic fields.

In the original \QUIET\ analysis work, two pipelines were developed
independently for cross-validation purposes, one implementing a
pseudo-$C_{\ell}$ framework \citep{hivon2002,quiet2011},
and the other implementing a brute-force maximum-likelihood framework
\citep{Tegmark1997,bond1998,quiet2011}. A major advantage of the
latter is that it provides both unbiased sky maps and corresponding
dense pixel-pixel covariance matrices, which are useful for comparison
and inverse-noise-variance co-addition with external data sets. The
following analysis is based entirely on the maximum-likelihood
pipeline.

\subsection{Observations and data selection}
\label{sec:data}

The full, unfiltered data set consists of more than 10,000 hours of observations
taken from the Chajnantor plateau in Chile between 2008 October and 2010
December, covering two frequency bands (Q and W, with center
frequencies 43 and 95\,GHz, respectively) and six separate fields on
the sky, not counting various calibration targets. Four of these
fields were selected as the primary low-foreground patches from which
CMB constraints were derived \citep{quiet2011,quiet2012}. To constrain
polarized foregrounds at the same frequencies, two high-foreground
fields targeting the Galactic plane were also included in the
observation schedule, and these measurements are the subject of this paper.
These fields were observed when no primary CMB
targets were available, effectively filling in right ascension
``gaps'' in the observing schedule. Their positions on the sky are indicated in
Figure~\ref{fig:patches}, together with corresponding patches of the 
polarized synchrotron and thermal-dust maps recently published by \Planck\ 
\citep{planck2014-a12}. Adopting the notation introduced in
\citet{quiet2011}, we will refer to the two Galactic fields as G-1 and
G-2. The line-of-sight in G-1 cuts through the Centaurus arm as well as the tangent region of the Norma spiral arm \citep{garcia2014}, while G-2 contains the Galactic center region. Thus, these fields cover the two most populated areas of the Galactic disk in terms of molecular gas (and therefore dust).

\begin{figure}[t]
\centering
\mbox{\includegraphics[width=0.48\linewidth,clip=true,trim=0.7in 0in 0.3in 0.15in]{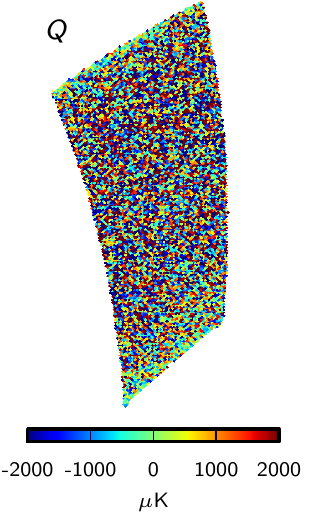}
\includegraphics[width=0.48\linewidth,clip=true,trim=0.7in 0in 0.3in 0.15in]{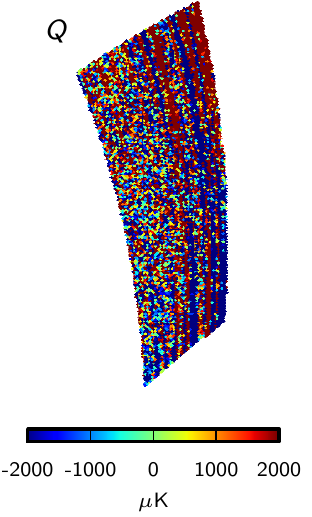}}
\vspace{-0.1in}
\caption{Example of scan cut due to excessive striping. Both maps show the data taken by a single detector module during a single CES. The left-hand map shows a normal CES (CES id no. 1808), while the right-hand map exhibits striping aligned with the scanning strategy (CES id no. 1826). All CES containing at least one such feature were cut from the analysis. Both example CES are taken from the W-band (95$\,$GHz) observations of field G-1.}
\label{fig:badces}
\end{figure}

Summary characteristics are provided for both fields in 
Table~\ref{tab:patch_summary}, including
positions, sky area, effective noise levels and basic data-quality
statistics. Regarding
systematic uncertainties, we include only the absolute responsivity and
polarization angle in Table~\ref{tab:patch_summary}, as these are the
most important ones for foreground analysis. We discuss the level of 
temperature-to-polarization leakage in our maps in 
\S~\ref{sub:quietmaps}. For a complete discussion
of systematic uncertainties relevant for B-mode analysis, we refer the
interested reader to \citet{quiet_instrument}. 

The basic observing block of the \QUIET\ scanning strategy was a
so-called {\em constant-elevation scan} (CES), in which the sky drifts
through the field-of-view while the telescope performs a simple
periodic azimuth slew of fixed amplitude. For the four CMB fields and
the G-1 field, the scan width was $15\deg$,
while for G-2 it was $10\deg$.  Once the target field center drifted
through the field-of-view by the same amount, the observing elevation
was changed either to the opposite edge of the same field, thereby
tracking the field on the sky in a set of discrete elevation steps, or
to another field. The typical duration of a single CES
was 30--60 minutes, depending on field size and elevation. In addition, 
the instrument was rotated about the optical axis (we refer to this as a 
{\em deck rotation}) in steps of $45\deg$, typically once a week, for 
a total of 8 angles. The combination of
natural cross-linking from sky rotation and frequent deck rotations
provided excellent modulation properties, suppressing many important
instrumental systematic effects \citep{quiet_instrument}.

\begin{table}[t] 
\begingroup 
\newdimen\tblskip \tblskip=5pt
\caption{Summary of data selection$^{\rm a}$ \label{tab:dataselection}}
\nointerlineskip                                                                                                                                                                                     
\vskip -3.5mm
\footnotesize 
\setbox\tablebox=\vbox{ %
\newdimen\digitwidth 
\setbox0=\hbox{\rm 0}
\digitwidth=\wd0
\catcode`*=\active
\def*{\kern\digitwidth}
\newdimen\signwidth
\setbox0=\hbox{+}
\signwidth=\wd0
\catcode`!=\active
\def!{\kern\signwidth}
\newdimen\decimalwidth
\setbox0=\hbox{.}
\decimalwidth=\wd0
\catcode`@=\active
\def@{\kern\signwidth}
\halign{ \hbox to 1.4in{#\leaderfil}\tabskip=1em&
  \hfil#\hfil\tabskip=2em&
  \hfil#\hfil\tabskip=2em&
  \hfil#\hfil\tabskip=2em&
  \hfil#\hfil\tabskip=0em\cr
\noalign{\doubleline}
\omit& \multispan2 \hfil Q-band \hfil& \multispan2 \hfil W-band\hfil\cr
\noalign{\vskip -3pt}
\omit&\multispan2\hrulefill&\multispan2\hrulefill\cr
\noalign{\vskip 2pt}
\omit \hfil Description\hfil& G-1& G-2& G-1& G-2\cr
\noalign{\vskip 3pt\hrule\vskip 5pt}
Total CES count& 295& 189& 568& 352\cr
Automatic cuts& **1& *13& *27& **8\cr
Poor pointing& **8& **0& **4& **4\cr
Short duration& **3& **0& *14& **0\cr
Excessive striping& **5& **2& *62& *27\cr
\noalign{\vskip 4pt}             
CES count after cuts& 278& 174& 461& 313\cr
\noalign{\vskip 4pt}
Observing time&     303\,h& 92\,h& 595\,h&273\,h\cr
\phantom{Observing} efficiency& *94\%& *91\%& *83\%& *90\%\cr
\noalign{\vskip 5pt\hrule\vskip 3pt}
}}
\endPlancktable 
\tablenote {{a}} Table lists the number of CES cut from the final data set by the cut criteria described in \S~\ref{sec:data}.\par
\endgroup
\end{table}

As described in \citet{quiet2011,quiet2012}, the CMB-oriented
\QUIET\ data-reduction process was based on a fully-blind analysis
philosophy, in which all data cuts, filters, and main processing steps
were defined and finalized before inspecting any final science
results, including power spectra and cosmological parameters. In this
process, each CES-diode (i.e., the CES time-stream from an individual
detector diode) was scanned for signs of contamination, and removed
from the data set if any problematic issues were identified. The
selection criteria assumed a low signal-to-noise ratio in any single
CES, and relied strongly on accurate noise and $\chi^2$ modeling.
This assumption, while valid for observations of the
low-foreground CMB sky, does not hold for the Galactic patches
considered in this paper. Rather, the amplitude of the Galactic
polarization signal is sufficiently high that the strongest signals
are visible even in a single CES, in particular at 43\,GHz. Under
the default CMB-targeted selection criteria, a large fraction of the
Galactic observations would be automatically excluded for this reason
alone, and the default pipeline is therefore not suitable for
Galactic fields. On the other hand, the same high signal-to-noise ratio also
implies that the fractional errors due to noise-modeling errors
are much less important for Galactic than for CMB analysis.

For these reasons, we adopt the following simplified data-selection
criteria in this paper: First, we apply the default selection pipeline
to eliminate obviously void scans, but exclude any tests that depend
directly on fits of noise quality. The CES removed in this step
include those affected by faulty hardware, and those for which the Moon
was within the telescope's sidelobes. Second, we manually remove scans
with poor pointing (i.e., scans that did not hit the main target
field) or short duration. Likewise we remove a small, discrete
set of scans that on visual inspection was found to exhibit
large-amplitude striping aligned with the scanning strategy.
The cause of this latter effect is unknown. It is illustrated in
Figure~\ref{fig:badces} through two 
single-CES, single-detector-module maps from the W-band observations
of field G-1. The left-hand panel shows a normal map, while the
right-hand panel shows a contaminated map. The data containing
the highest proportion of these CES are the W-band observations of
field G-1, in which they make up 10\,\% of the full data
set. Although a substantial fraction of these could be saved through
more aggressive filtering, considering the high signal-to-noise
ratio of these Galactic sky maps we prefer to minimize systematic
effects rather than instrumental noise, and conservatively remove
all CES that contain at least one striped single-detector map.

Table~\ref{tab:dataselection}
summarizes the data-selection statistics, both in terms of individual
cuts and total observing efficiency. In total, 392 (868) observation
hours are included in the final Q-band (W-band) maps,
corresponding to an acceptance rate of about 90\%, depending on field
and frequency. For comparison, the acceptance rate obtained in the
CMB-oriented \QUIET\ analyses was 70--73\,\%
\citep{quiet2011,quiet2012}.

\subsection{Mapmaking}
\label{sec:mapmaking}

Given a set of time-ordered data, we employ full maximum-likelihood
mapmaking to estimate unbiased sky maps, $\mathbf{m}$, by solving the
normal equations \citep[e.g.,][]{Tegmark1997,keskitalo:2010}
\begin{equation}
\mathbf{m} = 
\left(\mathbf{P}^{\textrm{T}}\tilde{\mathbf{N}}^{-1}\tilde{\mathbf{F}}\mathbf{P}\right)^{-1} 
\mathbf{P}^{\textrm{T}}\tilde{\mathbf{N}}^{-1} \tilde{\mathbf{F}} \tilde{\mathbf{d}}.
\label{eq:mlmapmaking}
\end{equation}
Here symbols marked by $^\sim$ denote pure time-domain objects, while
$\mathbf{P}$ and $\mathbf{m}$ denote (at least partially) map-domain
objects.  Specifically, $\mathbf{P}$ represents the pointing matrix,
as defined by the default \QUIET\ pointing model 
\citep{quiet2012}; $\tilde{\mathbf{N}}$ is the time-domain noise
covariance matrix, given by a $1/f$ noise model; $\tilde{\mathbf{F}}$
represents a general time-domain filter; and $\tilde{\mathbf{d}}$
denotes the actual time-ordered data.  The corresponding map-domain
noise covariance matrix is
\begin{equation}
\mathbf{N} =
\left(\mathbf{P}^{\textrm{T}}\tilde{\mathbf{N}}^{-1}\tilde{\mathbf{F}}\mathbf{P}\right)^{-1}
\left(\mathbf{P}^{\textrm{T}}\tilde{\mathbf{F}}^{\textrm{T}}\tilde{\mathbf{N}}^{-1}\tilde{\mathbf{F}}\mathbf{P}\right)
\left(\mathbf{P}^{\textrm{T}}\tilde{\mathbf{N}}^{-1}\tilde{\mathbf{F}}\mathbf{P}\right)^{-1}.
\label{eq:mlcovmat}
\end{equation}
Conversion between measured voltages and thermodynamic temperature
units, bandpass integration, and polarization-angle definitions are
all implicitly described by the pointing matrix, $\mathbf{P}$. For
full details and conventions, see
\citet{quiet2011,quiet2012,quiet_instrument}.

We use a HEALPix\footnote{http://healpix.sourceforge.org}
$N_{\textrm{side}}=512$ grid with
$7\arcm\times7\arcm$ pixels for our maps, sufficient to
support the $12\parcm8$ FWHM angular resolution of the \QUIET\ W-band
channel. The \QUIET\ Q-band channel has angular
resolution of $27\parcm3$ FWHM, and could in principle be pixelized
with $14\arcm\times14\arcm$ pixels; however, for consistency we pixelize
both channels with the same grid. The \WMAP\ polarization sky maps 
also use $N_{\textrm{side}}=512$ grids. 

The total number of observed Q-band (W-band) pixels is 47,288
(52,391) and 45,162 (56,216) for G-1 and G-2,
respectively. However, many of these pixels are observed only a few
times, and therefore have low signal-to-noise ratio. In order
to reduce the cost of subsequent matrix operations and data volumes,
and acknowledging the fact that we later will co-add our maps with
\WMAP\ maps, we apply a mask to each full map, removing
any pixels for which the effective \QUIET\ noise rms is more than
1.5 times the corresponding \WMAP\ noise rms. The resulting masks are
shown as black and white contours in Figure~\ref{fig:patches} for both
Q- and W-band.

While the \QUIET\ CMB analysis used several complementary
time-domain filters, the requirements for foreground observations are
somewhat different (see above). For these new maps, we have
found that a mildly-apodized high-pass filter with a cut-off
frequency of $0.5\,f_{\textrm{scan}}$ is sufficient to produce maps
with no obvious artifacts, where
$f_{\textrm{scan}}\approx0.1\,\textrm{Hz}$ is the scanning frequency
of the telescope. The only exception is a specific large-scale feature
in the G-1 field due to poor cross-linking. However, as
described below, rather than imposing a stronger time-domain filter in
this case, we project out all harmonic modes with $\ell\le10$ in the pixel
domain, to avoid excessive anisotropic filtering. No low-pass filters
are applied, in view of the fact that Galactic features tend to be
strongly localized and full angular resolution is particularly important.

\subsection{Co-addition with external data sets}
\label{sec:coadd}

\QUIET\ is for all practical purposes insensitive to physical modes
with wavelengths comparable to the size of the observed
field. The \QUIET\ field diameter of $\lambda\,\sim 20\deg$ thus suggests a 
loss of sensitivity for modes of $\ell \lesssim 18$\footnote{In the \QUIET\ CMB analyses, a lower limit of $\ell=25$ was chosen for CMB power spectrum estimation from \QUIET\ data \citep{quiet2011,quiet2012}.}. 
Although not vital for CMB power spectrum estimation, these
modes are important for astrophysical foreground inference. We
therefore co-add \QUIET\ with an external
large-scale experiment in order to produce optimal all-scale
maps. Algorithmically, the co-addition is given by an inverse-variance-weighted sum of the form
\begin{equation}
\mathbf{m}_{\textrm{tot}} = \left(\sum_i \mathbf{N}^{-1}_{i}\right)^{-1}
\left(\sum_i \mathbf{N}^{-1}_{i} \mathbf{m}_{i}\right),
\label{eq:coadd}
\end{equation}
where the sums run over experiments, and $\mathbf{N}_i$ represents the
noise covariance matrix for the $i$'th experiment. The covariance
matrix of the final map is
\begin{equation}
\mathbf{N}_{\textrm{tot}} = \left(\sum_i \mathbf{N}^{-1}_{i}\right)^{-1}.
\label{eq:coadd_N}
\end{equation}

If some set of $N$ modes (which may be organized column-wise into an
$N_{\textrm{pix}}\times N$ matrix $\mathbf{V}$) happens to be affected
by instrumental systematic errors in a given experiment, they can be 
projected out from the corresponding experiment
covariance matrix before co-addition. This is most easily done by
means of the Sherman-Morrison-Woodbury formula \citep[e.g.,][]{woodbury},
\begin{equation}
\mathbf{N}_i \rightarrow \mathbf{N}_i - \mathbf{N}_i
\mathbf{V}\left(\mathbf{V}^{\textrm{T}}\mathbf{N}_i^{-1}\mathbf{V}
\right)^{-1} \mathbf{V}^{\textrm{T}} \mathbf{N}_i.
\label{eq:smw}
\end{equation}
Effectively, this operation assigns infinite variance to all modes in
$\mathbf{V}$, ensuring that those modes do not contribute
to the final map. In practice, we will use this
operation to project out the largest-scale modes to which we can be certain that
\QUIET, due to its finite field size, has no sensitivity, by letting 
$\mathbf{V}$ consist of all spherical harmonics with $\ell\le10$.

We use the notation described above to define an instrument-specific weight
operator, $\mathbf{F}_i$, of the form
\begin{equation}
\mathbf{F}_i = \left(\sum_j \mathbf{N}^{-1}_j\right)^{-1}
\mathbf{N}^{-1}_{i},
\label{eq:coadd_filter}
\end{equation}
which simply measures the relative weight carried by experiment $i$ of
each mode in the final map. For instance,
$\mathbf{F}_{\textrm{Q}}\mathbf{m}_\textrm{Q}$ is the contribution
from \QUIET\ to the total map, $\mathbf{m}_{\textrm{tot}}$. Note that
the sum over these operators is unity, $\sum_{i} \mathbf{F}_{i} =
\mathbf{1}$, ensuring that the final map will be unbiased irrespective
of instrument-specific filtering, as long as each individual map is
inherently unbiased.

Although \QUIET, \WMAP, and \Planck\ all nominally observe at Q-band,
they do have slightly different bandpasses and effective frequencies, as 
listed in Table~\ref{tab:patch_summary}. To account for
these differences, we rescale the \WMAP\ and \Planck\ maps to the
nominal \QUIET\ frequency $\nu_{\textrm{Q}}$ before co-addition, assuming a
synchrotron-type power-law index across the bands. Explicitly, the
scaling factor for converting a map from frequency $\nu_{i}^{\mathrm{eff}}$ to 
$\nu_{\textrm{Q}}^{\mathrm{eff}}$ is
\begin{align}
\gamma_i =
\frac{g(\nu_{\mathrm{Q}}^{\mathrm{eff}})}{g(\nu_{i}^{\mathrm{eff}})}
\left(\frac{\nu_{\mathrm{Q}}^{\mathrm{eff}}}{\nu_{i}^{\mathrm{eff}}}\right)^{\beta},
\label{eq:freqscale}
\end{align}
where
\begin{equation}
  g(\nu) = \frac{(e^{x}-1)^2}{x^2 e^x}, \quad x = \frac{h\nu}{k_\textrm{B}
    T_{\textrm{CMB}}}
\end{equation}
is the conversion factor between brightness and differential
thermodynamic temperature. Here, $h$ and $k_\mathrm{B}$ denote the Planck and Boltzmann constants, and $T_{\mathrm{CMB}} = 2.7255\,\mathrm{K}$ is the CMB monopole temperature. In order to avoid
circularity in the analysis process, we adopt the synchrotron spectral
index values reported by \citet{fuskeland2014} for these re-scaling
factors, not those that will be derived from the \QUIET\ maps
themselves in \S~\ref{sec:physics}. Specifically, Fuskeland et
al.\ partitioned the whole sky (excluding bright compact objects 
and a region around the Galactic center of radius 1\deg) into 24
regions, and estimated the synchrotron spectral index for each region
from the \WMAP\ K and Ka-band polarization sky maps. For G-1, we adopt
the mean of their regions 23 and 24 (see Figure~1 in
\citealp{fuskeland2014}), resulting in $\beta_{\textrm{G-1}}=-2.93 \pm 0.01$,
while for G-2 we use the mean of regions 15 and 24, resulting in 
$\beta_{\textrm{G-2}}=-3.00 \pm 0.009$.  For \WMAP, these spectral indices
translate into scaling factors for G-1 and G-2 of 0.84 and 0.83,
respectively. For \Planck\ the corresponding factors are 1.067
and 1.069. If we instead were to adopt the spectral indices derived in
\S~\ref{sec:physics} from \QUIET\ (i.e., $\beta=-3.12\pm0.06$),
these numbers would change by 1.2\,\% and 0.7\,\% for \WMAP, and
by 0.4\,\% and 0.3\,\% for \Planck.  The impact of the precise
value of the assumed spectral index is small compared to the intrinsic
absolute responsivity uncertainty of 6\,\% in the
\QUIET\ observations \citep{quiet2011}. For W-band, the difference
between the \WMAP\ and \QUIET\ frequencies is negligible, and we omit
any re-scaling in this case.  A \Planck\ W-band polarization map is
not yet available \citep{planck2014-a01}.

\setlength{\fboxsep}{0pt}

\begin{figure}[t]
\subfigure{\includegraphics[width=0.5\linewidth, clip=true, trim=0.65in 0.1in 0.2in 0in]{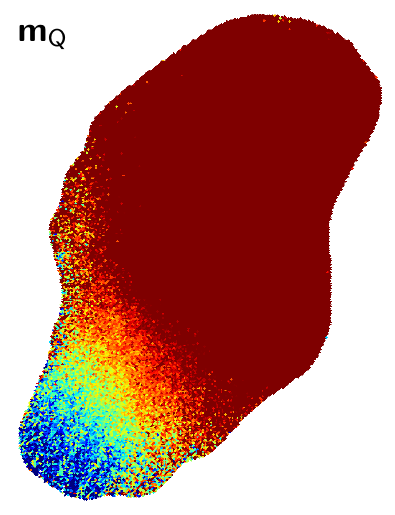}}\hspace{-0.05in}
\subfigure{\includegraphics[width=0.5\linewidth, clip=true, trim=0.65in 0.1in 0.2in 0in]{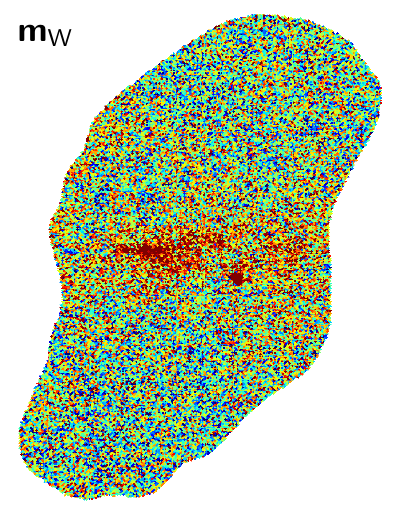}}

\vspace{-0.2in}
\subfigure{\includegraphics[width=0.5\linewidth, clip=true, trim=0.65in 0.1in 0.2in 0in]{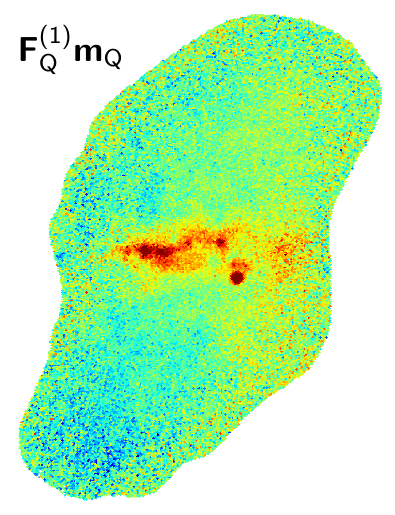}}\hspace{-0.05in}
\subfigure{\includegraphics[width=0.5\linewidth, clip=true, trim=0.65in 0.1in 0.2in 0in]{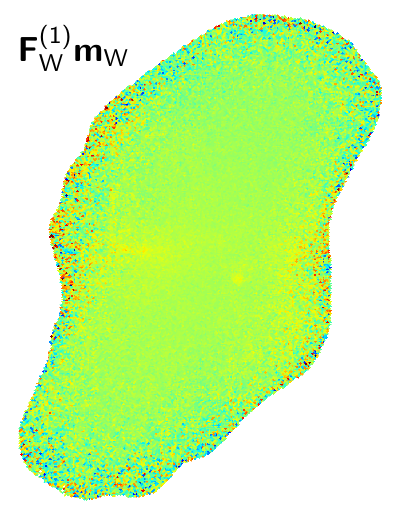}}

\vspace{-0.2in}
\subfigure{\includegraphics[width=0.5\linewidth, clip=true, trim=0.65in 0.15in 0.2in 0in]{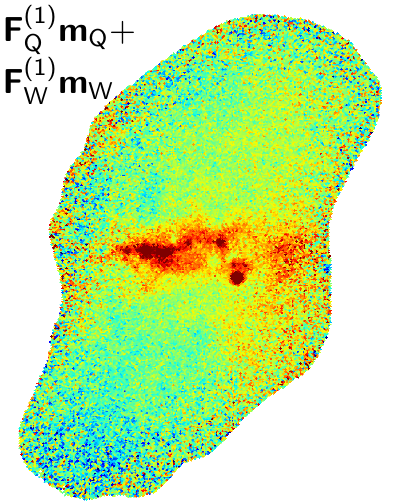}}\hspace{-0.05in}
\subfigure{\fbox{\includegraphics[width=0.5\linewidth, clip=true, trim=0.63in 0.15in 0.22in 0in]{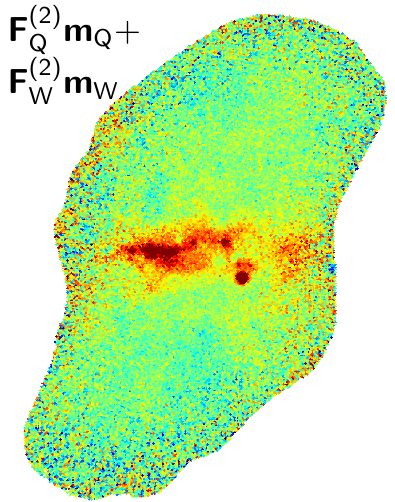}}}

\vspace{-0.1in}
\subfigure{\includegraphics[width=\linewidth, clip=true, trim=0in 0in 0in 0in]{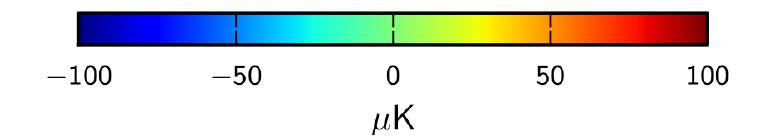}}
\vspace{-0.1in}
\caption{Example of map co-addition, applied to the Q-band \QUIET\ and \WMAP\ Stokes $Q$ maps of the G-1 field. The top row shows the \QUIET\ maximum-likelihood map, $\mathbf{m}_{\mathrm{Q}}$, and the \WMAP\ 9-yr map, $\mathbf{m}_{\mathrm{W}}$. The middle row shows the contribution to the co-added map from each data set, computed by applying the corresponding weight operators, defined in \S~\ref{sec:coadd}, to each map: $\mathbf{F}^{(1)}_{\mathrm{Q}}\mathbf{m}_{\mathrm{Q}}$ and $\mathbf{F}^{(1)}_{\mathrm{W}}\mathbf{m}_{\mathrm{W}}$. The co-added map, shown in the bottom left panel, is the sum of the two contributions. The framed panel (bottom right) shows an equivalent co-added map made using a version of the \QUIET\ map where all modes of $\ell\le 10$ have been discarded prior to co-addition; see \S~\ref{sec:codemo} for further details.}
\label{fig:codemo}
\end{figure}

\begin{figure*}[t]
\vspace{-0.05in}
\centering
\subfigure{\includegraphics[width=0.32\linewidth, clip=true, trim=0.3in 0in 0.05in 0.1in]{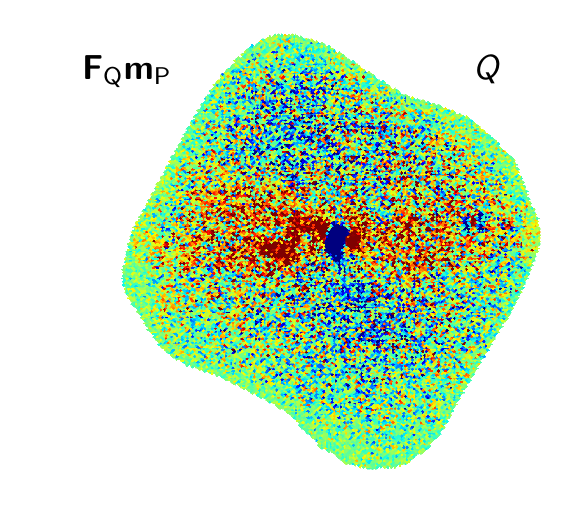}}
\subfigure{\includegraphics[width=0.32\linewidth, clip=true, trim=0.3in 0in 0.05in 0.1in]{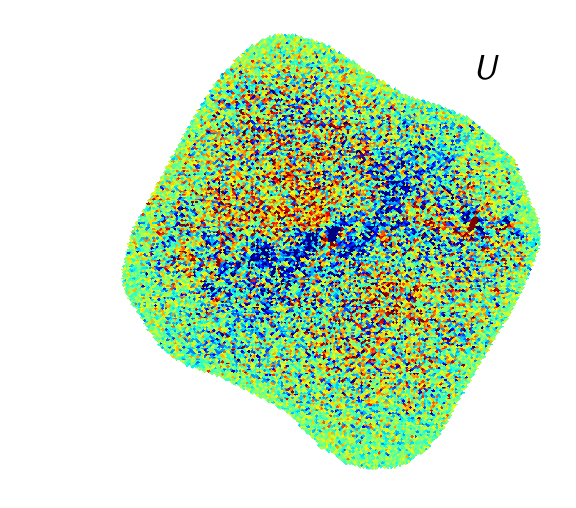}}
\subfigure{\includegraphics[width=0.32\linewidth, clip=true, trim=0.3in 0in 0.05in 0.1in]{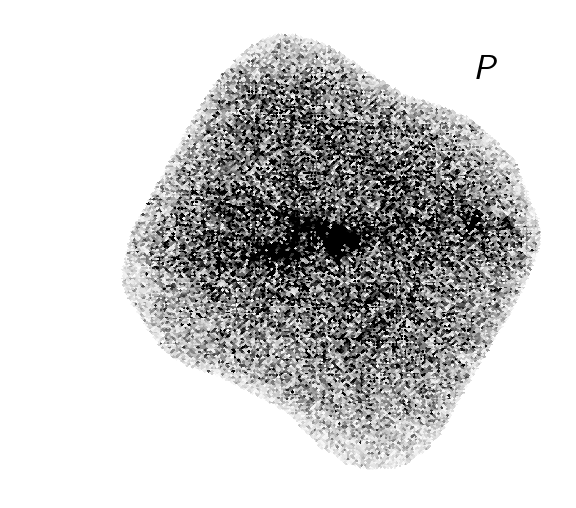}}

\vspace{-0.33in}
\subfigure{\includegraphics[width=0.32\linewidth, clip=true, trim=0.3in 0in 0.05in 0.1in]{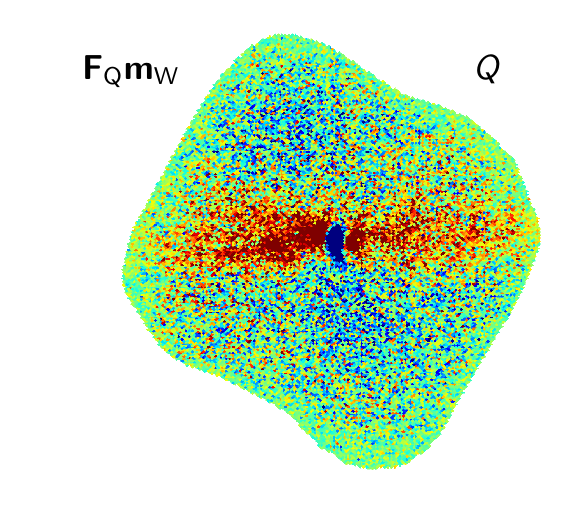}}
\subfigure{\includegraphics[width=0.32\linewidth, clip=true, trim=0.3in 0in 0.05in 0.1in]{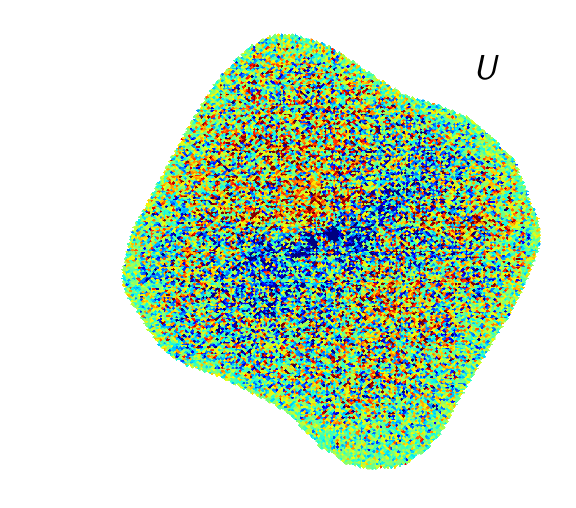}}
\subfigure{\includegraphics[width=0.32\linewidth, clip=true, trim=0.3in 0in 0.05in 0.1in]{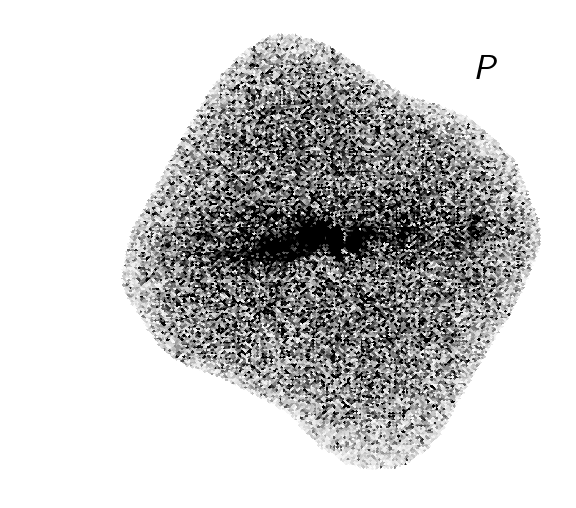}}

\vspace{-0.33in}
\subfigure{\includegraphics[width=0.32\linewidth, clip=true, trim=0.3in 0in 0.05in 0.1in]{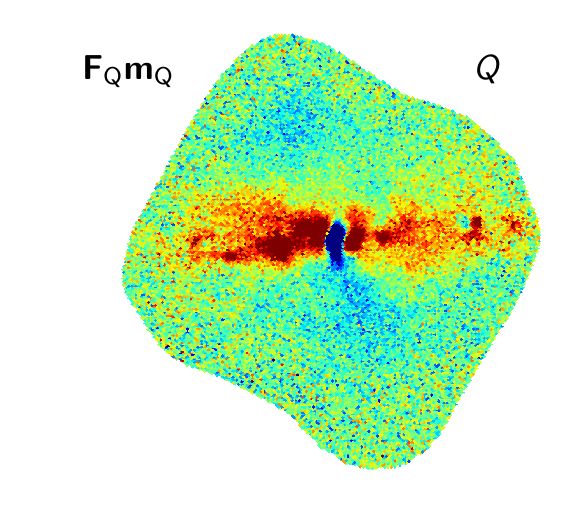}}
\subfigure{\includegraphics[width=0.32\linewidth, clip=true, trim=0.3in 0in 0.05in 0.1in]{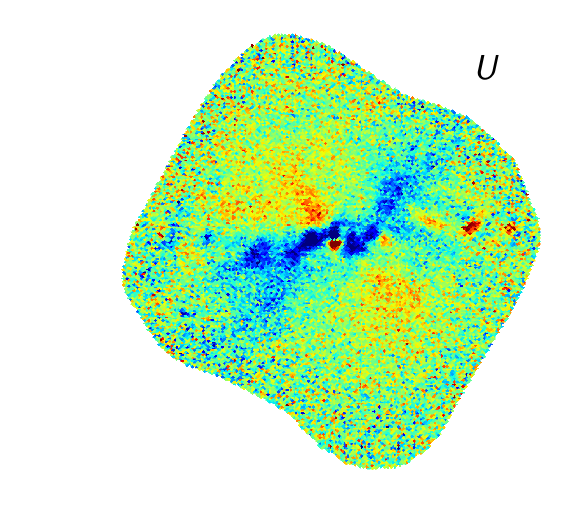}}
\subfigure{\includegraphics[width=0.32\linewidth, clip=true, trim=0.3in 0in 0.05in 0.1in]{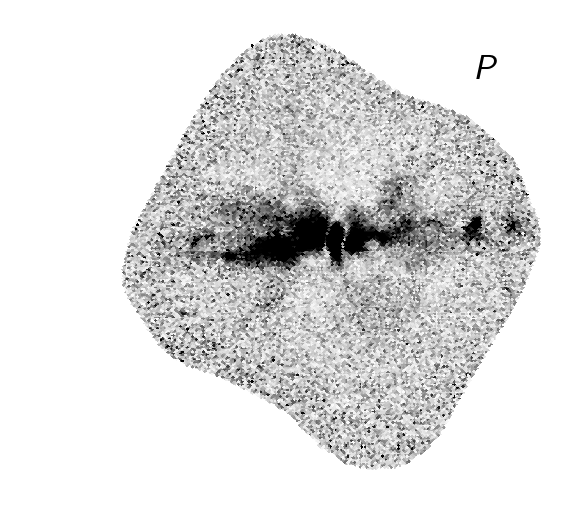}}

\vspace{-0.33in}
\subfigure{\includegraphics[width=0.32\linewidth, clip=true, trim=0.3in 0in 0.05in 0.05in]{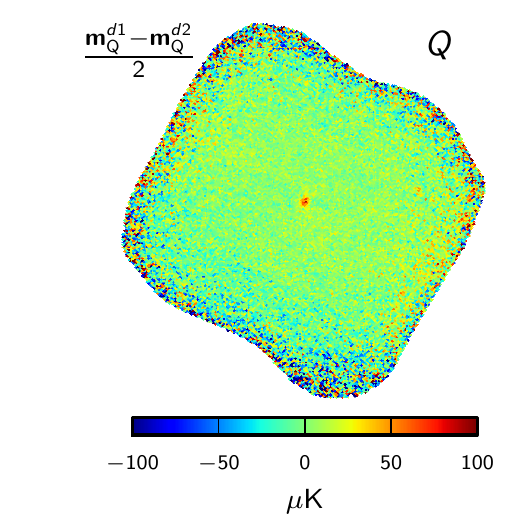}}
\subfigure{\includegraphics[width=0.32\linewidth, clip=true, trim=0.3in 0in 0.05in 0.05in]{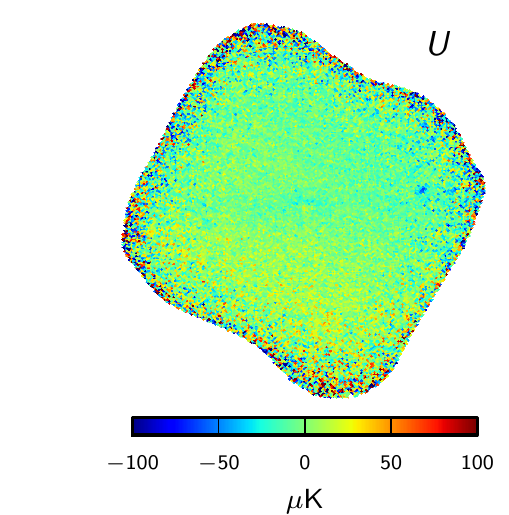}}
\subfigure{\includegraphics[width=0.32\linewidth, clip=true, trim=0.3in 0in 0.05in 0.05in]{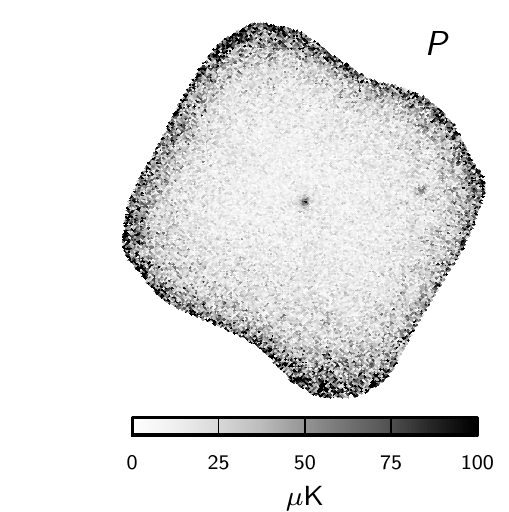}}
\vspace{-0.05in}
\caption{Inverse-noise-variance-weighted Q-band (43\,GHz) maps of the G-2 field (centered at Galactic coordinates $(l,b)=(0\deg,0\deg)$), for \Planck, \WMAP, and \QUIET. Columns show, from left to right, Stokes $Q$, Stokes $U$, and polarization amplitude $P=\sqrt{Q^2 + U^2}$. The top row shows the \Planck\ map $\mathbf{m}_{\mathrm{P}}$, filtered to only contain the small-scale modes observable by \QUIET, as determined by the \QUIET\ weighting operator $\mathbf{F}_{\mathrm{Q}}$ defined in \S~\ref{sec:coadd}. Rows 2 and 3 show the equivalent maps for \WMAP\ and \QUIET, respectively.  The bottom row shows the half-difference null maps of the deck-split \QUIET\ data; see \S~\ref{sub:quietmaps} for full details.}
\label{fig:compmaps_gcQ}
\end{figure*}

\subsection{Detailed analysis of Q-band G-1 field}
\label{sec:codemo}
Before presenting the results from our analysis, it is useful to gain
some intuition regarding the co-addition process described above. 
We therefore present the combination of the \QUIET\ and \WMAP\ 
Q-band G-1 maps in Figure~\ref{fig:codemo}. The top row shows the raw 
maps from each experiment separately. The \QUIET\ map is 
dominated by an essentially unconstrained mode with a
gradient extending from the upper right to lower left. Qualitatively similar features appear in all
\QUIET\ maximum-likelihood maps, but with an amplitude that varies
strongly from field to field.
In terms of how well the largest scales are constrained,
the G-1 field is by far the worst of all six \QUIET\ fields,
whereas G-2 is one of the best. The fundamental difference
between these two extreme cases lies in their degree of cross-linking
coupled to the size and shape of the field.  G-1 is neither a primary CMB field
nor a particularly useful calibration source. As a result, it 
was observed through a relatively small set of scanning
directions, from upper right to lower left edge in 
Figure~\ref{fig:codemo}. Moreover, only half the amplitude 
was scanned within a single CES, resulting in two only partially overlapping and almost
independent CES sets. Consequently, large-scale modes aligned with this direction are poorly
constrained.

The G-2 field, on the other hand, covers the Galactic center, including 
the Galactic center arc, the strongest polarized object within the
QUIET patches, and therefore is a particularly useful calibration
source, both for absolute responsivity and for pointing reconstruction
\citep{quiet_instrument}. As a result, this field was observed 
from many different angles,  leading to a more symmetric map.
Additionally, the G-2 field was smaller to reduce the noise per sky area, and could 
be scanned from edge to edge within a single CES. Thus, even
the large-scale modes are quite well constrained in G-2, and the raw G-2 
map shows only weak evidence for spurious large-scale gradients.

\begin{figure}[t]
\centering
\mbox{\includegraphics[width=\linewidth, clip=true, trim=0.3in 0in 0in 0.1in]{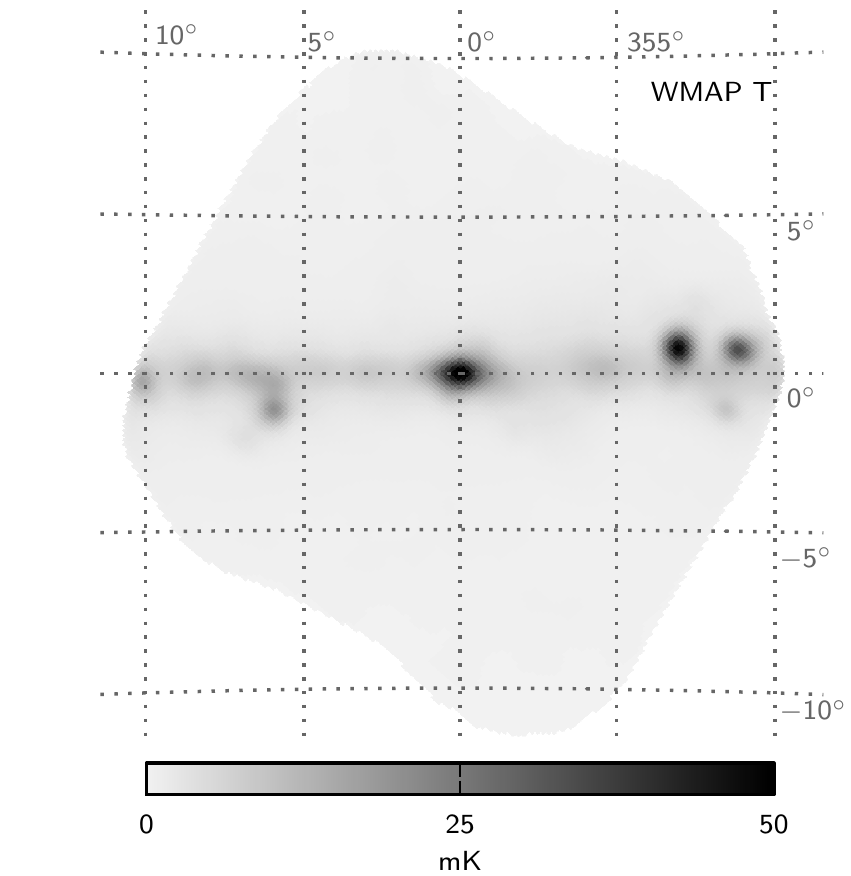}}
\mbox{\includegraphics[width=\linewidth, clip=true, trim=0.3in 0in 0in 0.1in]{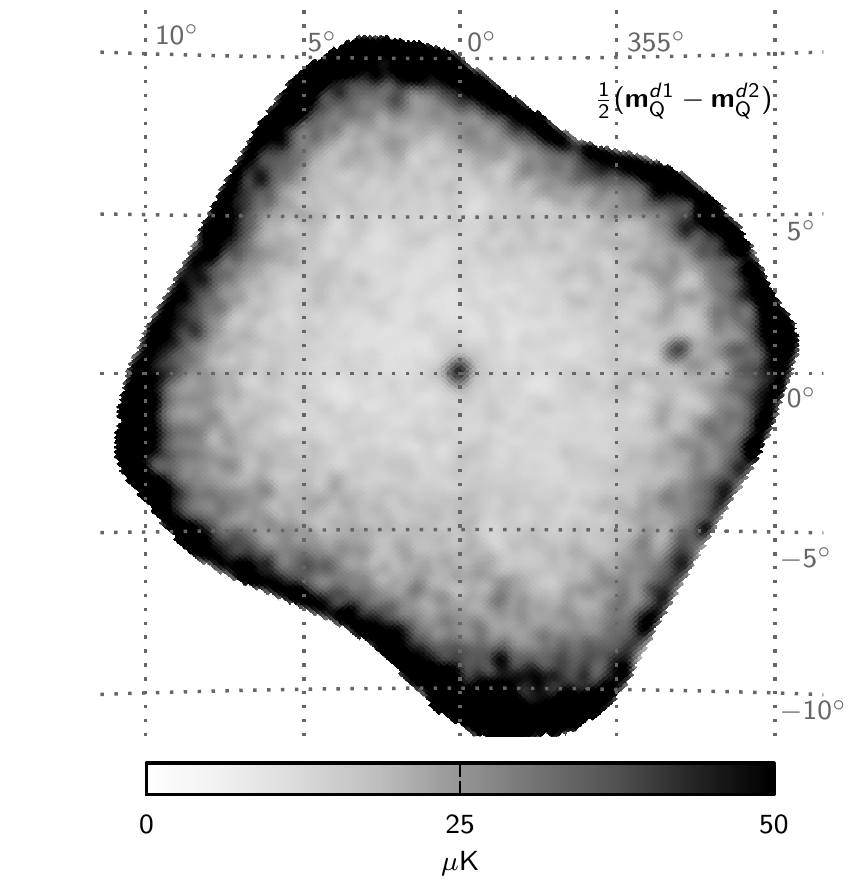}}
\caption{$I$-to-$Q$/$U$ leakage in the \QUIET\ Q-band (43\,GHz) G-2 field (centered on Galactic coordinates $(l,b)=(0\deg,0\deg)$). The top panel shows the WMAP9 Q-band temperature data. The bottom panel shows the half-difference map $(\mathbf{m}_{\mathrm{Q}}^{d1} - \mathbf{m}_{\mathrm{Q}}^{d2})/2$ of \QUIET\ data, split by deck-angle (rotation angle of optical axis), in polarization amplitude $P=\sqrt{Q^2 + U^2}$. Both maps have been smoothed to $40'\,$FWHM. Grid cell width is $5\deg$.}
\label{fig:leakcomp}
\end{figure}

Substantial benefits can be obtained by co-adding data from \QUIET\ with some large-scale experiment such as
\WMAP\ and/or \Planck.  The middle row of Figure~\ref{fig:codemo} shows the inverse-variance-weighted component
maps for \QUIET\ and \WMAP, $\mathbf{F}_{\textrm{Q}}\mathbf{m}_{\textrm{Q}}$ and
$\mathbf{F}_{\textrm{W}}\mathbf{m}_{\textrm{W}}$, as estimated from
Equation~\ref{eq:coadd_filter}.  \QUIET\ dominates the small-scale features in terms of signal-to-noise
ratio, while \WMAP\ dominates the large-scale modes. The 
previously dominating gradient in the raw \QUIET\ map is effectively suppressed,
and the weighted sum of the two contributions is shown in the bottom
left panel.

Nevertheless, a non-zero gradient is present even in the co-added map,
albeit at a greatly reduced level.  Neither \WMAP\ nor
\Planck\ observe this mode; it is clearly due to an instrumental
effect in \QUIET, perhaps ground pick-up
\citep{quiet_instrument}. Irrespective of its origin, but confident it
is an artifact in the \QUIET\ observations, we apply an additional
real-space filter that projects out all harmonic modes with
$\ell\le10$ from the \QUIET\ map, as described by
Equation~\ref{eq:smw}. We therefore adopt these few modes entirely
from \WMAP, rather than estimating them in terms of a weighted mean of
the two experiments. The result is shown in the lower right panel of
Figure~\ref{fig:codemo}, and this map appears astrophysically more
reasonable than the unfiltered version in the lower left panel. 
We evaluate the need for this filtering by comparing the rms of the \QUIET\
maps before and after applying it. 
In Q-band, filtering decreases the rms of the G-1 maps by more than 50\%,
whereas the corresponding value for field G-2 is a mere 2\%.
Similar results
are found for the W-band maps. Hence we conclude that such filtering is
prudent in the case of field G-1, but unneccessary for the far better 
constrained case of field \mbox{G-2}. All co-added maps for G-1 presented in the 
following have been derived using \QUIET\ maps pre-filtered in this way.

\section{Sky maps}
\label{sec:maps}

We are now ready to present the Galactic field sky maps as observed by
\QUIET. In order to avoid excessive repetition, we will focus our
discussion around the Q-band G-2 field, i.e., around the Galactic
center region at 43\,GHz. Corresponding plots and discussions for
the remaining three fields (G-1 at Q-band, and both G-1 and G-2 at
W-band) are given in Appendix~\ref{sec:support}.

\subsection{QUIET-only maps}
\label{sub:quietmaps}
The third row of Figure~\ref{fig:compmaps_gcQ} shows the inverse-variance-filtered
\QUIET\ G-2 maps, $\mathbf{F}_{\mathrm{Q}}\,\mathbf{m}_{\textrm{Q}}$. As described in the next section, we
choose for now to include \QUIET\ and \WMAP, but not \Planck, in the weighted sum
defined by Equations~\ref{eq:coadd}--\ref{eq:coadd_filter}. 
Thus, the modes that are weighted down by the $\mathbf{F}_{\mathrm{Q}}$ operator are
those for which \WMAP\ has lower instrumental noise than \QUIET,
as measured by the respective noise-covariance matrices.
This translates to the large-scale modes, as illustrated in \S~\ref{sec:codemo}.

The bottom row shows corresponding (half-difference) null-maps. These are 
derived by first dividing the full time-ordered \QUIET\ data set according 
to the angle of deck rotation, with one set consisting of data taken with 
deck angles $0\deg$, $90\deg$, $180\deg$, and $270\deg$, and the other with 
deck angles $45\deg$, $135\deg$, $225\deg$, and $315\deg$. Independent maps
are made from each subset, which are then subtracted.
In the absence of systematic errors, any such
null-map should contain instrumental noise only, and, as
already mentioned, the \QUIET\ analysis is fundamentally dependent on
understanding null-maps. In the original CMB-oriented analysis,
more than 20 different data splits were included.  In this paper
we focus on the deck-angle split alone, because it is the most 
stringent test for Galactic fields. Noise-weighted mean and standard
deviation values of the deck-split null-maps of all four fields, given in
Table~\ref{tab:patch_summary}, show that these maps are consistent with the
expected Gaussian distribution. The only significant excesses in the Q-band
G-2 null-map are two small-scale features, one toward the very
Galactic center, the other toward a compact object at Galactic
coordinates $(l,b)=(353\pdeg17,0\pdeg76)$ that is identified as
PCCS1 030 G353.17+00.76 in the \Planck\ Catalogue of Compact Sources
\citep{planck2011-1.10}.

The most likely explanation for these excesses is so-called 
temperature-to-polarization ($I$-to-$Q$/$U$) leakage. Each \QUIET\ MMIC
module contains four detector diodes, two measuring 
$Q$, two measuring $U$, as defined by the local
detector coordinate system \citep{quiet_instrument}.  Based on
sky-dips (i.e., elevation nods designed to monitor relative gain
variations) and lunar and Galactic observations, \citet{quiet2011}
found that the instantaneous temperature-to-polarization leakage for
the Q-band detectors was about 1\,\% in $Q$ and 0.2\,\%
in $U$.  Modulation by both sky and deck rotations
effectively suppresses this effect in final maps. The deck-angle null-test
shown in Figure~\ref{fig:compmaps_gcQ} therefore provides a very
strict upper limit on the net final leakage.

To quantify this effect more accurately, we compare the null-map
polarization amplitude with the \WMAP\ Q-band temperature map in
Figure~\ref{fig:leakcomp}, both smoothed to an effective resolution of
$40\arcm$ FWHM to reduce noise.  Comparing the two maps visually, the
qualitative correlation between the polarization excess and the
temperature signal is obvious. Furthermore, we find that the peak
value of the polarization amplitude in the null map at the Galactic
center is about 40$\,\mu\textrm{K}$, while the corresponding peak
temperature amplitude is 60\,mK. Thus, the net $I$-to-$Q$/$U$ leakage
is about 0.07\% in the deck-split null map. In terms of total net
polarization amplitude, this deck-split leakage corresponds to less
than 4\,\% of the full polarization signal of the Galactic center
source.  Again, after averaging over all possible polarization
detector angles, these numbers will be significantly lower in the
final maps.

\subsection{Comparison with \Planck\ and \WMAP}
\label{sec:comparison}

The top two rows in Figure~\ref{fig:compmaps_gcQ} show the
\Planck\ and \WMAP\ maps, scaled to the \QUIET\ frequency as per
Equation~\ref{eq:freqscale}, and filtered with the \QUIET\ weight
operator, i.e., $\mathbf{F}_{\mathrm{Q}}\mathbf{m}_{\mathrm{P}}$ and
$\mathbf{F}_{\mathrm{Q}}\mathbf{m}_{\mathrm{W}}$. By removing the same
large-scale basis functions from each map, all three can be directly
compared without confusion from poorly constrained large-scale
modes. A quantitative comparison between the filtered Q-band \QUIET,
\WMAP\, and \Planck\ maps is given in Table~\ref{tab:patch_summary} in
the form of best-fit linear regression slopes \citep{petrolini:2014};
corresponding W-band results are not provided, due to the very low
signal-to-noise ratio of the \WMAP\ W-band sky map and
non-availability of the \Planck\ 100$\,$GHz map.

\begin{figure}[t]
\centering
\subfigure{\includegraphics[width=0.5\linewidth, clip=true, trim=0.3in 0in 0in 0.1in]{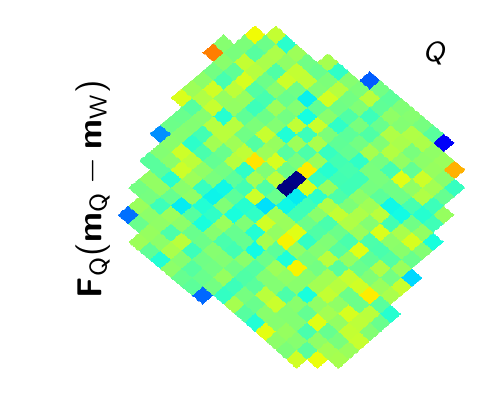}}\hspace{-0.1in}
\subfigure{\includegraphics[width=0.5\linewidth, clip=true, trim=0.3in 0in 0in 0.1in]{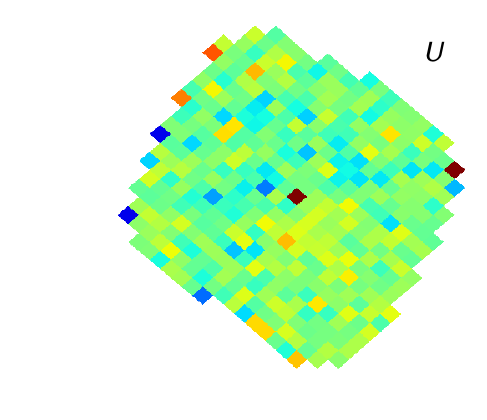}}

\vspace{-0.35in}
\subfigure{\includegraphics[width=0.5\linewidth, clip=true, trim=0.3in 0in 0in 0.1in]{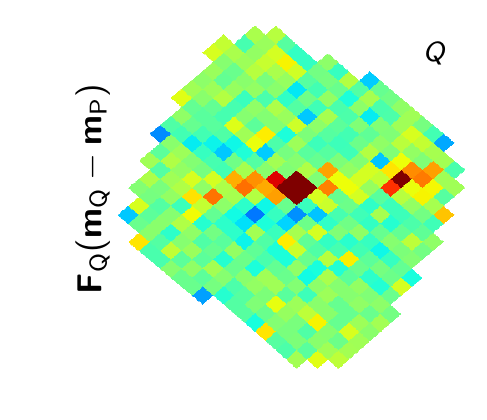}}\hspace{-0.1in}
\subfigure{\includegraphics[width=0.5\linewidth, clip=true, trim=0.3in 0in 0in 0.1in]{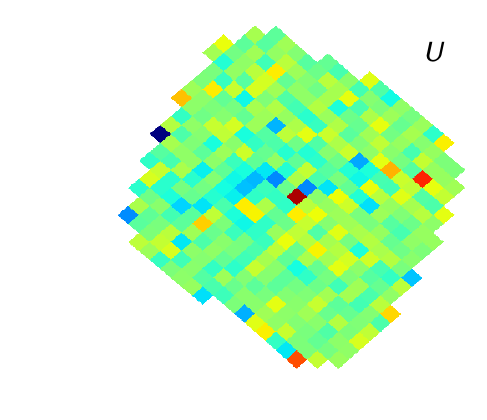}}

\vspace{-0.35in}
\subfigure{\includegraphics[width=0.5\linewidth, clip=true, trim=0.3in 0in 0in 0.1in]{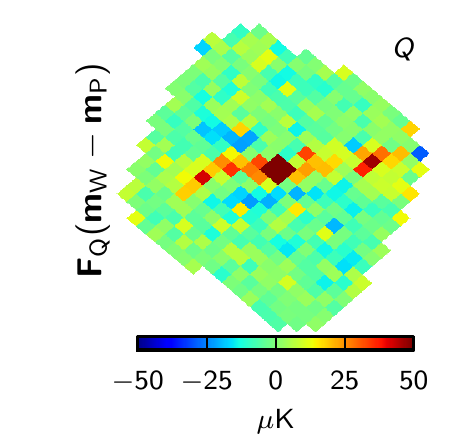}}\hspace{-0.1in}
\subfigure{\includegraphics[width=0.5\linewidth, clip=true, trim=0.3in 0in 0in 0.1in]{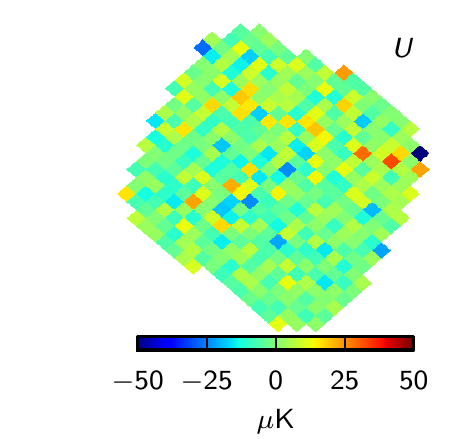}}
\vspace{-0.1in}
\caption{Pairwise differences of Q-band (43\,GHz) maps of the \mbox{G-2} (Galactic center) field, downgraded to HEALPix $N_{side}=64$ to suppress noise. All maps are weighted using the \QUIET\ weight operator $\mathbf{F}_{\mathrm{Q}}$, defined in \S~\ref{sec:coadd}, retaining only the small-scale modes observed by \QUIET\ in the differenced maps. Rows show, from top to bottom, \QUIET$-$\WMAP, \QUIET$-$\Planck, and \WMAP$-$\Planck. Columns show Stokes $Q$ and Stokes $U$.}
\label{fig:diffmaps_compare}
\end{figure}

\begin{figure}[t]
\centering
\includegraphics[width=\linewidth]{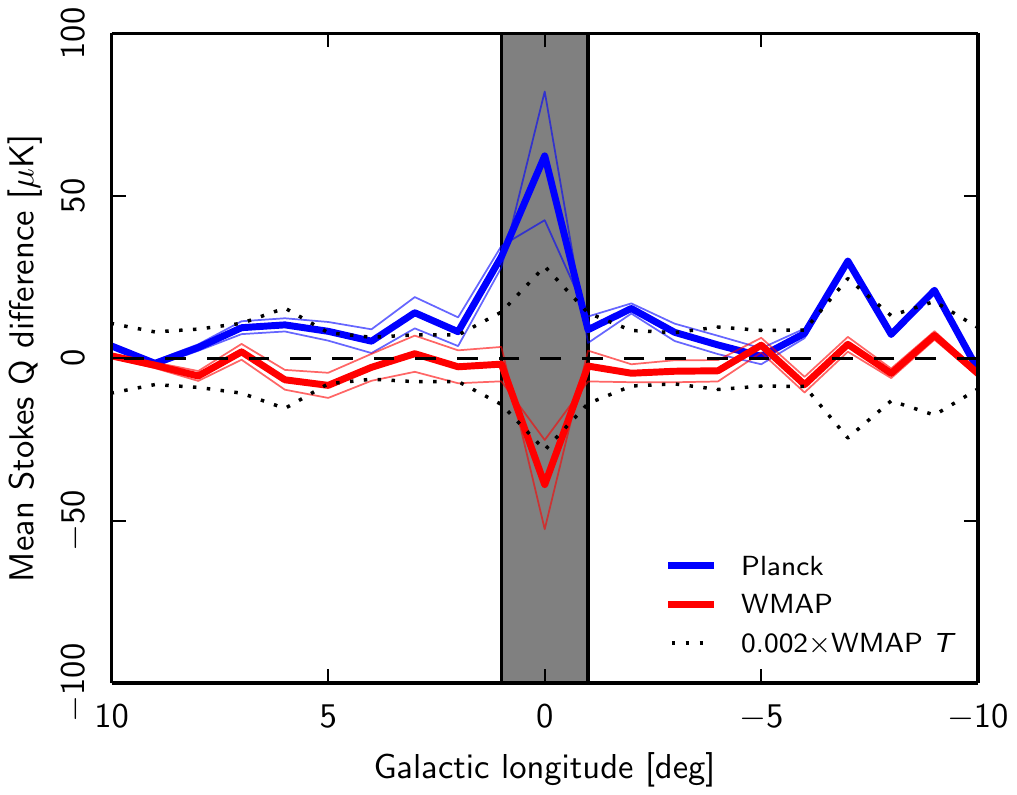}
\caption{Latitude-averaged difference between \QUIET\ and \WMAP\ (red)
  and between \QUIET\ and \Planck\ (blue) Q-band (43\,GHz) maps of
  field \mbox{G-2} (Galactic center), in Stokes $Q$, evaluated over a
  latitude band around the Galactic plane of $|b| \leq 1\pdeg5$. All
  maps are filtered with the \QUIET\ weight operator
  $\mathbf{F}_{\mathrm{Q}}$, defined in \S~\ref{sec:coadd}, retaining
  only the small-scale modes that are well constrained by \QUIET. The
  colored regions indicate the absolute \QUIET\ calibration
  uncertainty of $\pm$6$\,$\%. The dashed lines show the 
  latitude-band-averaged \WMAP\ Q-band temperature amplitude scaled by 
  a factor of $0.002$, providing a rough template of 0.2\,\%
  temperature-to-polarization leakage. The gray region marks an area
  in longitude $\pm 1\deg$ around the Galactic center within which all
  results are dominated by uncertainties in the foreground spectral
  index.}
\label{fig:long_diff}
\end{figure}

Several interesting features may be seen in these maps. First and
foremost, it is reassuring that all three experiments observe
the same broad structures, namely the positive Galactic plane and bright 
negative Galactic center in Stokes $Q$, and
the negative 'wings' in Stokes $U$. However, there
are noticeable differences as well, the most important of
which is the much lower noise of the
\QUIET\ maps.  While only broad features may be identified 
in the \Planck\ and \WMAP\ maps, even beam-sized features
may be picked out by eye in the \QUIET\ map.

A second important but more subtle difference is the apparent
amplitude of the Galactic plane in Stokes $Q$. Both \QUIET\ and
\WMAP\ appear to be slightly brighter than \Planck\ in the Galactic plane. This effect is
visually more striking in Figure~\ref{fig:diffmaps_compare}, where we
show pairwise difference maps between all three experiments, all
repixelized on a $55\arcm\times55\arcm$ grid ($N_{\textrm{side}}=64$)
in order to suppress instrumental noise. From top to bottom, the three
rows show \QUIET$-$\WMAP, \QUIET$-$\Planck, and \WMAP$-$\Planck. While
\QUIET\ and \WMAP\ are consistent, \Planck\ clearly shows a
deficit in the Galactic plane compared to the other two experiments.

\begin{figure*}
\centering
\mbox{\includegraphics[width=0.49\linewidth, clip=true, trim=0.4in 0in 0in 0.05in]{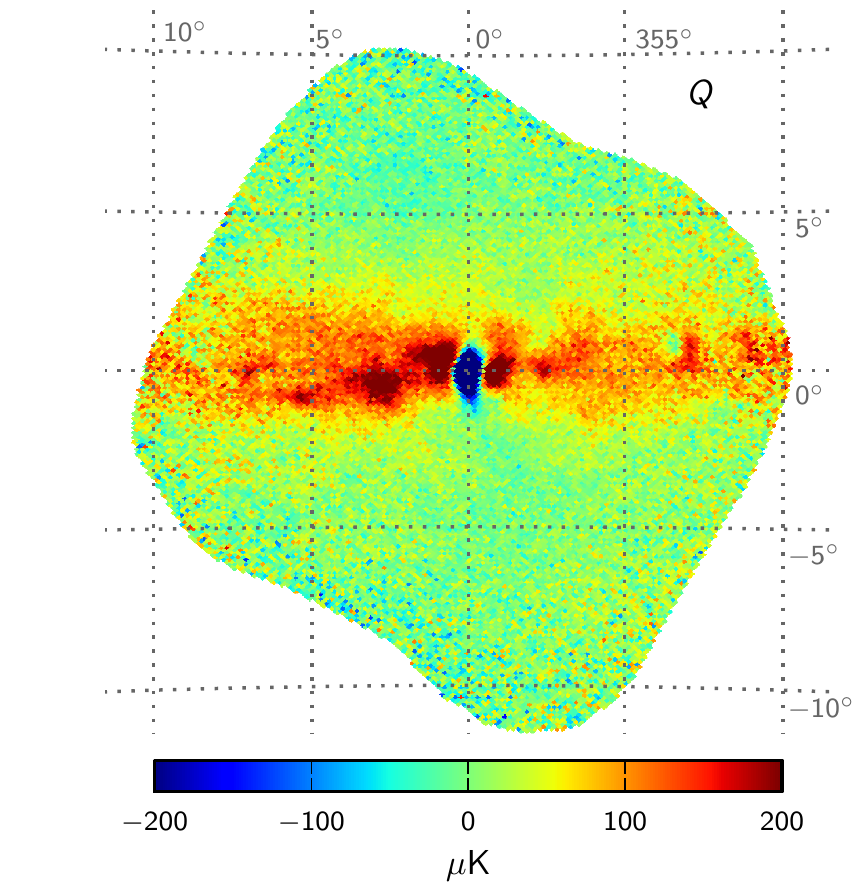}}
\mbox{\includegraphics[width=0.49\linewidth, clip=true, trim=0.4in 0in 0in 0.05in]{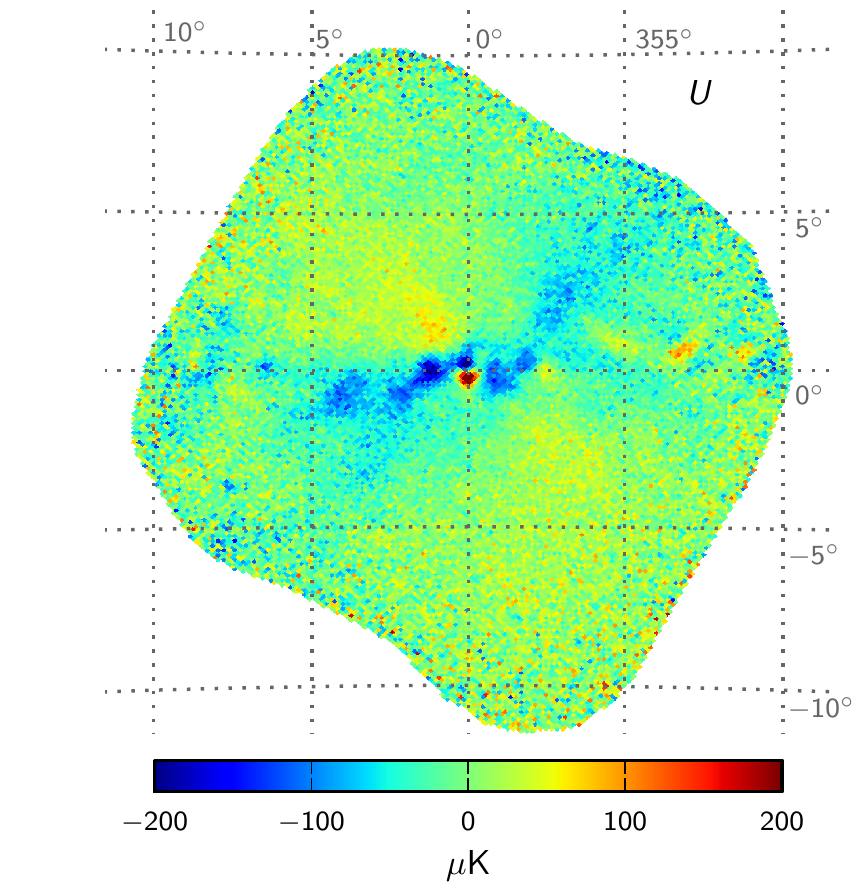}}
\caption{Final co-added \QUIET$+$\WMAP\, Q-band (43\,GHz) maps of field G-2 (centered at Galactic coordinates $(l,b)=(0\deg,0\deg)$). Using the weight operators defined in \S~\ref{sec:coadd}, these are expressed as $\mathbf{F}_{\mathrm{Q}}\mathbf{m}_{\mathrm{Q}} + \mathbf{F}_{\mathrm{W}}\mathbf{m}_{\mathrm{W}}$. Left and right panels show Stokes $Q$ and $U$, respectively. Grid cell side length is $5\deg$.}
\label{fig:finalmaps_gc_qband}
\end{figure*}

Another illustration of the same effect is provided in
Figure~\ref{fig:long_diff}, which shows the Stokes $Q$ differences
with respect to \QUIET\ of \Planck\ (blue curves) and \WMAP\ (red
curves) as a function of Galactic longitude, averaged over a
$|b|\le1\pdeg5$ latitude band around the Galactic plane. This
corresponds to the three center-most rows of pixels in
Figure~\ref{fig:diffmaps_compare}, although the evaluation was in fact
performed on the full-resolution maps. For comparison, we have also
plotted the corresponding mean of the \WMAP\ Q-band temperature map as
dotted lines, but scaled with a factor of 0.002. This signal would be
produced by temperature-to-polarization leakage of 0.2\,\%.  The thin
colored lines correspond to multiplying the \QUIET\ map by $\pm6\,$\%,
which is the \QUIET\ $1\sigma$ uncertainty in absolute calibration
\citep{quiet2011}. Finally, we have marked the Galactic center region
within $1\deg$ (i.e., the bright blue object in the Galactic center
seen in Figure~\ref{fig:compmaps_gcQ}) with a gray band. As noted in
\S~\ref{sec:method}, we assume a synchrotron spectral index of
$\beta_{\textrm{G-2}}=-3.00$ for this field, as estimated by
\citet{fuskeland2014}. However, the bright center object is not
included in this estimate, as its properties appear qualitatively
different from the surrounding environment. In addition, the amplitude
of this object is very large, reaching $2600\,\mu\textrm{K}$ at
$30\arcm$ scales, and any difference maps like those considered here
are therefore very sensitive to spectral index uncertainties. For
example, if the spectral index is $-2.7$ rather than $-3.0$, the
effective difference between \QUIET\ and \WMAP\ would be
$-40\,\mu\textrm{K}$, similar to what is seen in
Figure~\ref{fig:long_diff}. For now, we therefore exclude the central
$|l|\le1\deg$ region from our discussion, noting that further spectral
index estimation work is required before conclusions can be drawn for
this object.

Focusing on the remaining longitude region with $|l|~>~1\deg$ in
Figure~\ref{fig:long_diff}, we see again the good agreement between
\QUIET\ and \WMAP.  The thick red line fluctuates around zero with an
overall mean and standard deviation of $-1\pm3\,\mu\textrm{K}$. 
  In \S~\ref{sub:quietmaps} we derived an upper limit on the
  polarization amplitude uncertainty from temperature-to-polarization
  leakage of $\sim 4\,$\% in the \QUIET\ deck-angle null-map, which is
  further suppressed in the full map due to cross-linking. The
  uncertainty in the \QUIET\ maps due to such leakage is therefore well within
  the level indicated by the $\pm 6\,$\% uncertainty shown in the
  plot. In sum, we do not find any evidence for significant residual
temperature-to-polarization leakage in the full signal maps, either in
\QUIET\ or \WMAP.

For \Planck, we see a systematic positive excess, consistent with
Figure~\ref{fig:diffmaps_compare}. In principle, this excess could be
due to several different effects. However, its magnitude and spatial
pattern suggest temperature-to-polarization leakage, as discussed
extensively in \citet{planck2014-a03}. 
Compare the \QUIET$-$\Planck\ residual to the upper dotted line,
which indicates the mean \WMAP\ Q-band temperature signal as a
function of longitude, scaled by a factor of 0.002. Note in particular
the correlated structures between longitudes $l\approx-5\deg$ and
$-10\deg$.  The dotted line provides an approximate template of the
temperature-to-leakage signal. However, this template is only
approximate, since the detailed leakage pattern will additionally
depend on the \Planck\ scanning strategy and detector orientation,
effects not accounted for here. Note, though, that such scanning
strategy modulation can only reduce the correlation between the
observed residual and the simple leakage model, never enhance
it. Additionally, these features can not be due to intrinsic spectral
index variations (unlike the Galactic center), because \WMAP\ agree
very well with \QUIET\ in this region despite having a longer relative
frequency lever arm than \Planck.

Residuals at this level are consistent with the uncertainties
for temperature-to-polarization leakage given in
\citet{planck2014-a03}.  \citet{planck2014-a12} gives explicit leakage
corrections based on detailed astrophysical foreground modeling;
however, those models are necessarily associated with significant
uncertainties, as they depend sensitively on both instrument and
foreground models, and in particular on the assumed bandpass
properties of the instrument.  The \Planck\ 44$\,$GHz polarization map
fails a few null-tests \citep{planck2014-a01}, and is therefore not
used in the \Planck\ 2015 CMB likelihood \citep{planck2014-a13}. Thus,
the new high-sensitivity \QUIET\ maps presented in this paper
represent a unique opportunity to improve the \Planck\ leakage model
in future analyses, by virtue of providing a clean and direct
reference in the region of the sky with the highest signal-to-noise ratio.

\subsection{Co-added sky maps}

Given the qualitative differences between the \Planck\ and \QUIET\ maps reported above, we co-add the \QUIET\ and \WMAP\ maps in the current set of released maps, but not the \Planck\ maps. Co-addition with \Planck, and any other experiment observing the same field, can always be performed later.  Similarly, we co-add with the \WMAP\ W1, W2, and W3 differencing assembly maps at W-band, but not W4, since this
particular differencing assembly is known to have significantly worse
noise properties than the other three channels \citep{wmap9}.

Figure~\ref{fig:finalmaps_gc_qband} shows the final co-added Q-band
G-2 map in Stokes $Q$ and $U$. Comparing this to the filtered maps
shown in Figure~\ref{fig:compmaps_gcQ}, the most noticeable
differences appear, as expected, near the edges of the field, where
the \QUIET\ signal-to-noise ratio deteriorates. In the full co-added
map, the Stokes $Q$ amplitude remains high along the Galactic plane to
the very edge, where it tapers off in the filtered
version. Corresponding maps for the other data sets (Q-band G-1 and
W-band G-1 and G-2) are shown in Figure~\ref{fig:finalmaps_app} in the
appendix.

\section{Astrophysical implications}
\label{sec:physics}

\begin{figure*}[t]
\centering
\mbox{\includegraphics[width=0.46\linewidth]{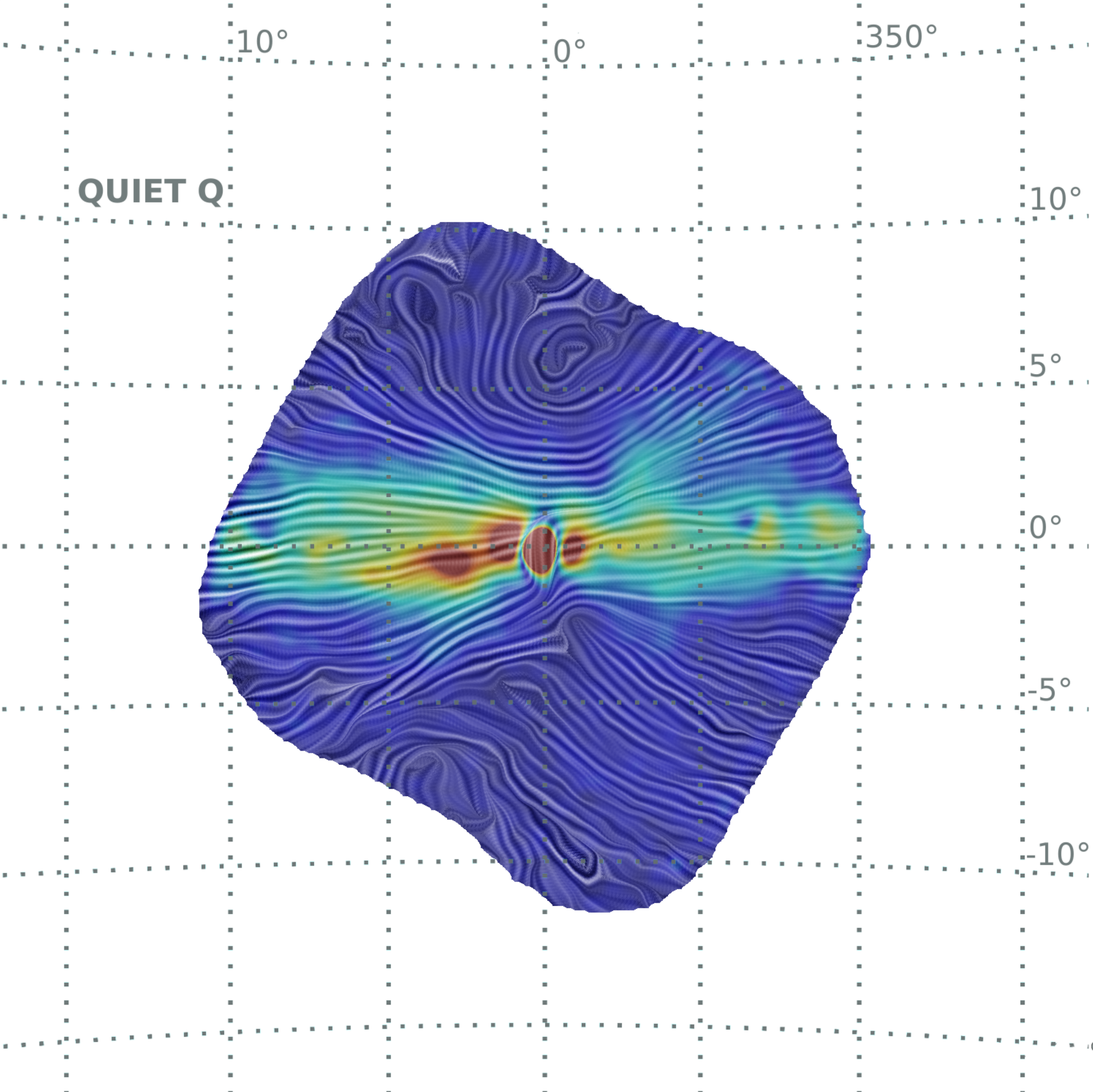}}\hspace*{1cm}
\mbox{\includegraphics[width=0.46\linewidth]{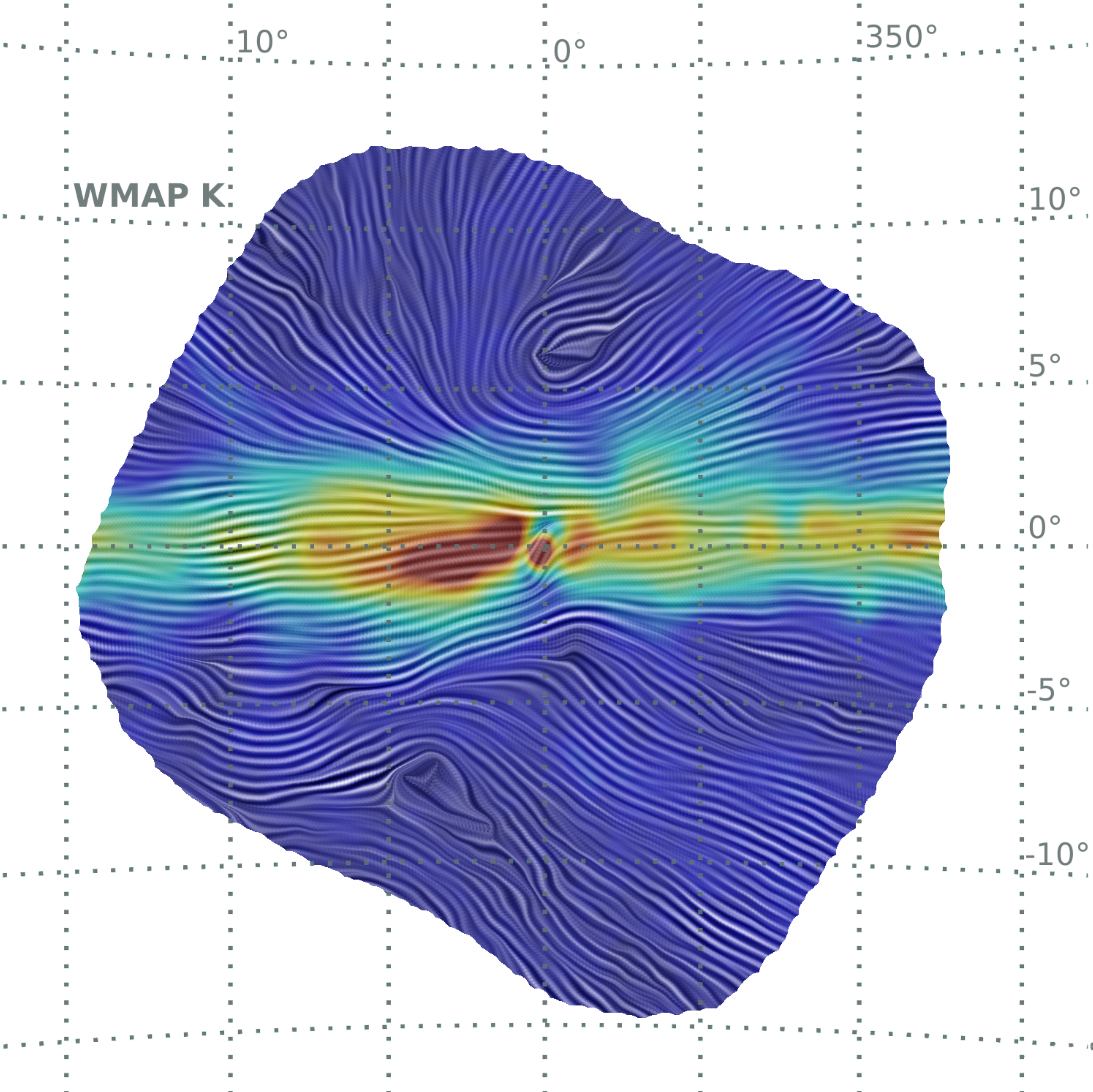}}
\hspace*{-0mm}\mbox{\includegraphics[width=0.30\linewidth]{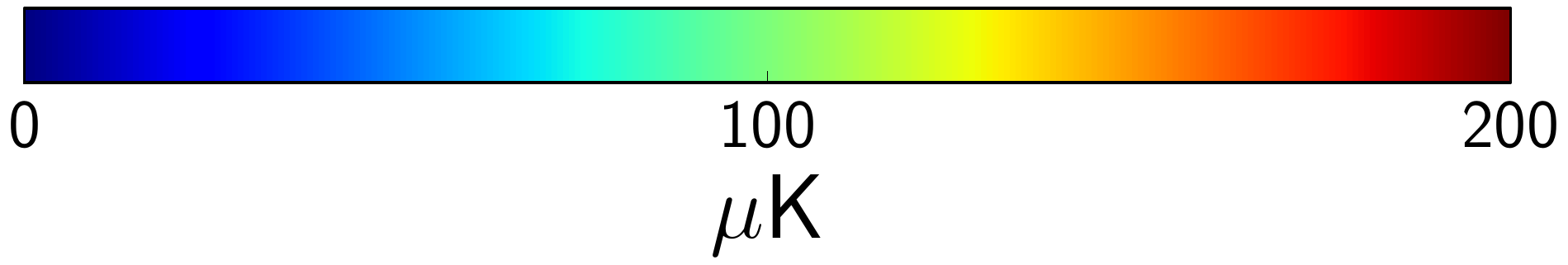}}
\hspace*{38mm}\mbox{\includegraphics[width=0.30\linewidth]{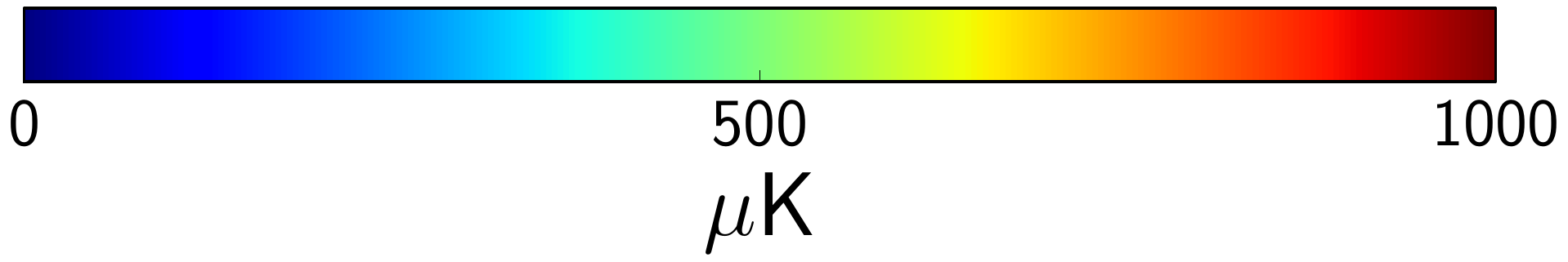}}

\vspace*{3mm}
\mbox{\includegraphics[width=0.46\linewidth]{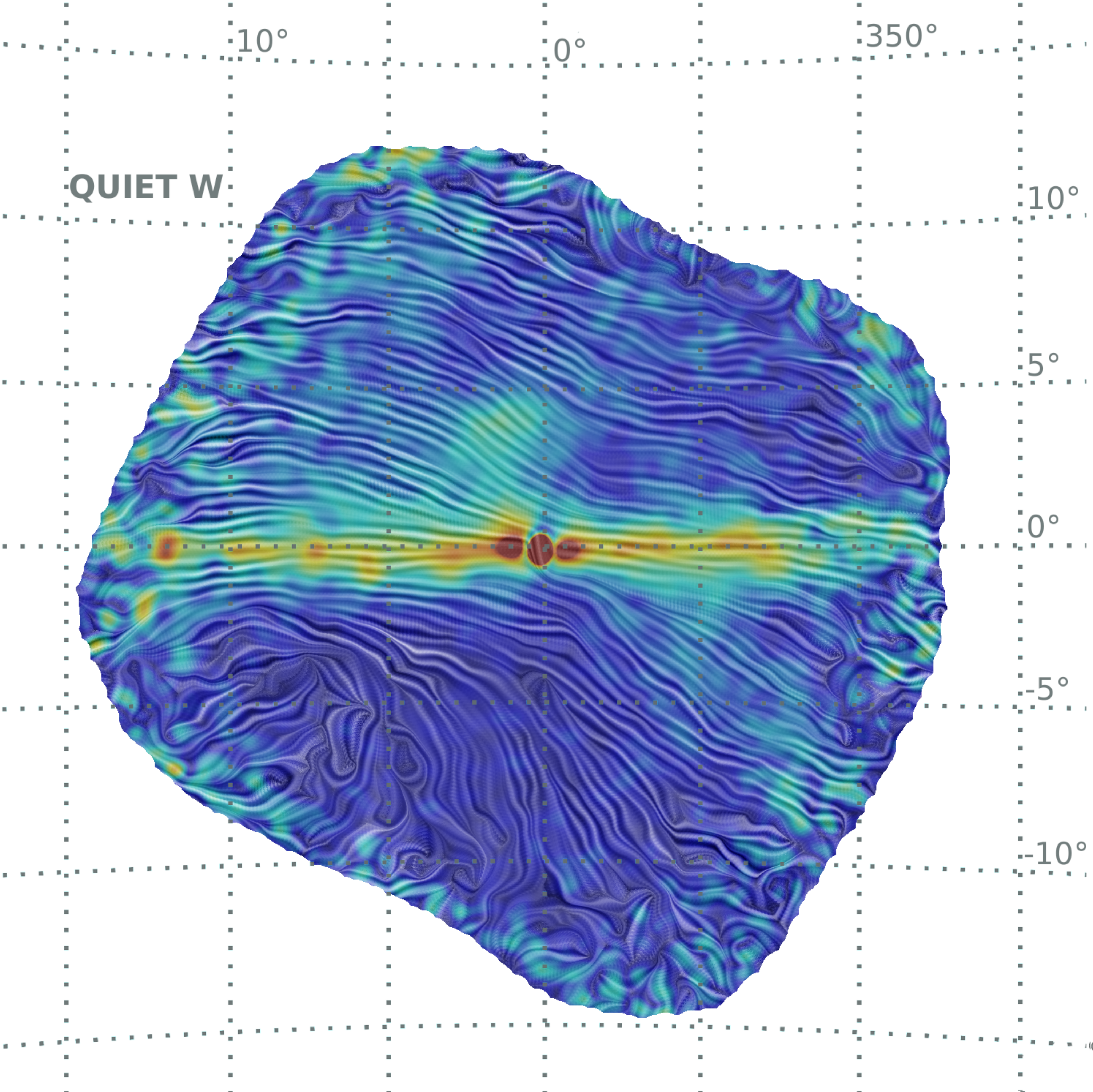}}\hspace*{1cm}
\mbox{\includegraphics[width=0.46\linewidth]{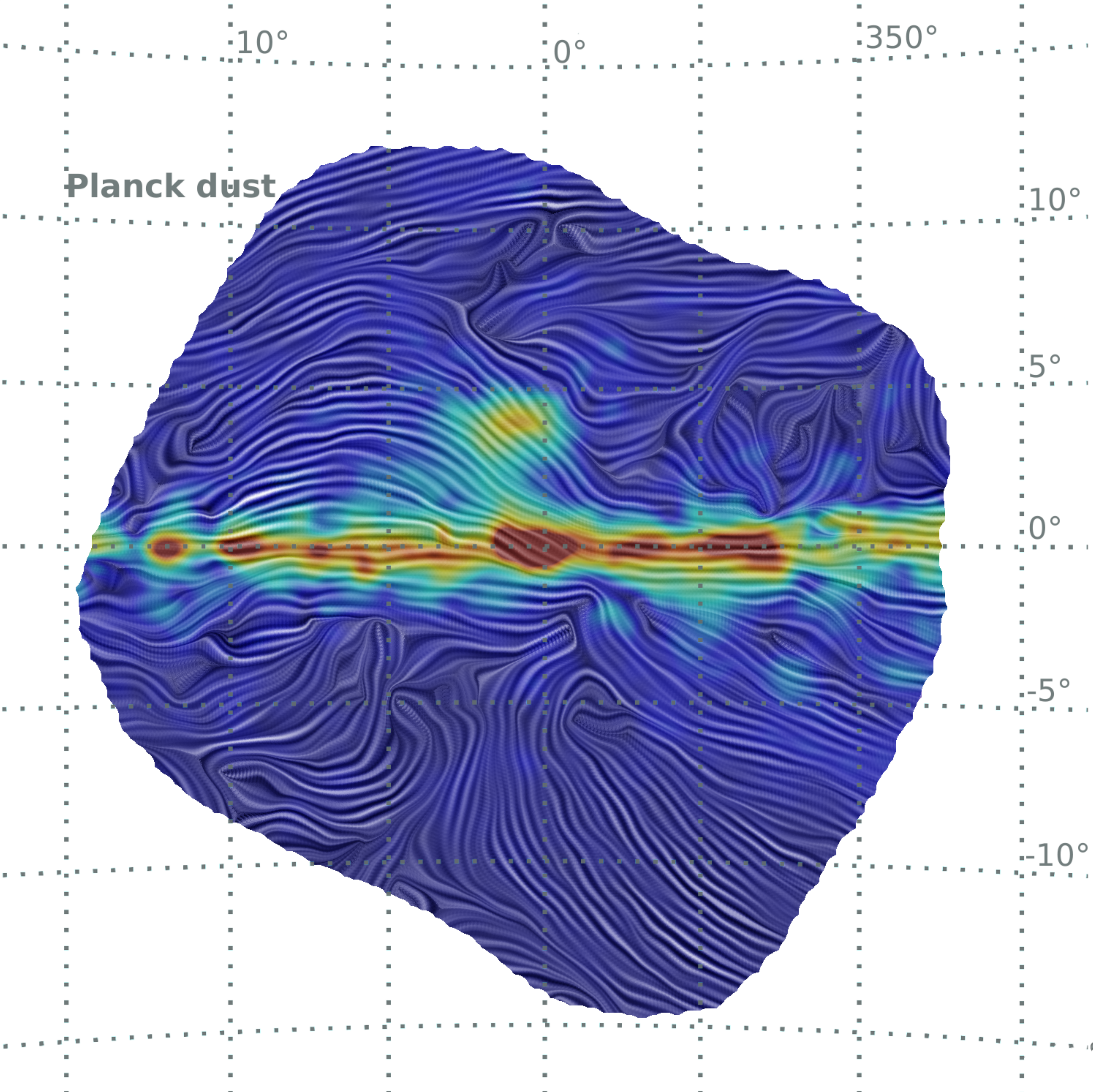}}
\hspace*{0mm}\mbox{\includegraphics[width=0.30\linewidth]{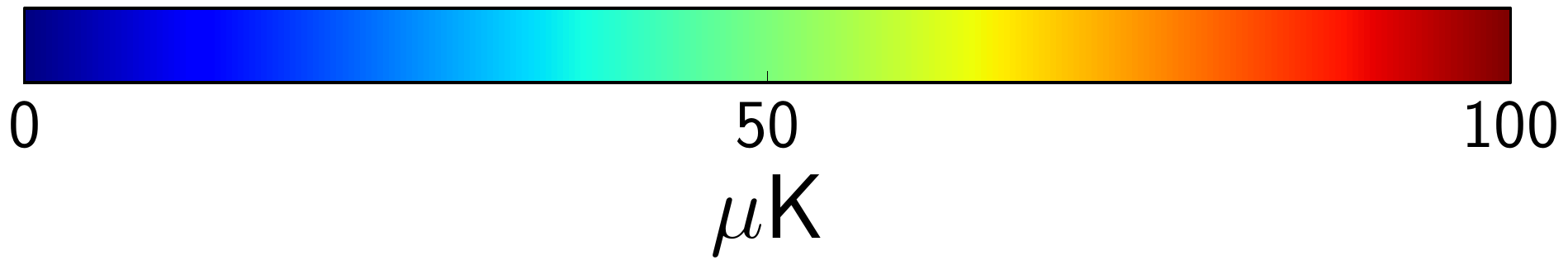}}
\hspace*{38mm}\mbox{\includegraphics[width=0.30\linewidth]{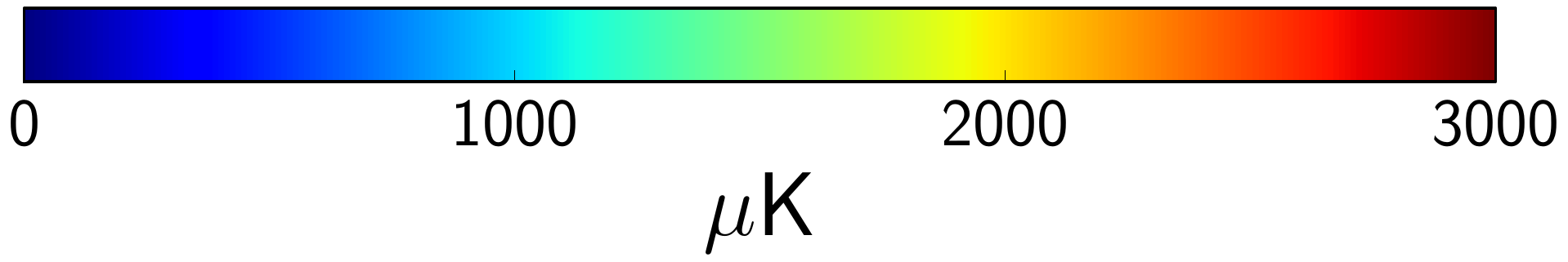}}
\caption{Comparison between the co-added \QUIET+\WMAP\ Q-band (43\,GHz, top left) and W-band (95\,GHz, bottom left) maps with the \WMAP\ K-band (23\,GHz, top right; \citealp{wmap9}) map and the \Planck\ thermal dust model (evaluated at 353\,GHz, bottom right; \citealp{planck2014-a12}), for field G-2 (centered on Galactic coordinates $(l,b)=(0\deg,0\deg)$). All plots are generated using the Line Integral Convolution algorithm \citep{cabral:1993}. The color scale indicates polarization amplitude, $P=\sqrt{Q^2+U^2}$, while the flow stripes indicate magnetic-field orientation, i.e., rotated 90$\deg$ with respect to the local polarization orientation.}
\label{fig:finalmaps_gcP}
\end{figure*}

\begin{figure*}
\centering
\mbox{\includegraphics[width=0.46\linewidth]{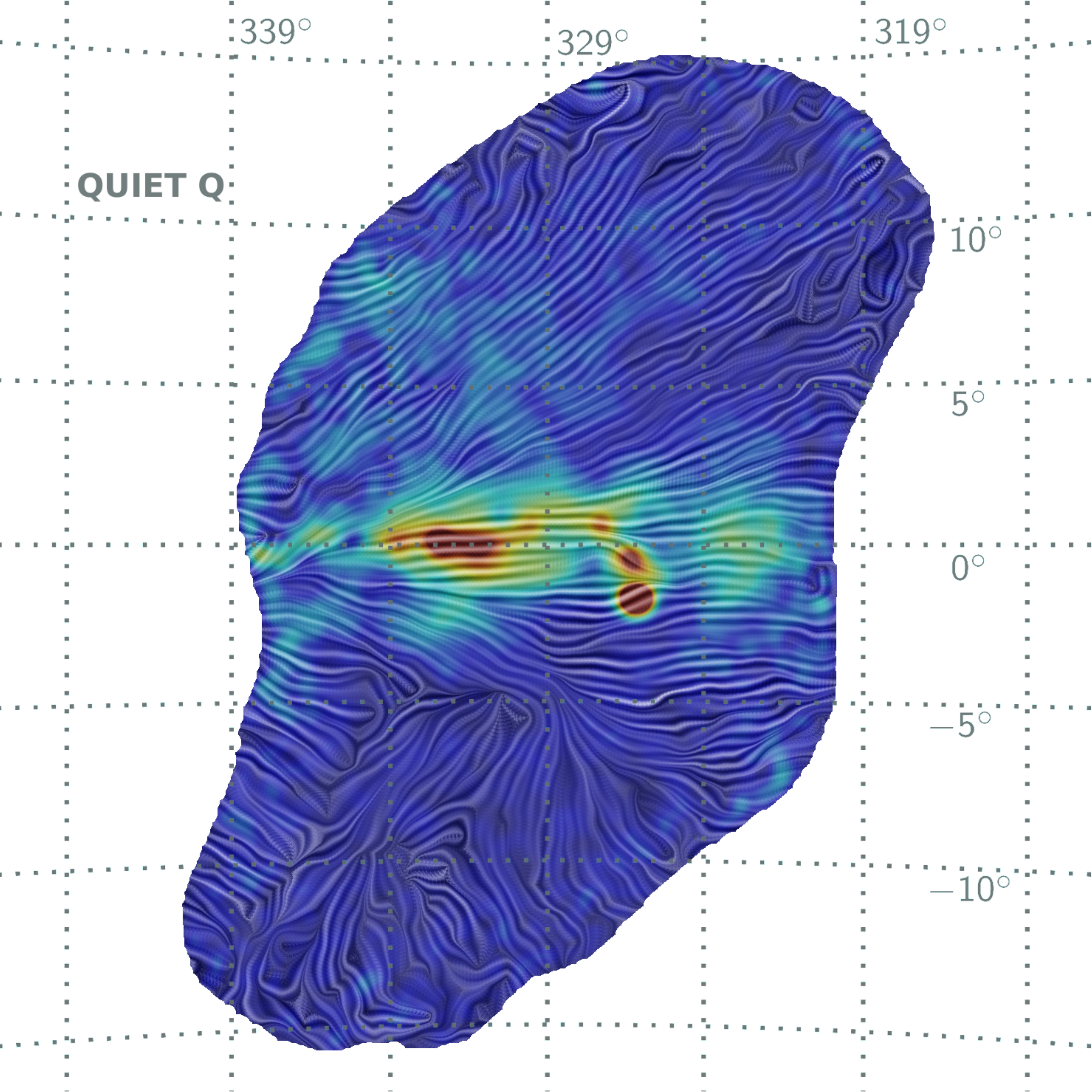}}\hspace*{1cm}
\mbox{\includegraphics[width=0.46\linewidth]{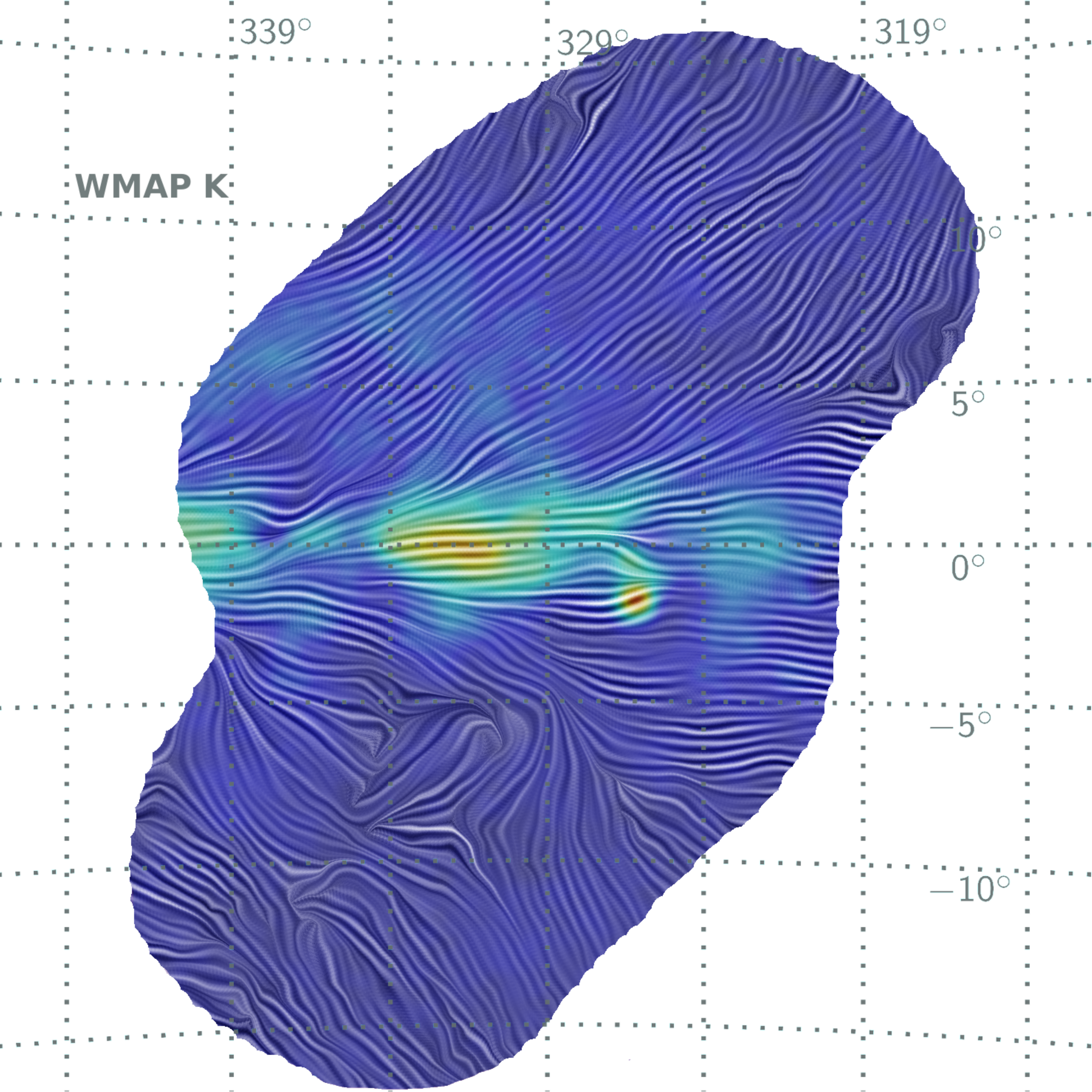}}
\hspace*{0mm}\mbox{\includegraphics[width=0.30\linewidth]{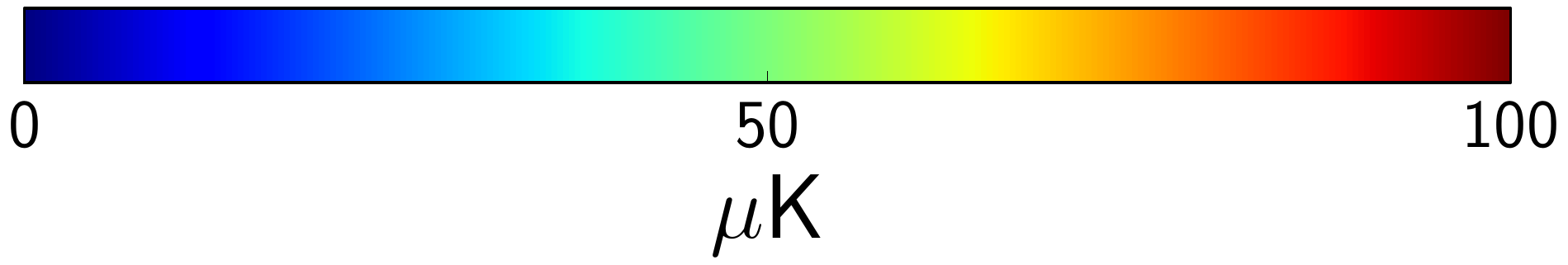}}
\hspace*{38mm}\mbox{\includegraphics[width=0.30\linewidth]{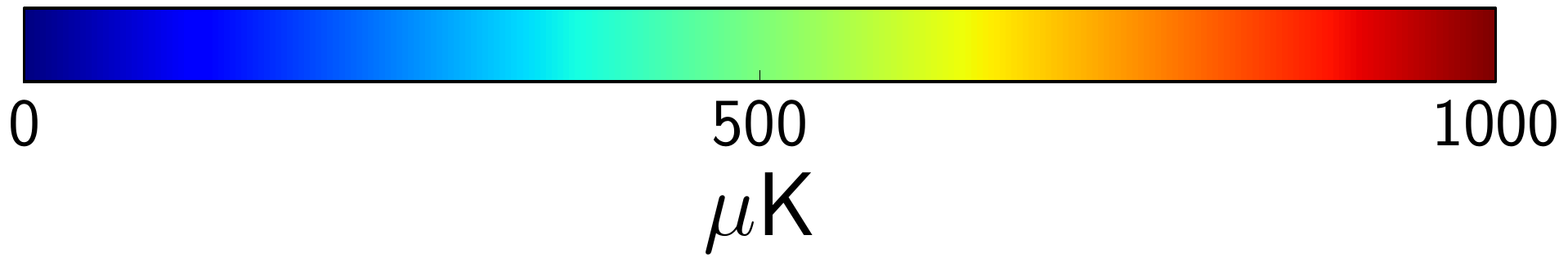}}

\vspace*{3mm}
\mbox{\includegraphics[width=0.46\linewidth]{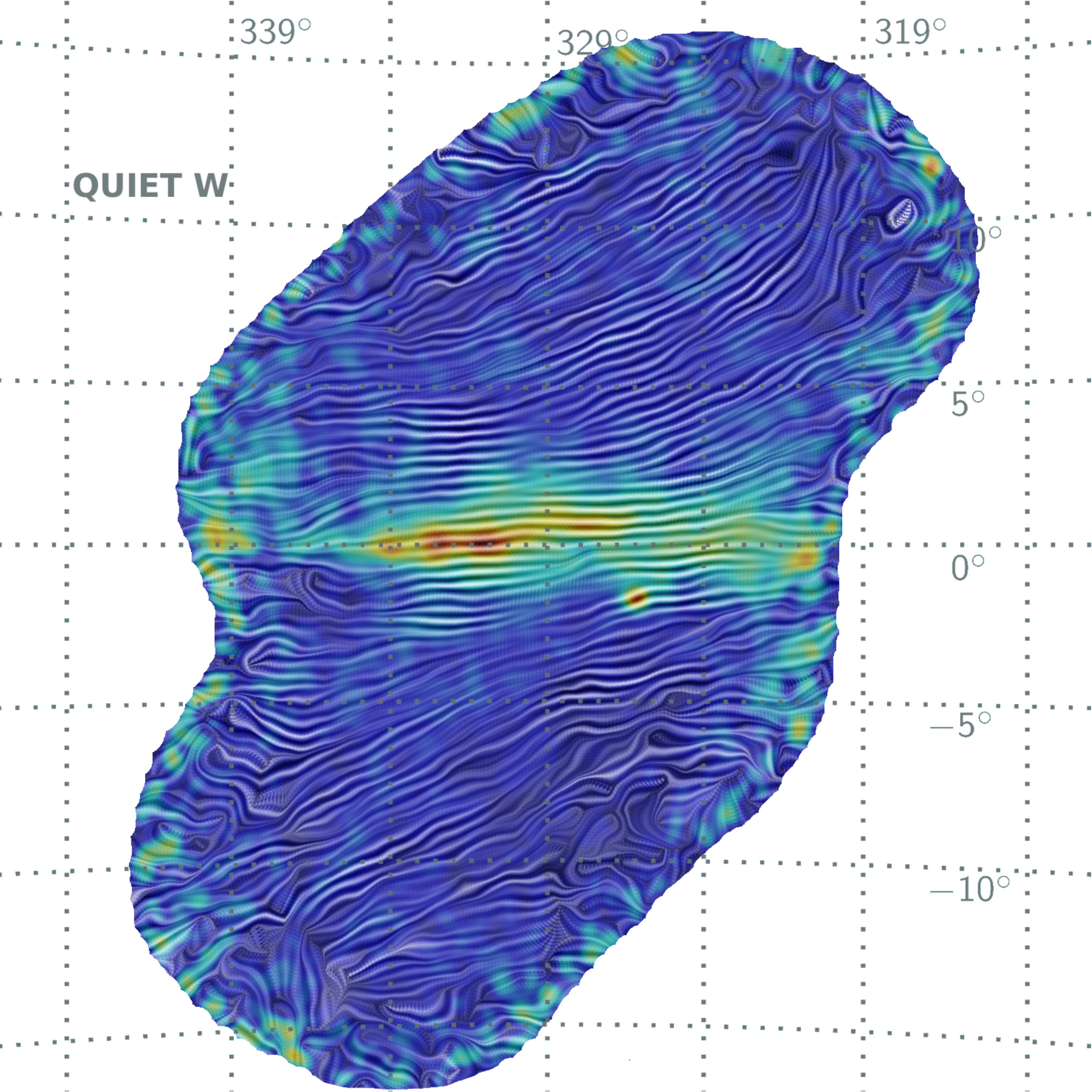}}\hspace*{1cm}
\mbox{\includegraphics[width=0.46\linewidth]{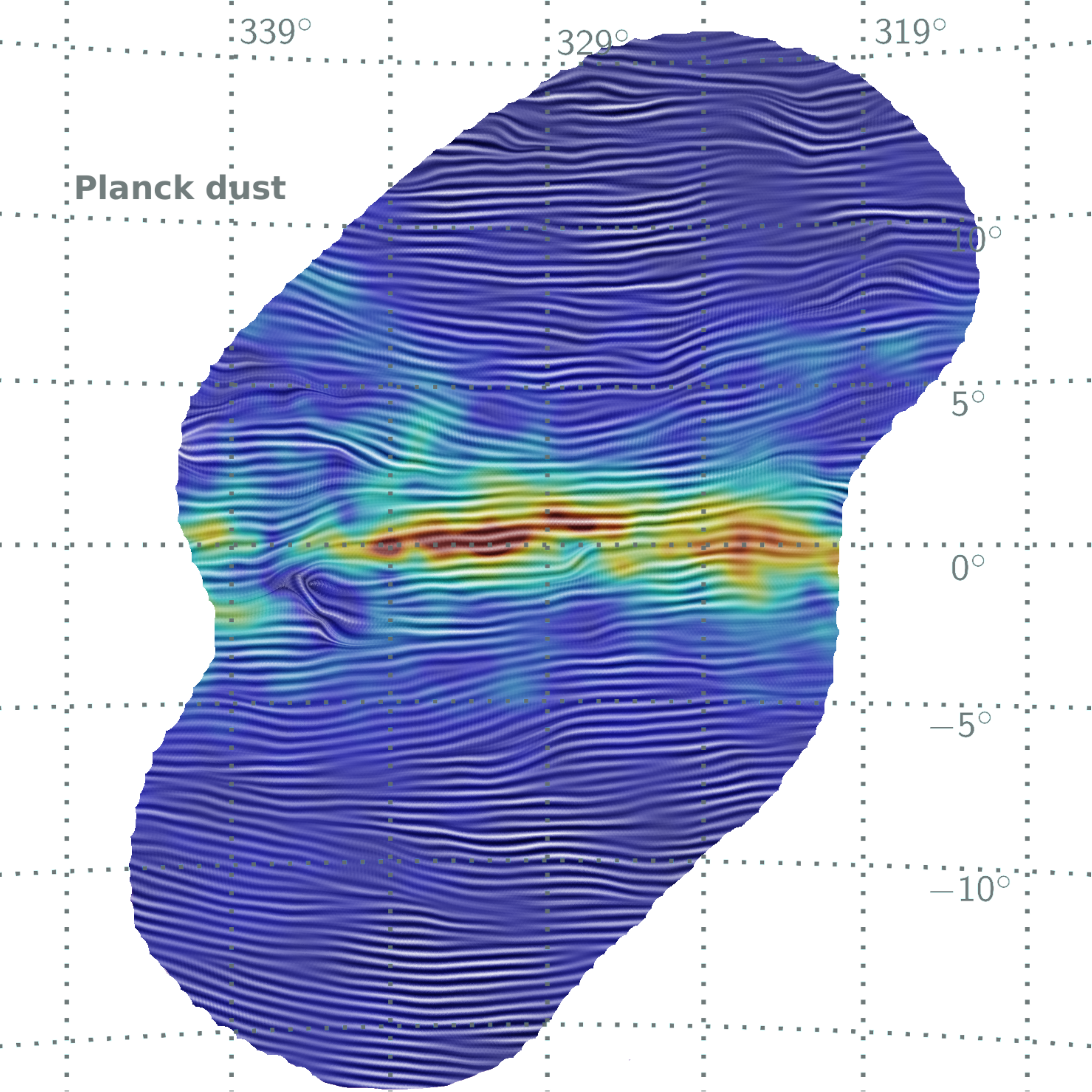}}
\hspace*{0mm}\mbox{\includegraphics[width=0.30\linewidth]{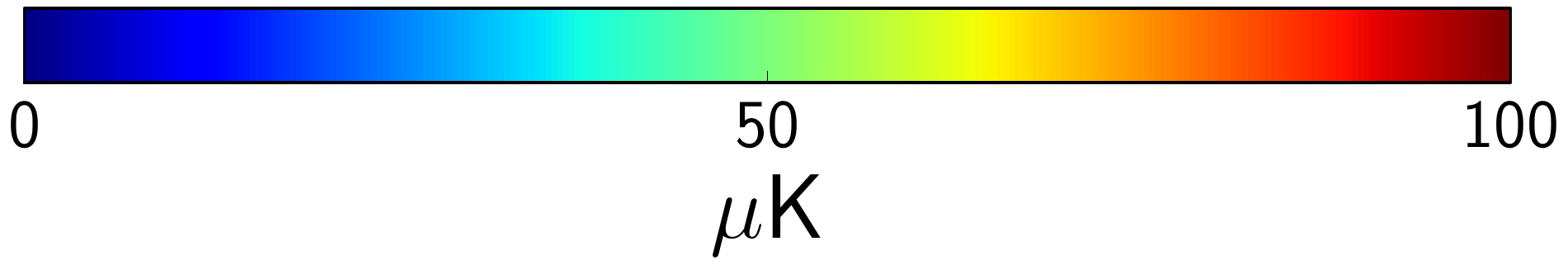}}
\hspace*{38mm}\mbox{\includegraphics[width=0.30\linewidth]{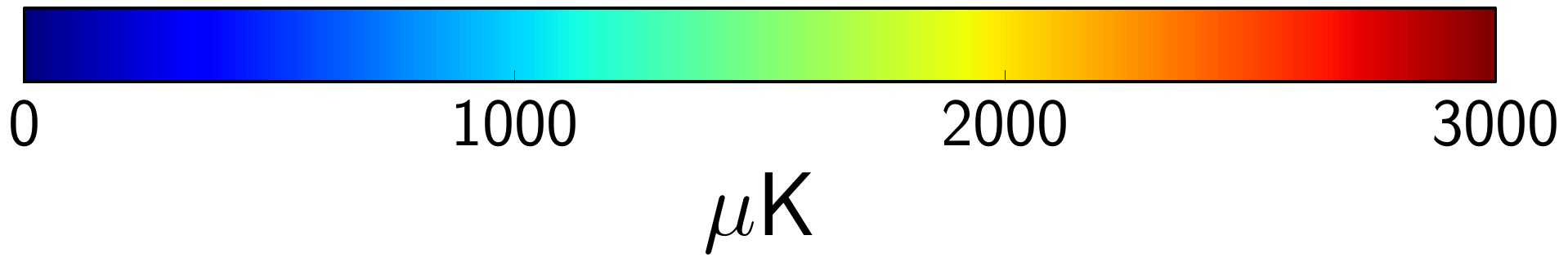}}
\caption{Same as Figure~\ref{fig:finalmaps_gcP}, but for G-1.} 
\label{fig:finalmaps_gbP}
\end{figure*}

The sky maps presented in \S~\ref{sec:maps} offer a fresh
view of astrophysical foregrounds at microwave frequencies.
In this section, we compare the
co-added \QUIET\ maps to \WMAP\ and \Planck, and estimate both
the polarized synchrotron spectral index and the Faraday rotation
measure in the Galactic center.

\subsection{Visual inspection of sky maps}
\label{sec:visual}

In Figures~\ref{fig:finalmaps_gcP} and \ref{fig:finalmaps_gbP} we
compare our maps with the \WMAP\ K-band map \citep{wmap9} and the
\Planck\ 2015 map of thermal dust polarization \citep{planck2014-a12}. All maps have been smoothed with a $30\arcm$ FWHM Gaussian kernel to reduce instrumental noise.  The map of thermal dust is evaluated at 353\,GHz, whereas the effective frequency of the K-band map is 22.4\,GHz for a synchrotron-like frequency spectrum
\citep{page:2003b}. The color scale indicates the polarization
amplitude, $P$, while the flow pattern
traces field lines rotated by $90\deg$ with respect to the measured EVPA (electric vector position angle), corresponding roughly to the magnetic field direction (this correspondence is exact under the assumption of optically thin synchrotron radiation).
All plots are 
generated using an implementation of the Line Integral Convolution algorithm
\citep{cabral:1993} called \texttt{Alice}, provided in the HEALPix package.

Figure~\ref{fig:finalmaps_gcP} shows the Galactic center field G-2.  The magnetic field structure in the \QUIET\ Q-band (top left panel) 
shows correlation with that in the \WMAP\ K-band map (top right panel), showing not only the high quality of the maps, but also 
indicating that the Q-band sky is dominated by synchrotron emission.  
According to the \Planck\ 2015 astrophysical baseline model summarized in Figure~51 of
\citet{planck2014-a12}, synchrotron emission should dominate over
thermal dust emission by about an order of magnitude at Q-band, and
our measurements are visually consistent with this picture.

\begin{figure}[t]
\centering
\mbox{\includegraphics[width=0.50\linewidth, clip=true, trim=0.7in 0in 0.2in 0.2in]{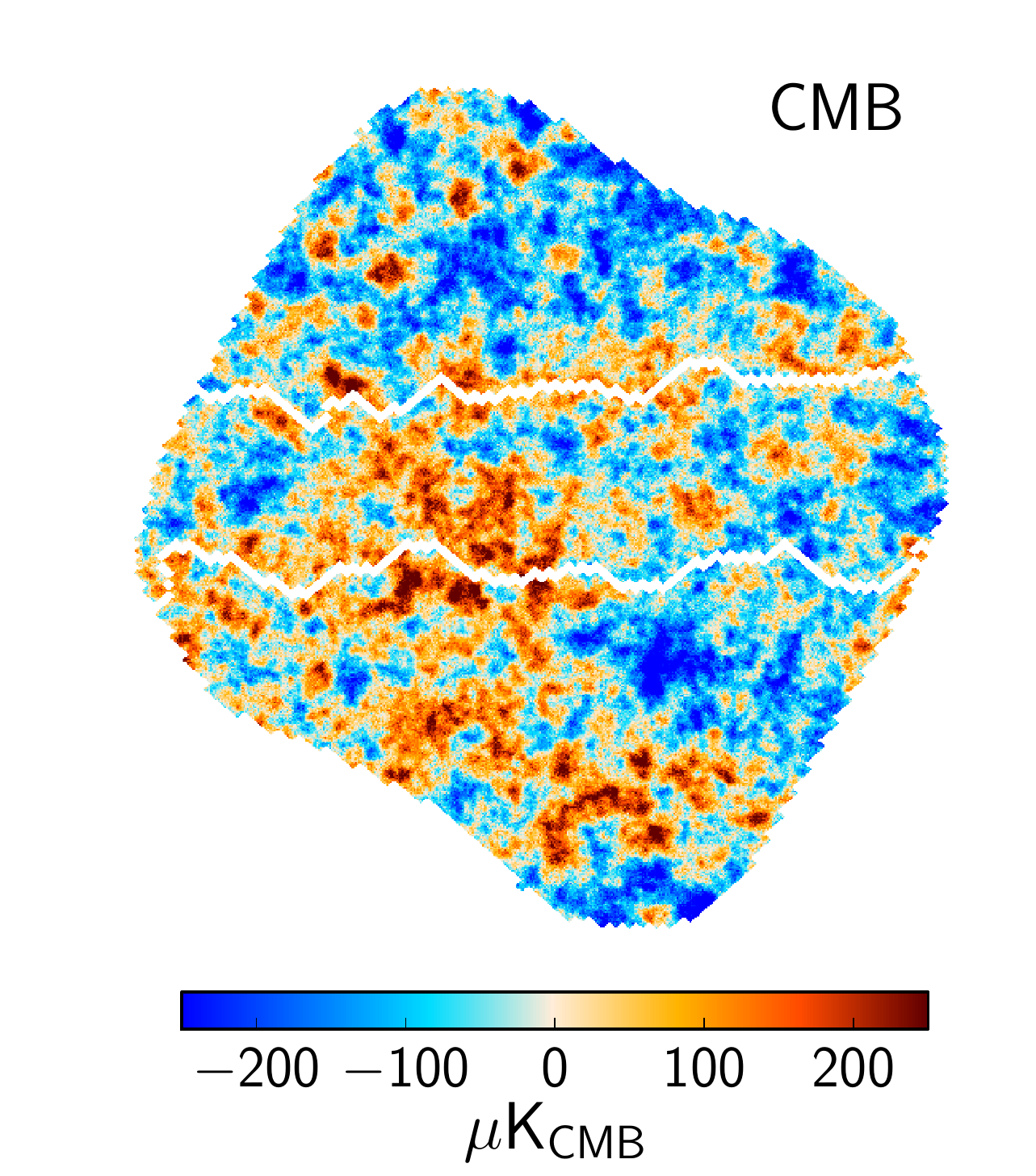}
      \includegraphics[width=0.50\linewidth, clip=true, trim=0.7in 0in 0.2in 0.2in]{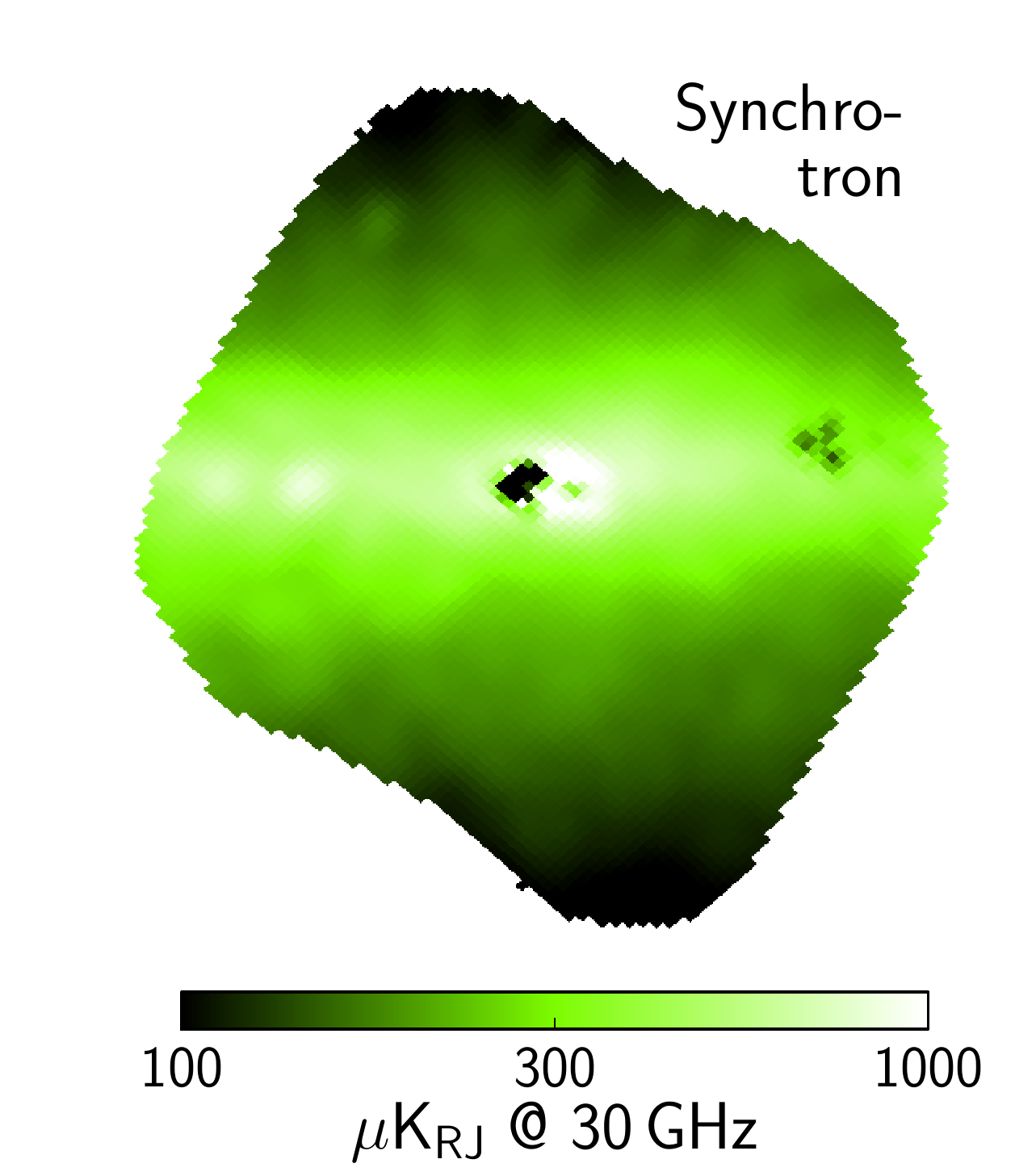}}
\mbox{\includegraphics[width=0.50\linewidth, clip=true, trim=0.7in 0in 0.2in 0.2in]{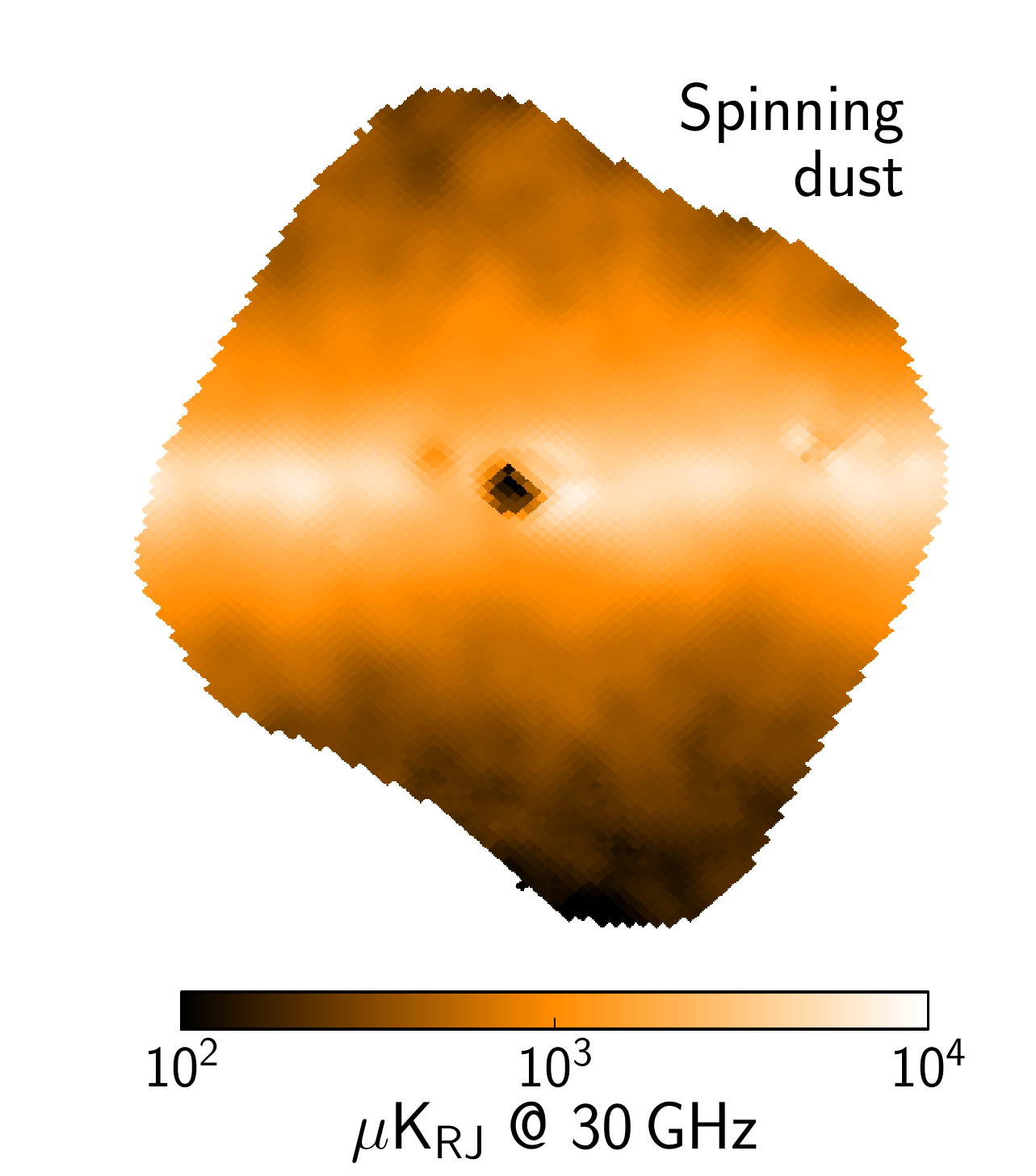}
      \includegraphics[width=0.50\linewidth, clip=true, trim=0.7in 0in 0.2in 0.2in]{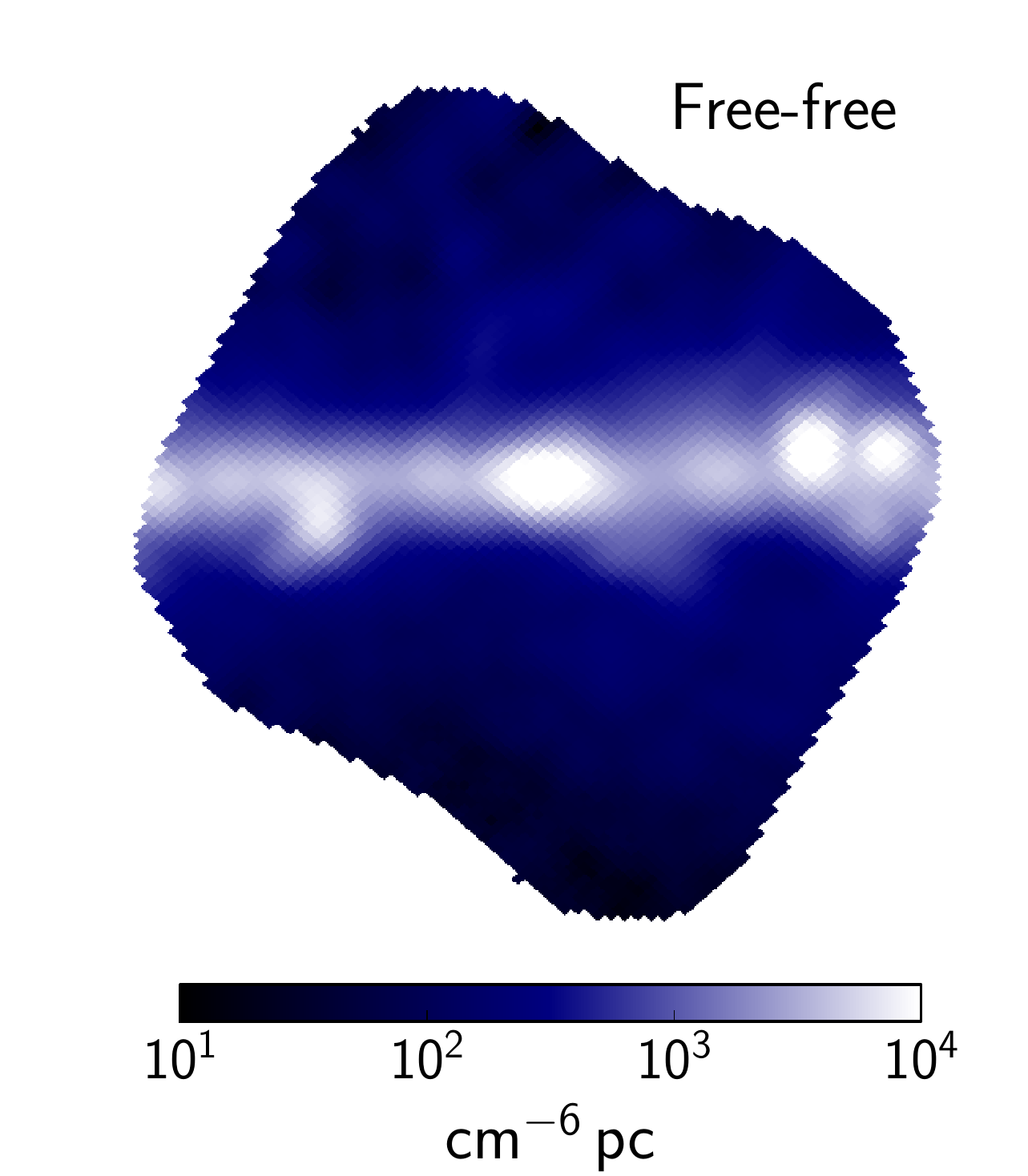}}
\mbox{\includegraphics[width=0.50\linewidth, clip=true, trim=0.7in 0in 0.2in 0.2in]{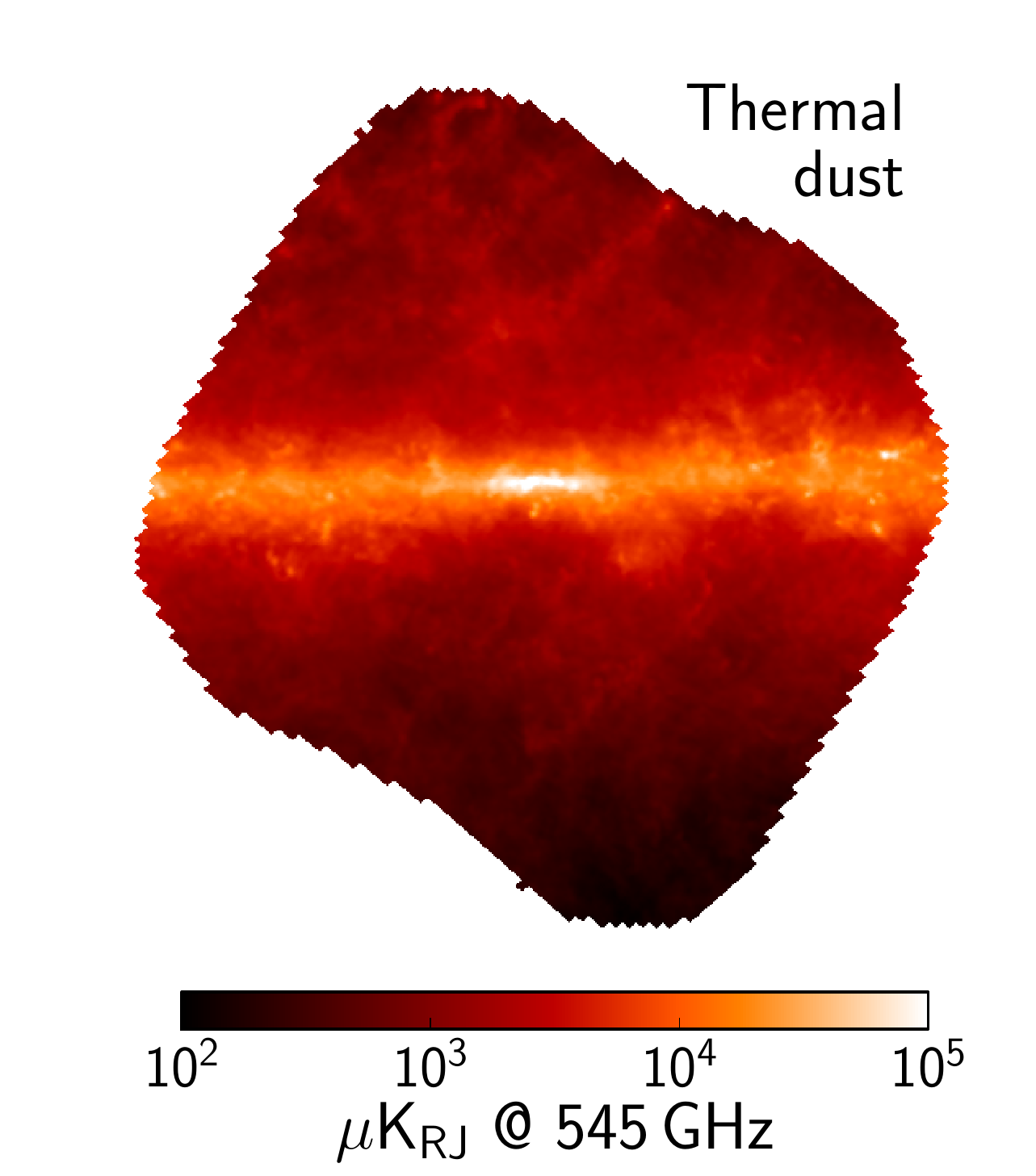}
      \includegraphics[width=0.50\linewidth, clip=true, trim=0.7in 0in 0.2in 0.2in]{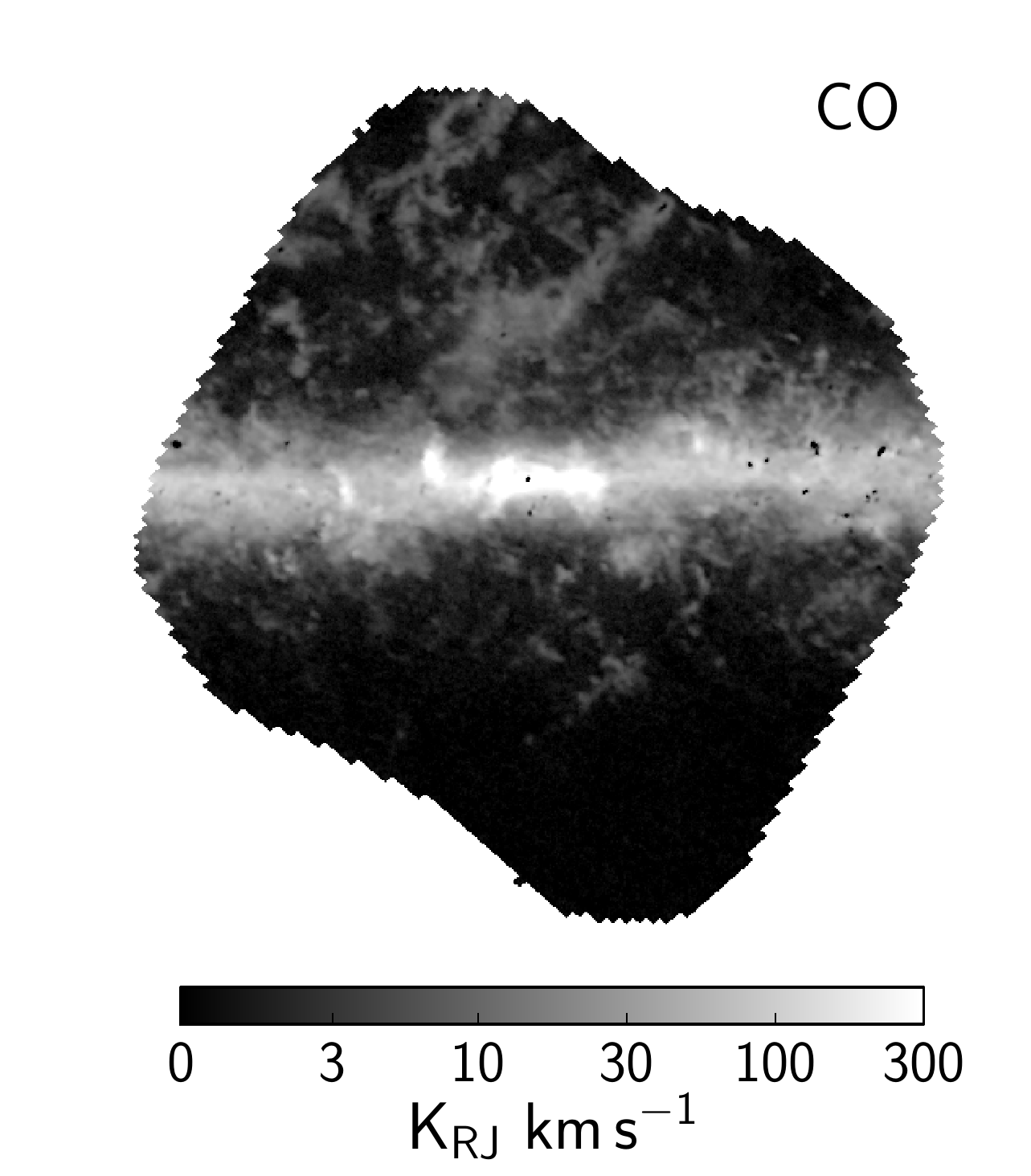}}
\caption{\Planck\ 2015 baseline astrophysical foreground reconstruction in
  intensity for the G-2 field, as estimated with \texttt{Commander}
  \citep{planck2014-a12}. From top to bottom and left to right, the
  panels show 1) CMB, 2) synchrotron, 3) spinning dust, 4) free-free,
  5) thermal dust, and 6) CO $J$=2$\rightarrow$1. The region inside
  the white boundary in the CMB map has been replaced with a
  constrained Gaussian realization as part of the \texttt{Commander} algorithm.}
\label{fig:commander_gc}
\end{figure}

\begin{figure}[t]
  \centering
\mbox{\includegraphics[width=0.50\linewidth, clip=true, trim=0.7in 0in 0.2in 0in]{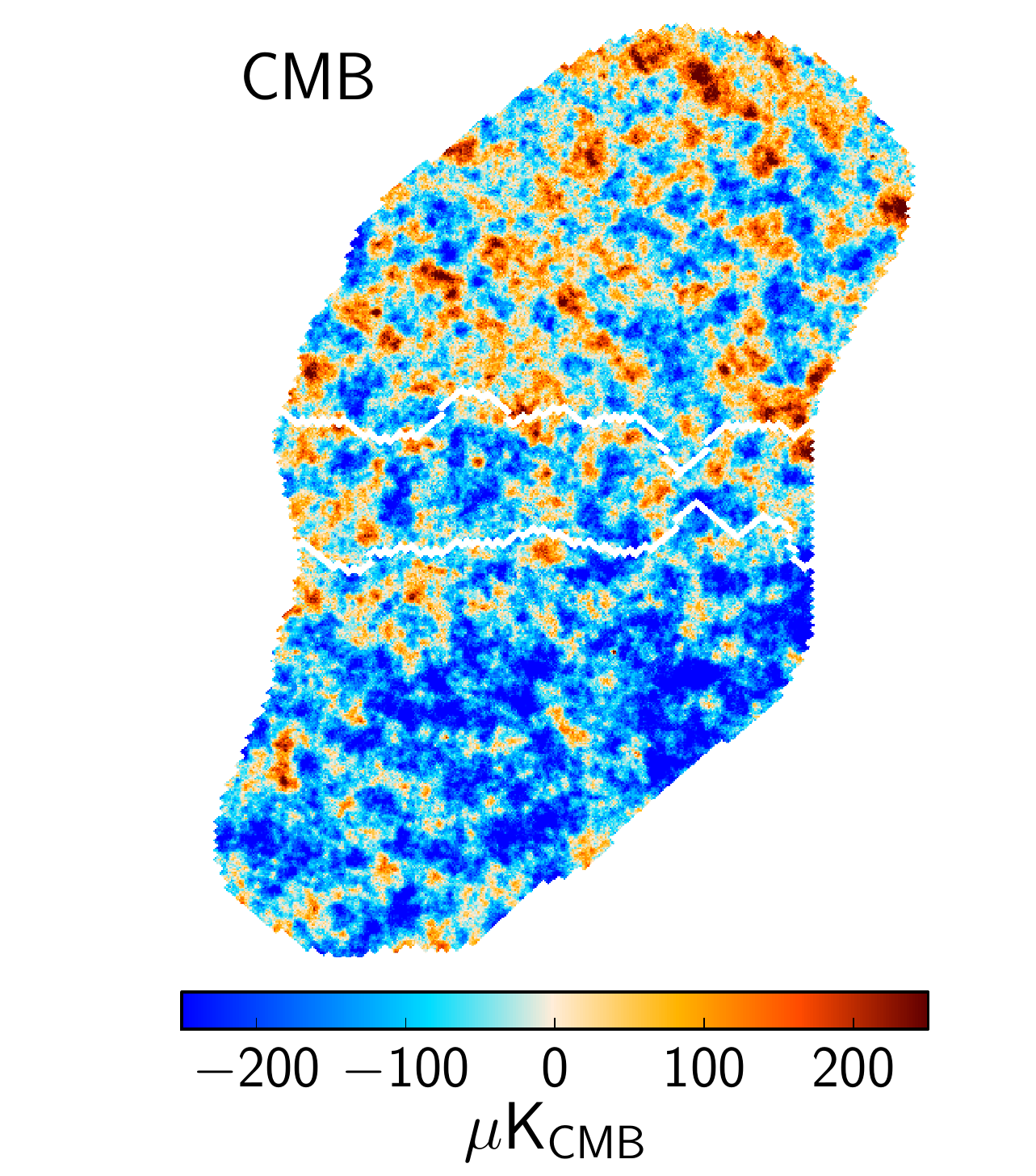}
      \includegraphics[width=0.50\linewidth, clip=true, trim=0.7in 0in 0.2in 0in]{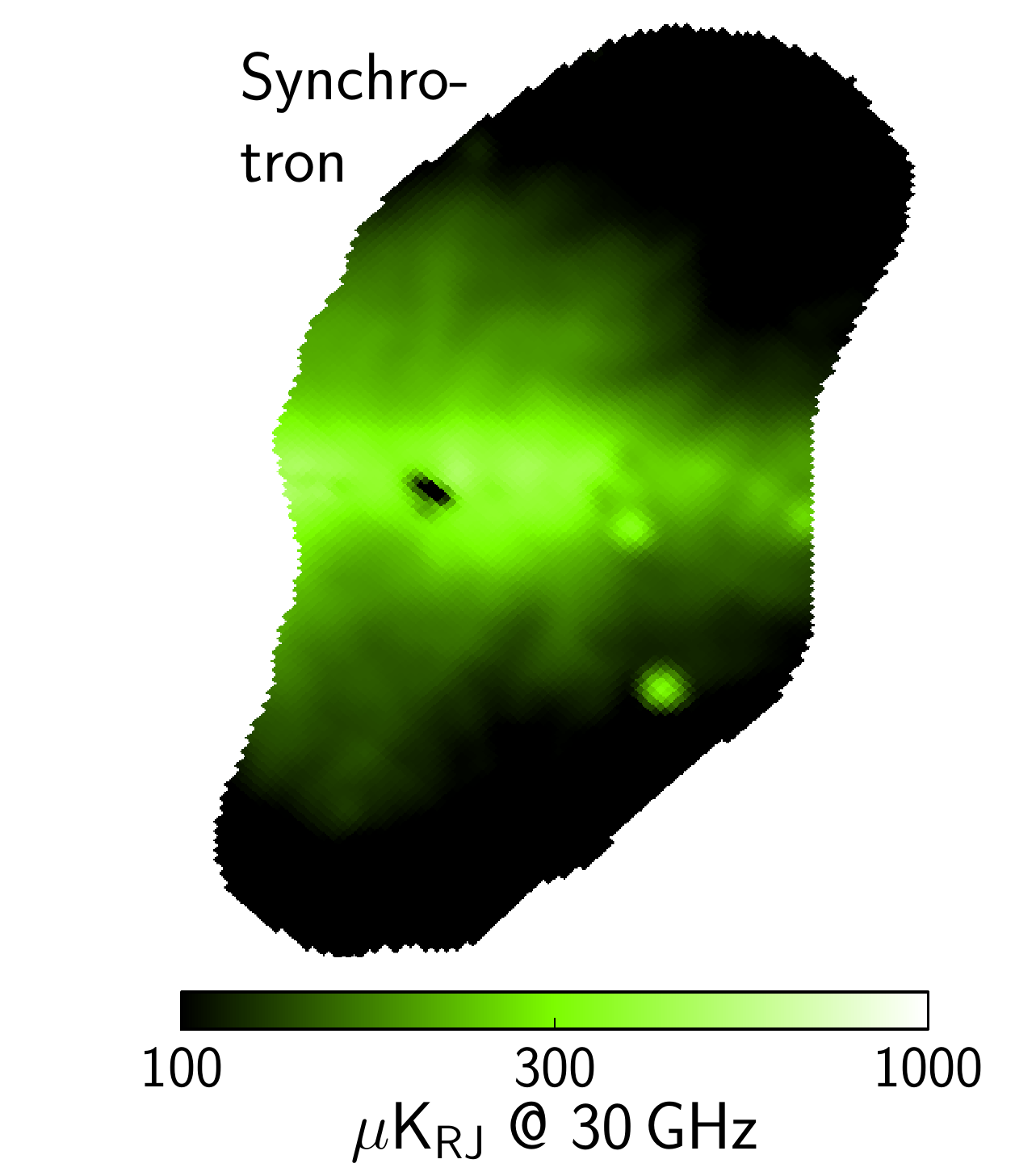}}
\mbox{\includegraphics[width=0.50\linewidth, clip=true, trim=0.7in 0in 0.2in 0in]{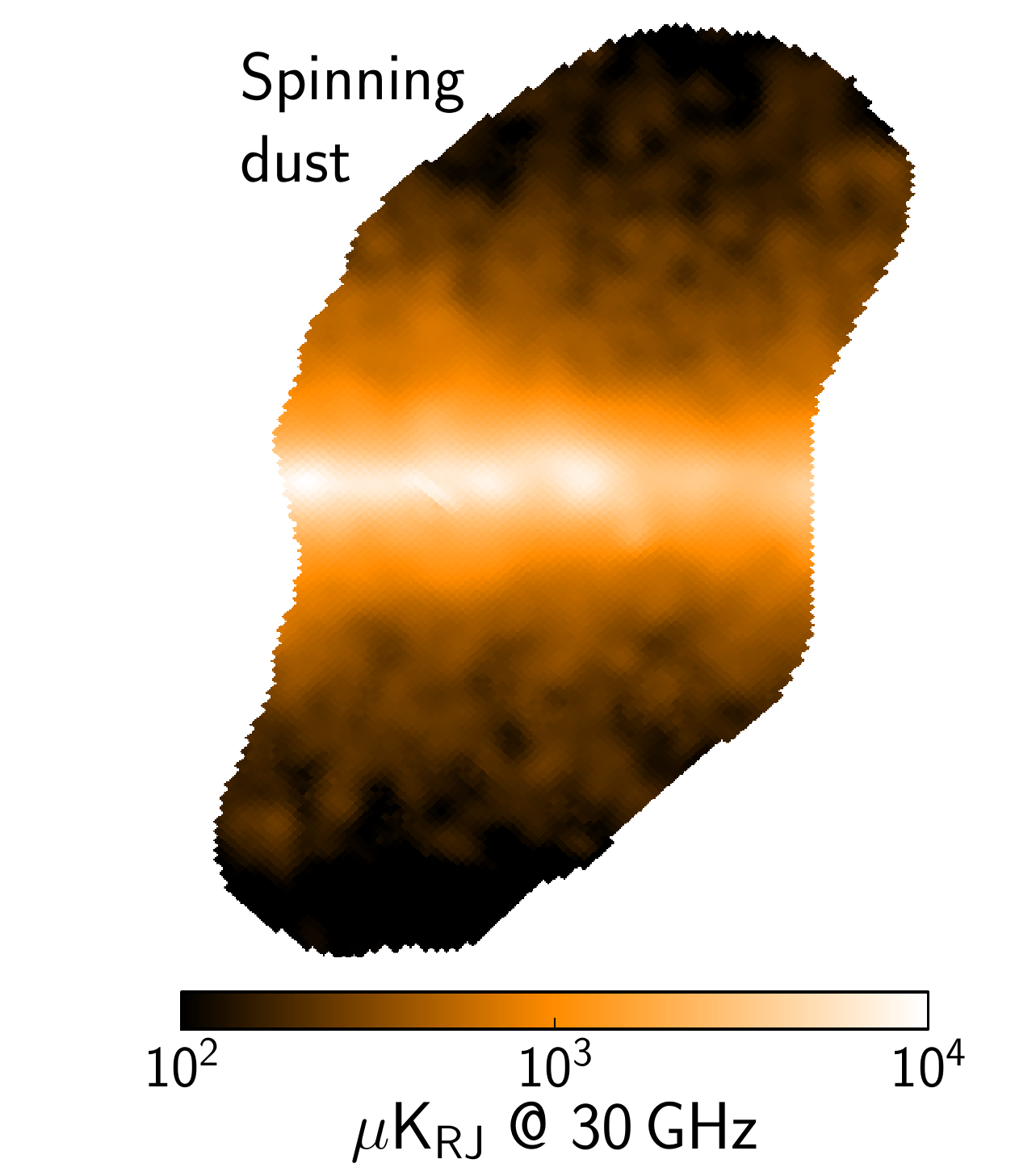}
      \includegraphics[width=0.50\linewidth, clip=true, trim=0.7in 0in 0.2in 0in]{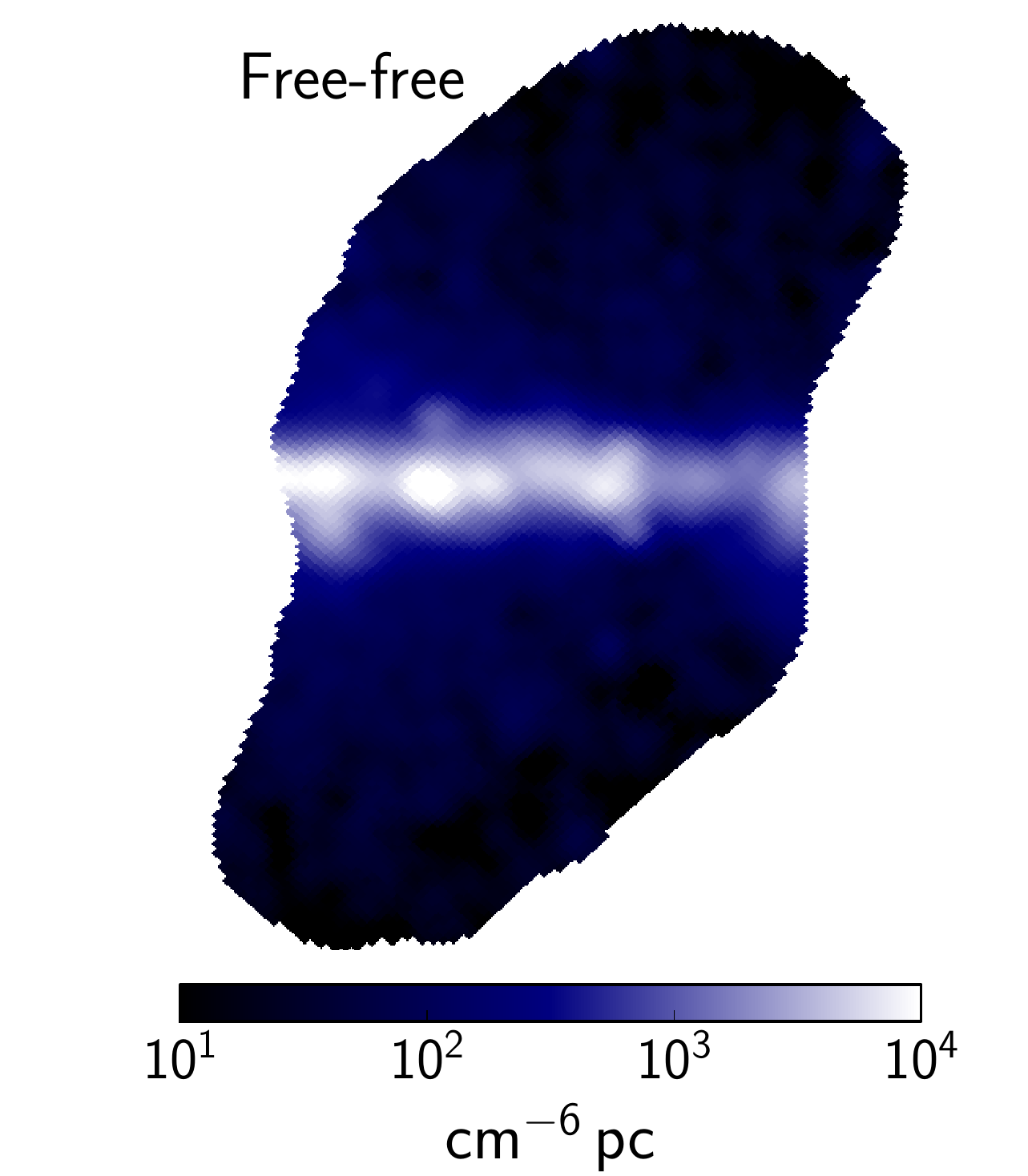}}
\mbox{\includegraphics[width=0.50\linewidth, clip=true, trim=0.7in 0in 0.2in 0in]{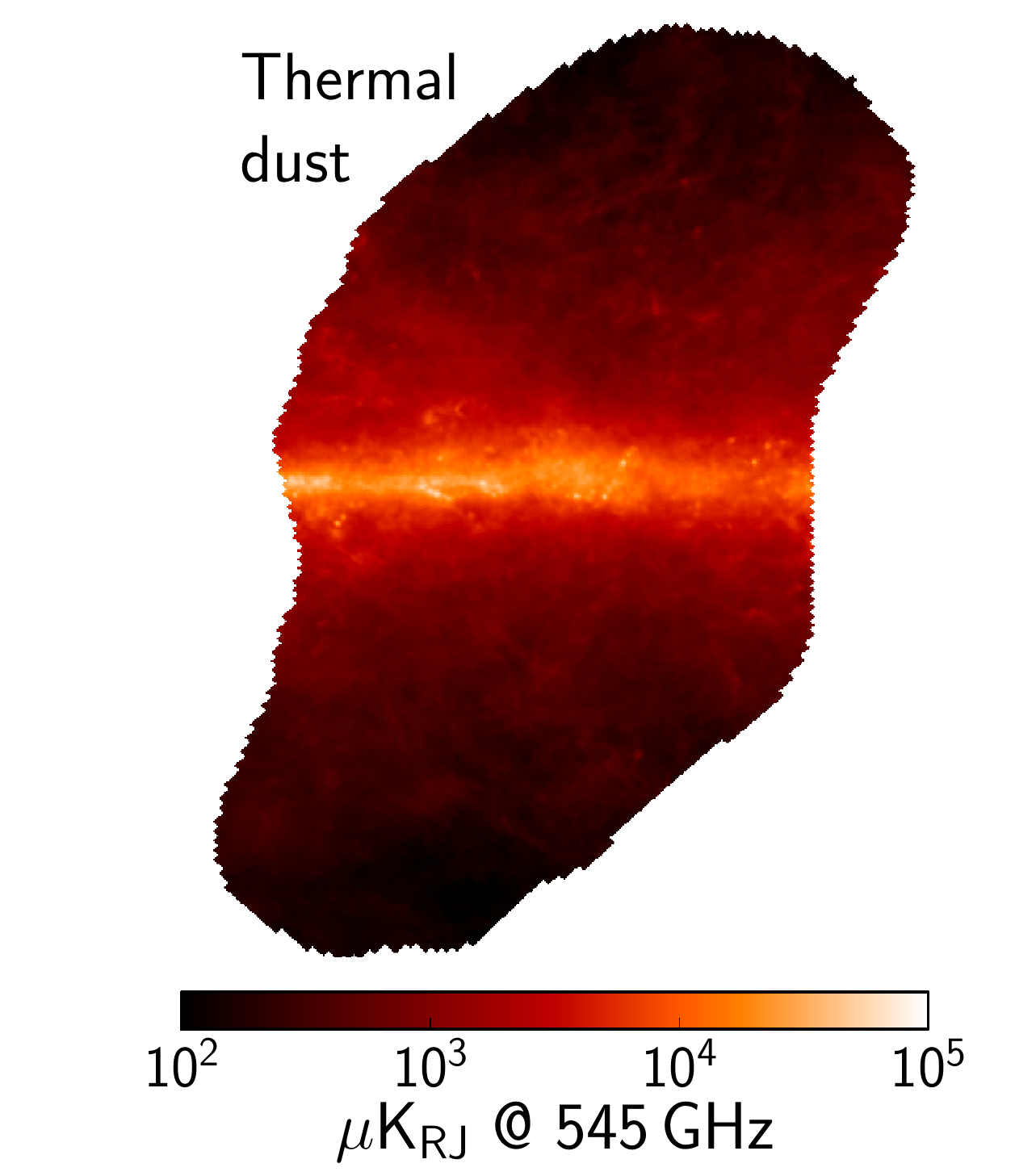}
      \includegraphics[width=0.50\linewidth, clip=true, trim=0.7in 0in 0.2in 0in]{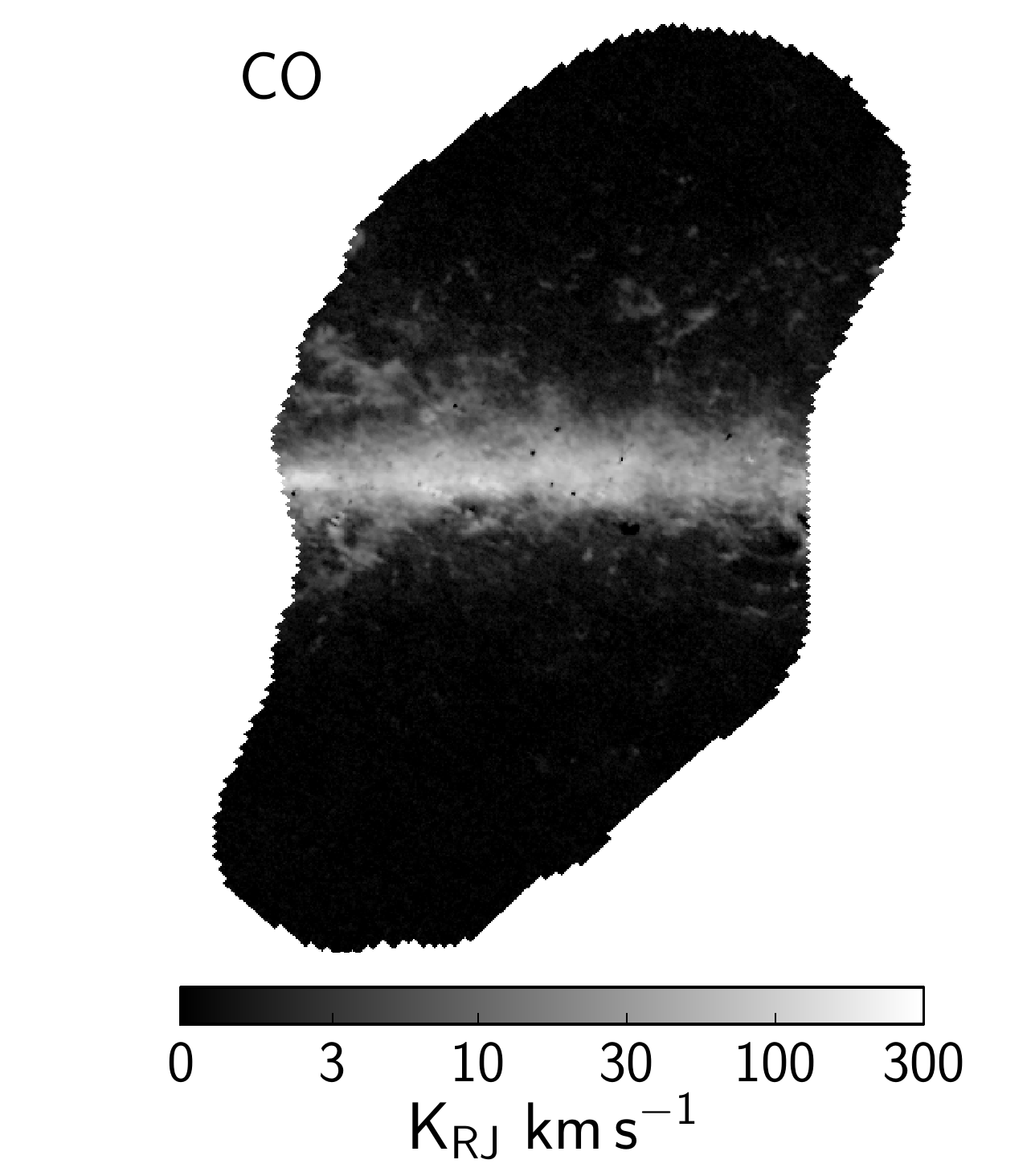}}
\caption{Same as Figure~\ref{fig:commander_gc}, but for field G-1.}
\label{fig:commander_gb}
\end{figure}

Three individual features are noteworthy in these low-frequency maps. First, as is well-known, \citep[e.g.,][]{page:2007} the orientation of the magnetic field at low Galactic latitudes is parallel to the Galactic plane. 
Second, the magnetic field lines to the north of the Galactic center
form a 'U' shape with an opening angle of $\sim 45\deg$.
These field lines correspond to the origin of the polarized filaments IX and XIV identified by \citet{Vidal2015a} in the \WMAP\ sky maps.  Third, the magnetic field lines in the Galactic center are rotated by an angle of almost $90\deg$ with respect to the Galactic plane in Q-band, while at K-band this angle is about $70\deg$. This is the expected signature of Faraday rotation, and we discuss it quantitatively in \S~\ref{sec:faraday}.

In the \QUIET\ W-band map (bottom left panel of
Figure~\ref{fig:finalmaps_gcP}), we also see that the
magnetic field is well ordered and parallel to the Galactic plane at
low latitudes. Comparison with the \Planck\ thermal dust map (bottom
right) suggests that the \QUIET\ map is dominated by
dust rather than by synchrotron emission.  Based on this qualitative
analysis alone, we conclude that the effective polarized
foreground minimum must lie between 43 and 95\,GHz, well separated
from either side.  Overall, the \QUIET\ measurements are in good
agreement with the \Planck\ determination that the foreground minimum
lies between 70--80\,GHz \citep{planck2014-a12}.

\begin{figure*}[t] 
  \centering
  \mbox{\includegraphics[width=0.5\linewidth]{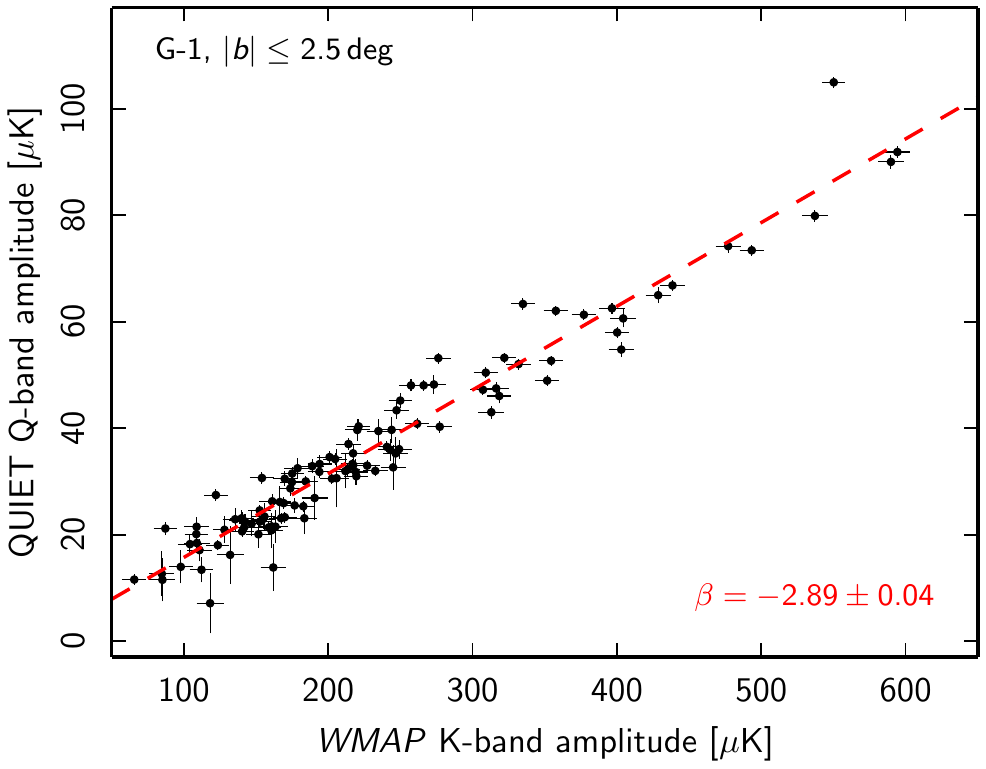}
        \includegraphics[width=0.5\linewidth]{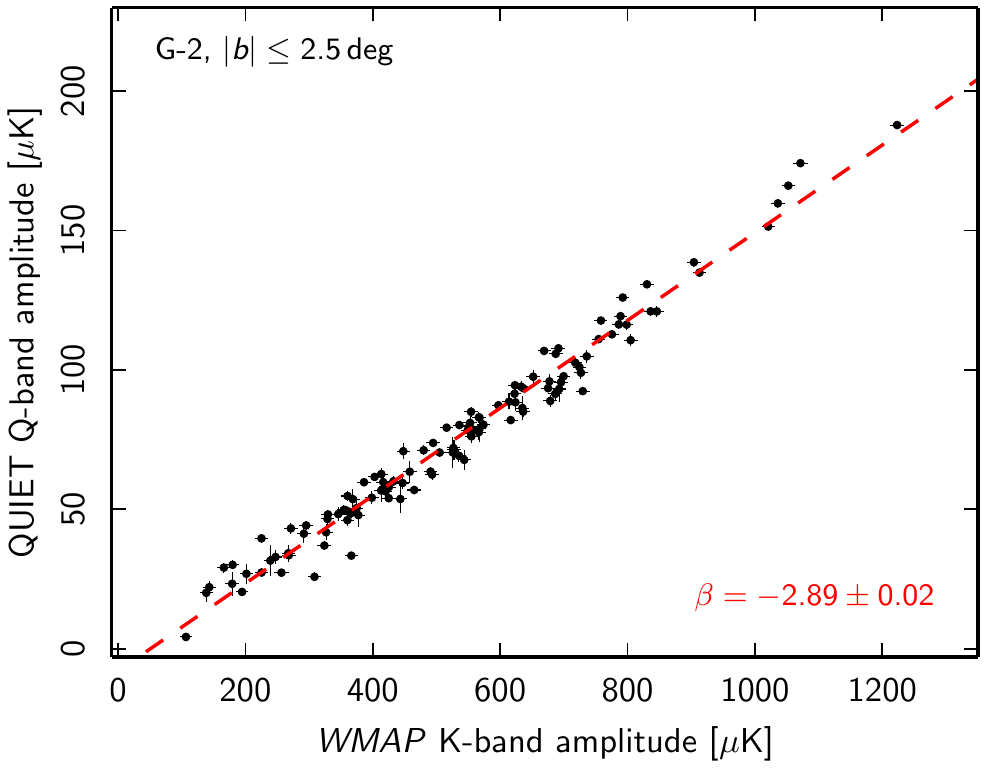}}
  \mbox{\includegraphics[width=0.5\linewidth]{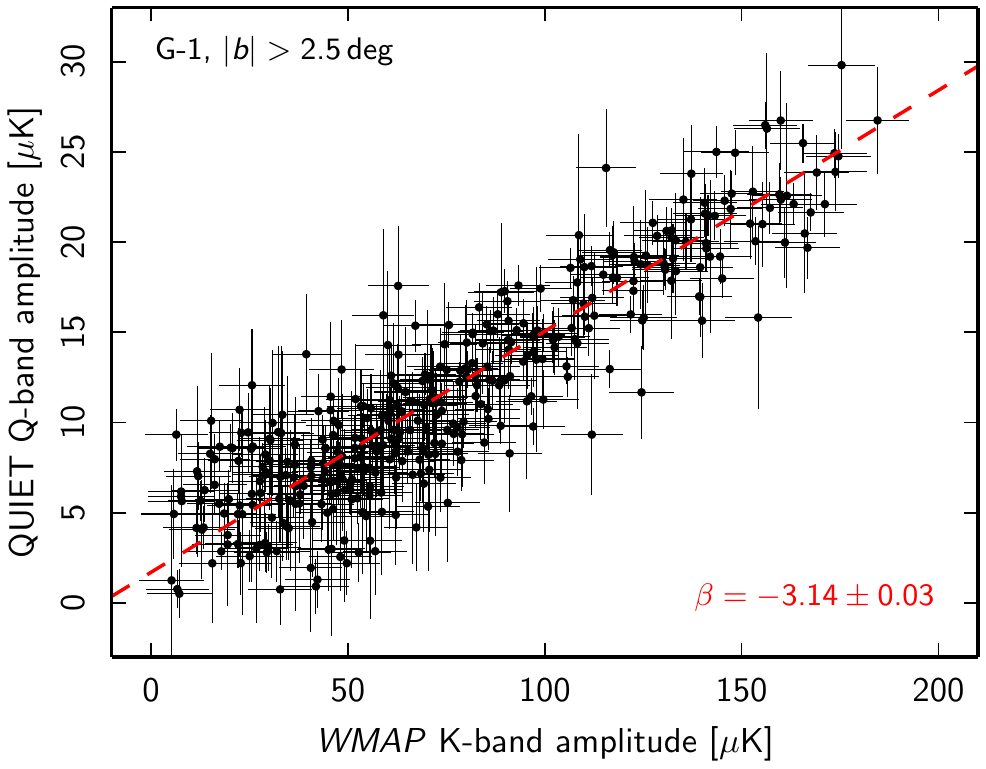}
        \includegraphics[width=0.5\linewidth]{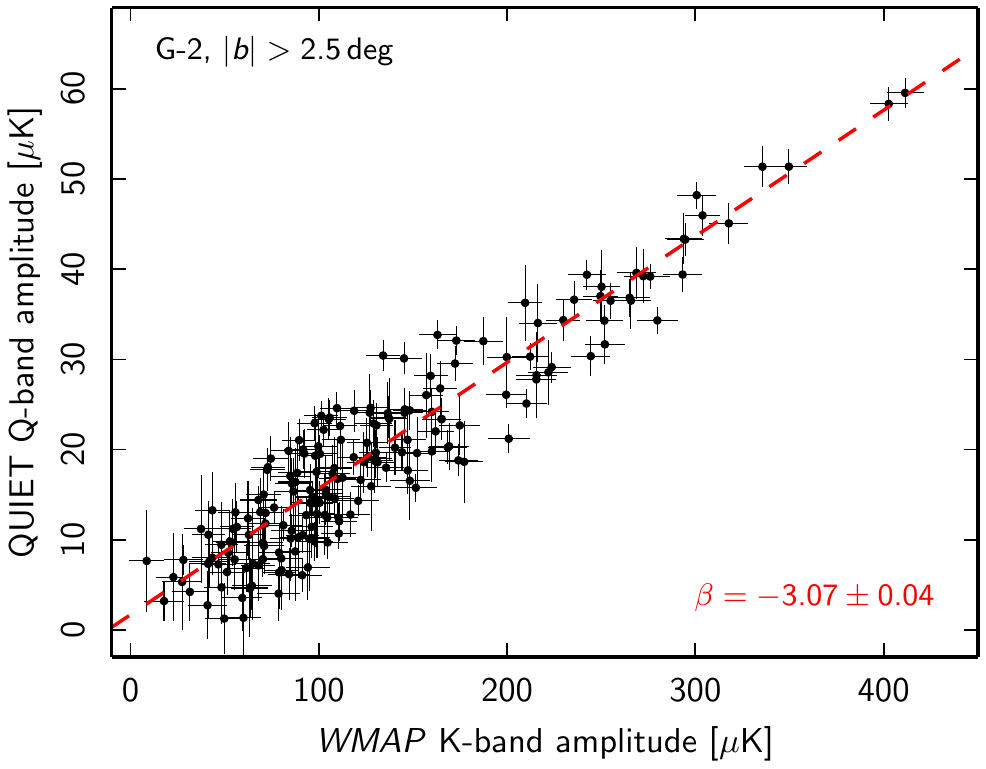}}\\
  \caption{Scatter plots between \WMAP\ K-band (23\,GHz) and \QUIET\
    Q-band (43\,GHz) polarization amplitudes for fields G-1 (left column) and G-2
    (right column), considering separately on-plane ($|b|\le2\pdeg5$;
    top row) and off-plane ($|b|>2\pdeg5$; bottom row) pixels. The
    dashed red lines indicate the best-fit power-law fit to each data
    combination.}
  \label{fig:scatter}
\end{figure*}

Figure~\ref{fig:finalmaps_gbP} shows the corresponding information for G-1, the field centered on $(l,b)=(329\deg,0\deg)$.  Although both maps have lower signal-to-noise ratio than those of G-2, the same
qualitative conclusions hold.  The Q-band map is clearly dominated by synchrotron emission and the W-band map is dominated by thermal dust emission.

Comparing the \QUIET\ W-band and the \Planck\ thermal dust maps, we
note the presence of a bright localized feature in the former at
Galactic coordinates $(l,b)=(326\deg,-2\deg)$ with no counterpart in
the thermal dust map. We identify this as the supernova remnant (SNR)
G326.3--1.8 (MSH\,15--56), as discussed by \citet{Green2009}. It is a
composite SNR that contains a shell with a relatively steep radio
spectrum and an interior plerion component with a flat spectrum 
\citep{Dickel2000,Weiler1988}. From the K, Ka, and Q bands, we
estimate its spectral index as $\beta\approx-2.7$, consistent
with a (relatively flat spectrum) synchrotron source. This value is
steeper than the spectrum of the SNR shell ($\beta=-2.34$) measured
by \citet{Dickel2000} between 0.408 and 14.7\,GHz. This new
measurement by QUIET might indicate a steepening of the polarized
spectrum with frequency. 

Figures~\ref{fig:commander_gc} and \ref{fig:commander_gb} show the \Planck\ baseline temperature reconstruction in our fields, allowing for direct comparison between our maps and individual astrophysical temperature components. From these we see that the bright G-1 source is indeed recognized as a synchrotron emitter in the \Planck\ model, with little or no counterpart in any other component.

\subsection{Spectral index of polarized emission}

\begin{table}[t]                                                                                                                                                   
\begingroup                                                                                                                                     
\newdimen\tblskip \tblskip=5pt
\caption{Polarized synchrotron spectral index between K and Q band$^{\rm a}$\label{tab:spec_index_quiet_wmap}}
\vskip -2mm
\footnotesize                                                                                                                                     
\setbox\tablebox=\vbox{                                                                                                                                                                             
\newdimen\digitwidth                                                                                                                       
\setbox0=\hbox{\rm 0}
\digitwidth=\wd0
\catcode`*=\active
\def*{\kern\digitwidth}
\newdimen\signwidth
\setbox0=\hbox{+}
\signwidth=\wd0
\catcode`!=\active
\def!{\kern\signwidth}
                                                                                                                                                                                                     
\newdimen\decimalwidth
\setbox0=\hbox{.}
\decimalwidth=\wd0
\catcode`@=\active
\def@{\kern\signwidth}
                                                                                                                                                                                                     
\halign{ \hbox to 1.0in{#\leaderfil}\tabskip=2em&
  \hfil#\hfil&
  \hfil#\hfil\tabskip=0em\cr
\noalign{\doubleline}
\omit\hfil Field\hfil& \hfil QUIET\hfil& \hfil \WMAP\hfil\cr
\noalign{\vskip 3pt\hrule\vskip 5pt}
G-1 full& $-2.91\pm0.01$& $-3.02\pm0.03$\cr
\phantom{G-1} $|b|\le2.5\deg$& $-2.89\pm0.04$& $-3.02\pm0.06$\cr
\phantom{G-1} $|b|>2.5\deg$& $-3.14\pm0.03$& $-3.17\pm0.08$\cr
\noalign{\vskip 4pt}
G-2 full& $-3.01\pm0.01$& $-3.02\pm0.02$\cr
\phantom{G-2} $|b|\le2.5\deg$& $-2.89\pm0.02$& $-2.93\pm0.03$\cr
\phantom{G-2} $|b|>2.5\deg$& $-3.07\pm0.04$& $-3.16\pm0.07$\cr
\noalign{\vskip 3pt\hrule\vskip 3pt}
}}
\endPlancktable 
\tablenote {{a}} Spectral indices are computed between \WMAP\ K-band
and either \QUIET\ (second column) or \WMAP\ (third column)
Q-band.\par
\endgroup
\vskip -2mm
\end{table}

In this section we determine the effective power-law index between
\WMAP\ K-band and \QUIET\ Q-band, which are heavily dominated by 
synchrotron emission. All maps are
smoothed to a common resolution of $1\deg$ FWHM, slightly larger than
the \WMAP\ K-band beam of 53$\arcm$. All spectral index estimates are
derived from the polarization amplitude, $P=\sqrt{Q^2 + U^2}$. The
associated bias from instrumental noise is corrected using the
asymptotic estimator \citep{Montier2015,Vidal2015b},
which is a generalization of the estimator first proposed by
\citet{Wardle1974} for the case where the uncertainties $\sigma_Q$ and
$\sigma_U$ are non-identical.

Many studies have reported a significant steepening in the synchrotron
spectral index at high Galactic latitudes compared to the Galactic
plane \citep[e.g.,][]{Kogut2007,fuskeland2014}.
As a first test, we therefore partition our fields into high ($|b|>2.5\deg$)
and low ($|b|\le2.5\deg$) latitudes, and
determine the spectral indices separately for each region. This
calculation is summarized in the scatter plots of Figure~\ref{fig:scatter} (based on polarization amplitudes) between \WMAP\ K-band and \QUIET\ Q-band for both fields (left and right
columns) and low and high latitudes (top and bottom rows). The red
dashed lines indicate the best-fit lines, corresponding to the spectral
index indicated in each panel.  We find results consistent with earlier measurements. In G-2 we derive
a Galactic plane spectral index of $\beta=-2.89\pm0.02$, which
steepens to $\beta=-3.07\pm0.04$ at high latitudes\footnote{All error
  bars include the uncertainty in the \QUIET\ absolute
  responsivity.}. For G-1, the corresponding numbers are
$\beta=-2.89\pm0.04$ and $\beta=-3.14\pm0.03$. In both cases, the
spectral index is about 0.2 steeper off the plane than in the plane.

\begin{table}[t!]                                                                                                                                             
\begingroup                                                                                                                                      
\newdimen\tblskip \tblskip=5pt
\caption{EVPA as a function of Galactic latitude$^{\rm a,b}$\label{tab:pol_ang}}
\nointerlineskip                                                                                                                                                                                     
\vskip -4mm
\scriptsize                                                                                                                                      
\setbox\tablebox=\vbox{                                                                                                                                                                            
\newdimen\digitwidth                                                                                                                          
\setbox0=\hbox{\rm 0}
\digitwidth=\wd0
\catcode`*=\active
\def*{\kern\digitwidth}
\newdimen\signwidth
\setbox0=\hbox{+}
\signwidth=\wd0
\catcode`!=\active
\def!{\kern\signwidth}
\newdimen\decimalwidth
\setbox0=\hbox{.}
\decimalwidth=\wd0
\catcode`@=\active
\def@{\kern\signwidth}
\halign{ \hfil#\hfil\tabskip=0.3em&
  \hbox to 1.5cm{#\leaderfil}\tabskip=0.6em&
  \hfil#\hfil\tabskip=0.5em&
  \hfil#\hfil\tabskip=1em&
  \hfil#\hfil\tabskip=0.5em&
  \hfil#\hfil\tabskip=0em\cr
  \noalign{\doubleline}
\noalign{\vskip -2pt}
\multispan2\hfil $b$\hfil&\multispan2\hfil $\chi$ in G-1\hfil&\multispan2\hfil$\chi$ in G-2\hfil\cr
\noalign{\vskip -3pt}
\multispan2\hrulefill&\multispan2\hrulefill&\multispan2\hrulefill\cr
\noalign{\vskip 2pt}
Min&\omit\hfil Max\hfil& \hfil Q-band\hfil& \hfil W-band\hfil& \hfil Q-band\hfil& \hfil W-band\hfil\cr
\noalign{\vskip 4pt\hrule\vskip 5pt}
+3\pdeg0&+5\pdeg0&    $!30\pdeg0\pm0\pdeg3$&$11\pdeg7\pm0\pdeg2$&$!*3\pdeg5\pm0\pdeg3$&$-20\pdeg7\pm0\pdeg2$\cr
+1\pdeg0&+3\pdeg0&    $!14\pdeg9\pm0\pdeg2$&$*7\pdeg5\pm0\pdeg2$&$!*0\pdeg3\pm0\pdeg1$&$-17\pdeg2\pm0\pdeg2$\cr
$-$1\pdeg0&+1\pdeg0&  $*-0\pdeg8\pm0\pdeg1$&$*4\pdeg6\pm0\pdeg1$&$!*7\pdeg0\pm0\pdeg1$&$*-4\pdeg7\pm0\pdeg1$\cr
$-$3\pdeg0&$-$1\pdeg0&$*-5\pdeg3\pm0\pdeg2$&$*5\pdeg0\pm0\pdeg2$&$!10\pdeg1\pm0\pdeg2$&$-12\pdeg9\pm0\pdeg2$\cr
$-$5\pdeg0&$-$3\pdeg0&$*-9\pdeg9\pm0\pdeg4$&$*3\pdeg2\pm0\pdeg2$&$*-4\pdeg7\pm0\pdeg3$&$-17\pdeg4\pm0\pdeg3$\cr
\noalign{\vskip 5pt\hrule\vskip 3pt}
}}
\endPlancktable
\tablenote {{a}} $\chi$ is measured counterclockwise relative to the Galactic north direction.\par
\tablenote {{b}} Errors are statistical only. See Table~\ref{tab:patch_summary} for systematic errors.\par
\endgroup
\vskip -2mm
\end{table}

Including all latitudes in the spectral index evaluation, we find {\bf
  $\beta_{\mathrm{G-1}}=-2.91\pm0.01$} and {\bf
  $\beta_{\mathrm{G-2}}=-3.01\pm0.01$}, as listed in
Table~\ref{tab:spec_index_quiet_wmap}. These are fully consistent with
the values adopted from \citet{fuskeland2014}
($\beta_{\textrm{G-1}}=-2.93$ and $\beta_{\textrm{G-2}}=-3.00$; see
\S~\ref{sec:coadd}), derived from \WMAP\ K- and Ka-band, that were used
to re-scale the \WMAP\ and \Planck\ maps to the effective
\QUIET\ Q-band frequency.


\subsection{Electric vector position angles (EVPA)}
\label{sec:polangle}

\begin{figure}[t] 
  \centering
  \includegraphics[width=0.49\linewidth]{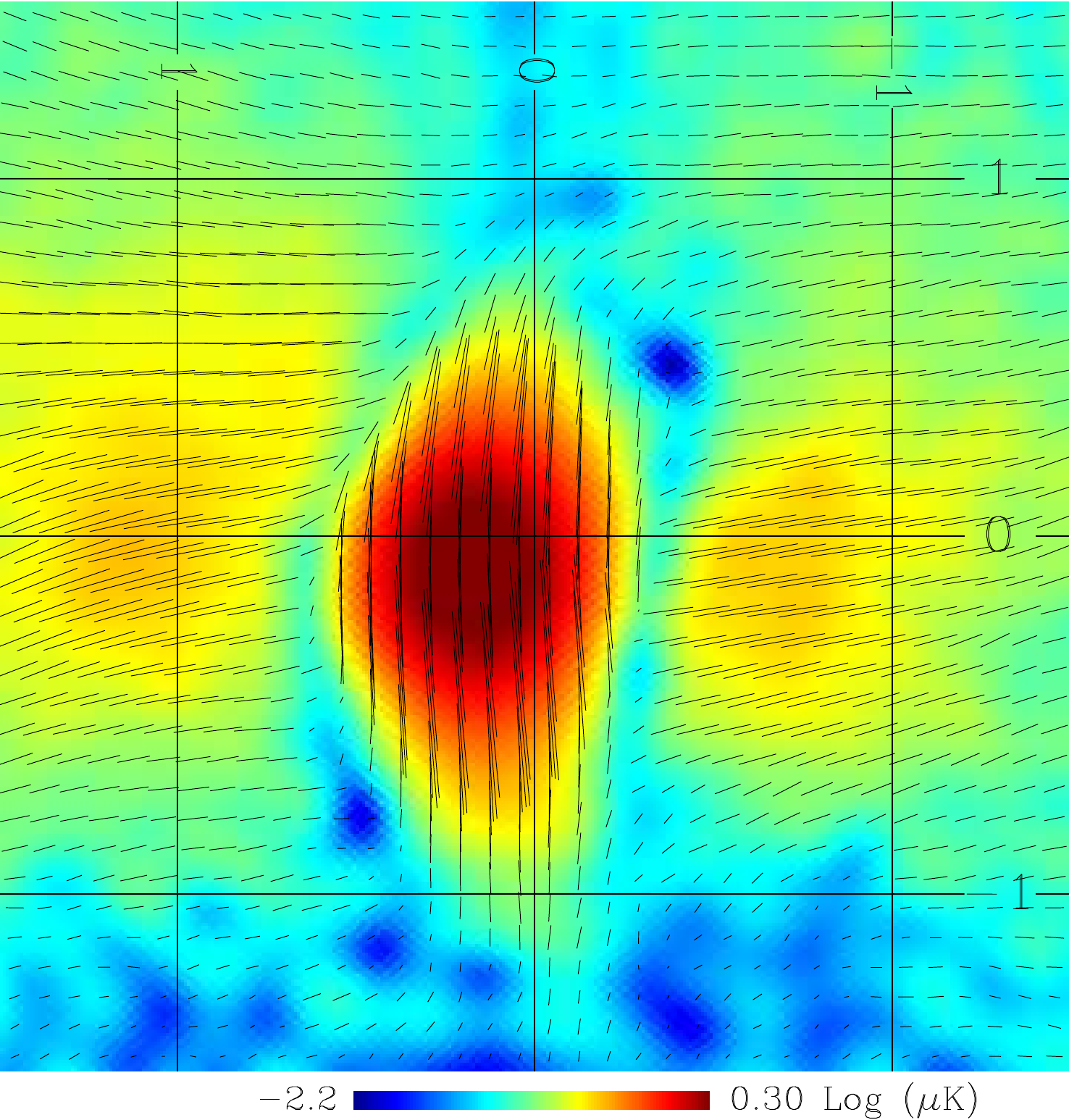}
  \includegraphics[width=0.49\linewidth]{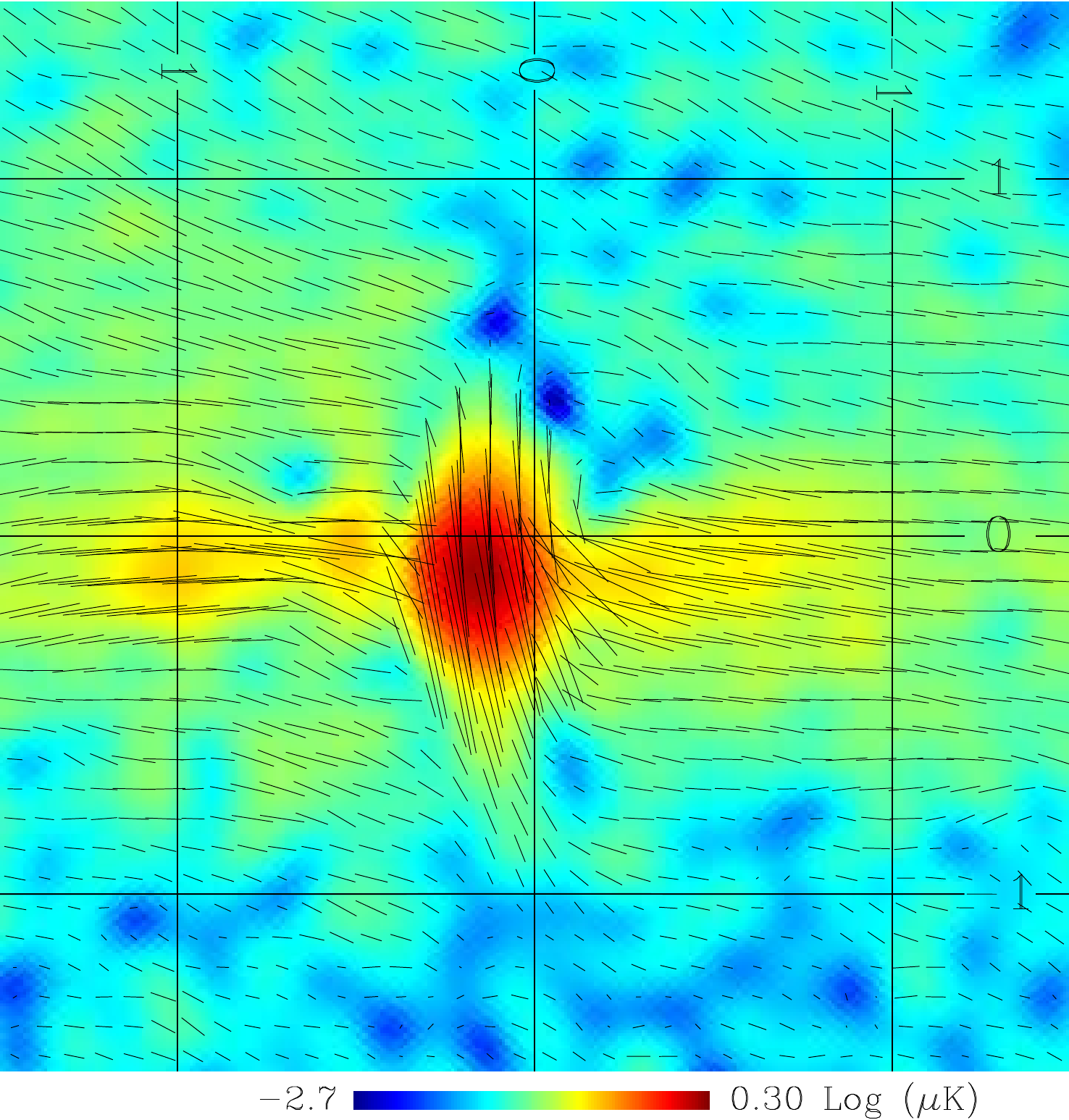}
  \caption{Polarization intensity maps of the Galactic centre region
  seen by QUIET at 43\,GHz ({\it left}) and 95\,GHz ({\it right}),
  at the original angular resolution of 27$'\!\!$.3 and 12$'\!\!$.8
  FWHM respectively.
  The vectors are rotated $90\deg$ with respect to the measured EVPA,
  indicating the orientation of the magnetic field. Both maps
  are displayed in Galactic coordinates with a grid size of 1\deg. }
  \label{fig:gc_maps}
\end{figure}

\begin{table}[t] 
\begingroup 
\newdimen\tblskip \tblskip=5pt
\caption{Galactic center source polarization parameters$^{\rm a}$\label{tab:ang_gc}}
\nointerlineskip                                                                                                                                                                                     
\vskip -4mm
\footnotesize 
\setbox\tablebox=\vbox{ %
\newdimen\digitwidth 
\setbox0=\hbox{\rm 0}
\digitwidth=\wd0
\catcode`*=\active
\def*{\kern\digitwidth}
\newdimen\signwidth
\setbox0=\hbox{+}
\signwidth=\wd0
\catcode`!=\active
\def!{\kern\signwidth}
\newdimen\decimalwidth
\setbox0=\hbox{.}
\decimalwidth=\wd0
\catcode`@=\active
\def@{\kern\signwidth}
\halign{ \hbox to 0.9in{#\leaderfil}\tabskip=1em&
  \hfil#\hfil\tabskip=1em&
  \hfil#\hfil\tabskip=1em&
  \hfil#\hfil\tabskip=0em\cr
\noalign{\doubleline}
\omit& Stokes $Q$& Stokes $U$& Pol angle\cr
\omit \hfil Band\hfil&      [mK]& [mK]& [\deg]\cr
\noalign{\vskip 4pt\hrule\vskip 5pt}
\QUIET\          Q & $-1.39 \pm  0.04$ & $ !0.01 \pm  0.02$ & $ *90.1 \pm   0.8$ \cr
\phantom{\QUIET} W & $-0.89 \pm  0.06$ & $ !0.67 \pm  0.03$ & $108.5 \pm   1.3$ \cr
\noalign{\vskip 4pt}
\WMAP\          K  & $-0.44 \pm  0.06$ & $ -0.72 \pm  0.03$ & $ *60.5 \pm   1.4$ \cr
\phantom{\WMAP} Ka & $-1.06 \pm  0.05$ & $ -0.38 \pm  0.03$ & $ *80.1 \pm   1.2$ \cr
\phantom{\WMAP} Q  & $-1.39 \pm  0.04$ & $ -0.17 \pm  0.03$ & $ *86.6 \pm   0.9$ \cr
\phantom{\WMAP} V  & $-1.62 \pm  0.03$ & $ !0.26 \pm  0.02$ & $ *94.6 \pm   0.5$ \cr
\phantom{\WMAP} W  & $-0.84 \pm  0.06$ & $ !0.56 \pm  0.04$ & $106.9 \pm   1.6$ \cr
\noalign{\vskip 5pt\hrule\vskip 4pt}
}}
\endPlancktable 
\tablenote {{a}}  Values correspond to the mean evaluated inside a
$20'$ diameter aperture centered at Galactic coordinates 
$(l,b)=(0\pdeg15,-0\pdeg1)$. Uncertainties in $Q$ and $U$ correspond
to the statistical fluctuations in the map, as measured in a larger
aperture around the source. See Table~\ref{tab:patch_summary} for systematic errors.\par
\endgroup
\end{table}

\begin{figure}[t] 
  \centering
  \includegraphics[width=\linewidth]{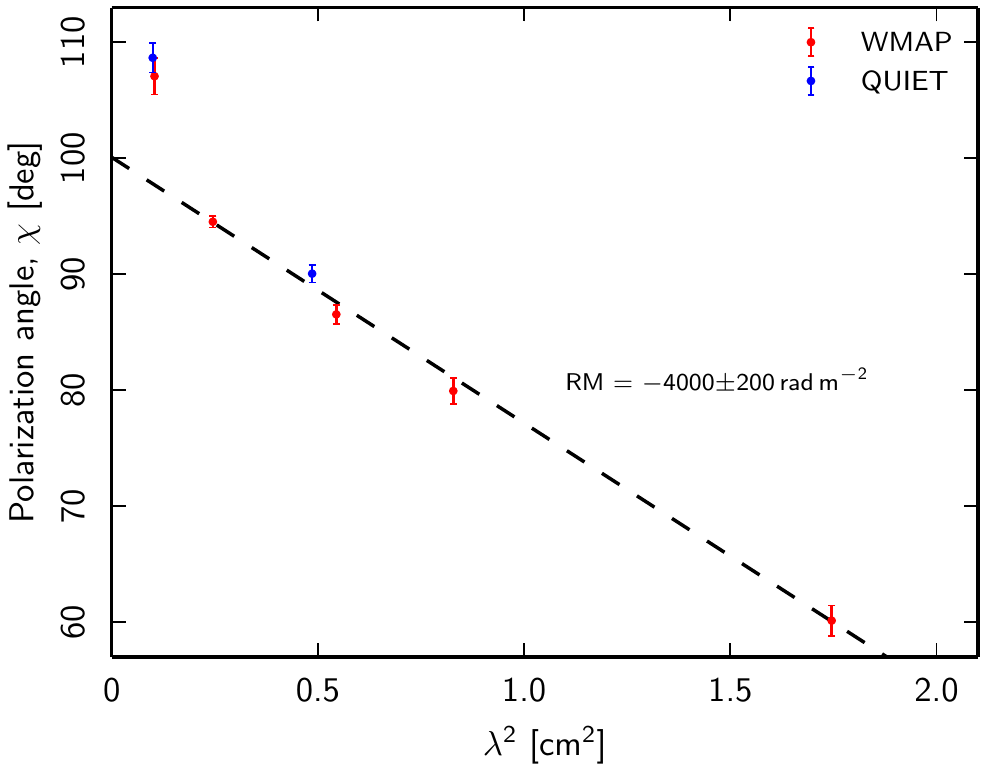}
  \caption{EVPA as a function of wavelength squared for
    the Galactic center arc for both \WMAP\ (red points),
    and \QUIET\ (blue points). The dashed line is a
    linear fit, as expected for a pure foreground Faraday screen
    resulting in an EVPA $\chi \propto
    \mathrm{RM}~\lambda^2$. We exclude the W-band points from the fit; see \S~\ref{sec:faraday} for explanation. The best-fit rotation measure is
    RM$\,=-4000\pm200$\,rad\,m$^{-2}$.}
  \label{fig:far_rot_gv}
  \vspace*{-1mm}
\end{figure}

As noted in \S~\ref{sec:visual} and seen in
Figures~\ref{fig:finalmaps_gcP} and \ref{fig:finalmaps_gbP}, the
observed orientation of the polarization vector (EVPA) is roughly perpendicular to the
Galactic plane in the millimeter wavelength range. Given the assumption of optically thin synchrotron radiation, the Galactic
magnetic field is therefore roughly parallel to the plane.
To quantify these alignments, we measure the EVPA, 
\begin{equation}
\chi = \frac{1}{2}\arctan(U/Q)\footnote{We remind the reader that
  QUIET uses the COSMO convention for the Stokes parameters, where
  Stokes {\em U} is replaced by $-U$ to follow the IAU convention. }
\end{equation}
\noindent defined as zero for polarization vectors aligned with Galactic meridians and
increasing counter-clockwise in Galactic coordinates.
Spatial variations in $\chi$ can be used to
constrain models of the Galactic magnetic field \citep[e.g.,][]{Jaffe2013}.

In Table~\ref{tab:pol_ang} we tabulate the mean EVPA as
a function of Galactic latitude for both G-1 and G-2, averaging over
latitude slabs of $2\deg$ width and masking bright
objects. Averaging over the entire fields, the Q-band EVPA 
is $-1\pdeg9 \pm 0\pdeg4$ (stat) $\pm 1\pdeg7$ (syst) for G-1 and $-10\pdeg8\pm 0\pdeg6$ (stat) $\pm 1\pdeg7$ (syst) for G-2. Similar results were reported by \citet{Bierman2011} for 100, 150,
and 220\,GHz derived from the BICEP observations.

\subsection{Faraday rotation at the Galactic center}
\label{sec:faraday}

The Galactic center patch G-2 includes the brightest polarized source
observed by QUIET. Figure~\ref{fig:gc_maps} shows the QUIET maps of the 
central region at 43 and 95\,GHz. The magnetic field of the central
source is well aligned and perpendicular to the Galactic plane.
In the Galactic plane itself, the magnetic field is parallel to the plane in both bands.
At Q-band, a ring with a minimum of
polarized emission is seen around the central source. This is due to
the cancellation of emission from the source and the Galactic plane with different polarization angles.

The polarization of the Galactic center has been studied in detail at
higher angular resolution. \citet{Haynes1992} mapped the region at
8.5\,GHz with a beam size $\sim$2$'\!\!$.8 using the Parkes
telescope. \citet{Tsuboi1995} also observed this region at 42.5\,GHz
(similar to \QUIET-Q band) with 39$''$ angular resolution.  From these
results, it is clear that the dominant source of polarized radio
emission originates from the Galactic center `arc'. This arc is a very
narrow non-thermal filament, $\approx 15\arcm$ in length, perpendicular to
the Galactic plane, located within a distance of about 20\arcm\ from
the Sgr\,A$^{*}$ radio source at the Galactic center \citep[see
  e.g.][]{Yusef-Zadeh1984,Pedlar1989}.

We measure the EVPA of the Galactic center arc as a
function of frequency from the \QUIET, \WMAP, and \Planck\ maps, all
smoothed to a common resolution of $1\deg$ FWHM, averaging over a
20\arcm\ diameter disk centered on the peak of the polarized emission at
$(l,b)=(0\deg\!\!.15,-0\deg\!\!.1)$.  The resulting values are
listed in Table~\ref{tab:ang_gc}, including individual Stokes
parameters and EVPAs. The latter are also plotted in
Figure~\ref{fig:far_rot_gv} as a function of wavelength squared,
$\lambda^2$. Due to Faraday rotation \citep[e.g.,][]{burn:1966}, the
observed EVPA is expected to follow $\chi =
\textrm{RM}\,\lambda^2$, where \textrm{RM} is the rotation measure.

Neglecting the W-band data points, we find a tight relation with the
expected form, with a best-fit rotation measure of
RM = $-4000\pm200\,\textrm{rad}\,\textrm{m}^{-2}$,
corresponding to a total rotation angle of $\approx35\deg$ between 23
and 60\,GHz.  \citet{Tsuboi1995} reported a value of
RM = $-3120\pm188$\,rad\,m$^{-2}$ between 10\,GHz and 42.5\,GHz within a
$2'$ beam centered at the Galactic center arc.  The difference between
our value and the one from \citet{Tsuboi1995} could be due to Faraday
depolarization at low frequencies, which biases the RM measurements
towards lower values \citep{Law2008}. Large absolute values of RM
similar to the one we reported here have also been measured close to
the Galactic center by \citet{Roy2005}.

The EVPAs for the Galactic center arc found in the
\QUIET\ sky maps are in good agreement with those reported by \WMAP,
both at Q- and W-band. However, the W-band angles differ significantly
from the expected Faraday rotation prediction by about 15\deg. The
reason for this is the contribution of dust polarization, which
dominates W-band, while synchrotron only dominates below $\sim$ 60\,GHz.

\section{Summary}
\label{sec:conclusions}

We have presented polarization measurements of the Galactic plane at
43 and 95\,GHz, as observed with QUIET between 2008 October and 2010
December. The resulting Galactic maps are the deepest published to
date at their respective frequencies, by a factor of 2--4 at Q-band
compared to \Planck\ and \WMAP, and by a factor of 5--6 at W-band
compared to \WMAP.  We find no significant evidence of
residual instrumental systematic errors in these high-signal-to-noise-ratio maps.
We derive a conservative upper limit on temperature-to-polarization leakage of $<0.07\,$\% in the Q-band, translating to a $\lesssim4\,$\% uncertainty in polarization amplitude at the Galactic center. For comparison, the uncertainty in absolute responsivity is 6\,\% for Q-band and 8\,\% for W-band.

Our maps agree very well with corresponding \WMAP\ polarization
observations in both Q- and W-band over the entire fields. Accounting for the different
effective frequencies and uncertainties in the synchrotron spectral
index, we find no compelling evidence for significant systematic
differences between the two.  At 44\,GHz, both \QUIET\ and \WMAP\ observe a stronger
polarization signal than \Planck\ along the Galactic plane.  One
potential explanation for this difference is
temperature-to-polarization leakage of $\simeq$$\,0.2\,$\% in the
\Planck\ data set, which possibly might be related to the
null-test failures for this particular channel already reported by
\citet{planck2014-a03}. Further work is needed to understand these
discrepancies in detail.

Exploiting the agreement between \QUIET\ and \WMAP, we have computed
inverse-noise-variance-weighted averages between the two experiments,
combining small-scale information from \QUIET\ with large-scale
information from \WMAP.  The resulting maps are publicly
available on LAMBDA, and should prove useful both for experimental
consistency checks, as exemplified in this paper, and for
understanding the physical properties of polarized foreground emission
at microwave wavelengths. In the current paper, we have presented a
few examples of such analyses, evaluating the spectral index of
synchrotron emission, the mean EVPA near the Galactic
plane, and the Faraday rotation measure of the Galactic center
source. A key result from this work is robust evidence for true
spatial variations in the synchrotron spectral index of diffuse
polarized emission along and off the Galactic plane.

\vspace*{-4mm}
\acknowledgements
 Bruce Winstein, who led the QUIET project, died in 2011, soon after
 observations concluded. The project's success owes a great debt to
 his intellectual and scientific leadership.

 Support for the QUIET instrument and operation comes through the NSF
 cooperative agreement AST-0506648. Support was also provided by NSF
 awards PHY-0355328, AST-0448909, AST-1010016, and PHY-0551142;
 KAKENHI 20244041, 20740158, and 21111002; PRODEX C90284; a KIPAC
 Enterprise grant; and by the Strategic Alliance for the
 Implementation of New Technologies (SAINT). This work was performed
 on the Abel cluster, owned and maintained by the University of Oslo
 and NOTUR (the Norwegian High Performance Computing Consortium).
 Portions of this work were performed at the Jet Propulsion Laboratory
 (JPL) and California Institute of Technology, operating under a
 contract with the National Aeronautics and Space Administration. The
 Q-band polarimeter modules were developed using funding from the JPL
 R\&TD program. HKE acknowledges an ERC Starting Grant under
 FP7.  CD and MV acknowledge support from an ERC Starting Grant
 (no.~307209). CD also acknowledges support from the STFC (U.K.).
 JZ gratefully acknowledges a South Africa National Research Foundation
 Square Kilometre Array Research Fellowship.
 LB acknowledges support from CONICYT Grant PFB-06.
 ADM acknowledges a Sloan Fellowship.

 PWV measurements were provided by the Atacama Pathfinder Experiment
 (APEX).
 We thank CONICYT for granting permission to operate within the
 Chajnantor Scientific Preserve in Chile, and ALMA for providing site
 infrastructure support.
 Field operations were based at the Don Esteban facility run by Astro-Norte.
 We are particularly indebted to the engineers
 and technician who maintained and operated the telescope: Jos\'e Cort\'es,
 Cristobal Jara, Freddy Mu\~noz, and Carlos Verdugo.

 In addition, we would like to acknowledge the following people for
 their assistance in the instrument design, construction,
 commissioning, operation, and in data analysis: 
 Augusto Gutierrez Aitken, Colin Baines, Phil Bannister, Hannah
 Barker, Matthew R. Becker, Alex Blein, Mircea Bogdan, Alison Brizius,
 Ricardo Bustos, April Campbell, Anushya Chandra, Sea Moon Cho, Sarah
 Church, Joelle Cooperrider, Mike Crofts, Emma Curry, Maire Daly,
 Fritz Dejongh, Joy Didier, Greg Dooley, Hans Eide, Pedro Ferreira,
 Jonathon Goh, Will Grainger, Peter Hamlington, Takeo Higuchi, Seth
 Hillbrand, Christian Holler, Ben Hooberman, Kathryn D. Huff, William
 Imbriale, Koji Ishidoshiro, Norm Jarosik, Pekka Kangaslahti,
 Dan Kapner, Oliver King, Eiichiro Komatsu, Jostein Kristiansen,
 Donna Kubik, Richard Lai, David Leibovitch, Kelly Lepo, Siqi Li,
 Martha Malin, Jorge May, Mark McCulloch, Jeff McMahon, Steve Meyer,
 Oliver Montes, David Moore, Makoto Nagai, Hogan Nguyen, Glen Nixon, Ian
 O'Dwyer, Gustavo Orellana, Stephen Osborne, Heather Owen, Stephen Padin, Felipe
 Pedreros, Ashley Perko, Joey Richards, Alan Robinson, Jacklyn
 Sanders, Dale Sanford, Yunior Savon, Kendrick Smith, Mary Soria, Alex
 Sugarbaker, David Sutton, Keith Vanderlinde, Liza Volkova, Ross
 Williamson, Edward Wollack, Stephanie Xenos, Octavio Zapata, Mark
 Zaskowski, and Joe Zuntz.

 Some of the results in this paper have been derived using the HEALPix
 package.

\bibliographystyle{apj}
\bibliography{quiet_papers}

\vspace*{-6mm}
\appendix
\section{Supporting figures}
\label{sec:support}

\begin{figure*}[t]
\centering
\mbox{
\includegraphics[width=0.36\linewidth, clip=true, trim=0.4in 0in 0.15in 0in]{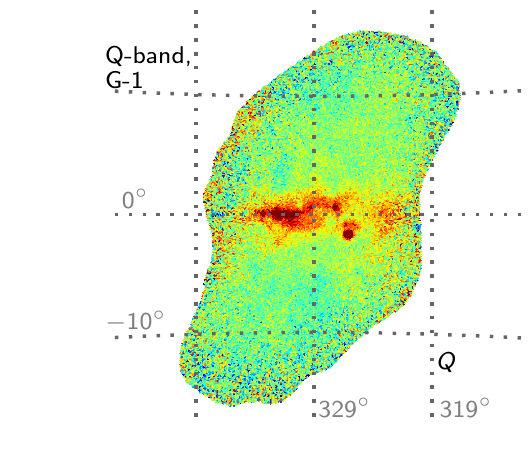}
\includegraphics[width=0.36\linewidth, clip=true, trim=0.45in 0in 0.1in 0in]{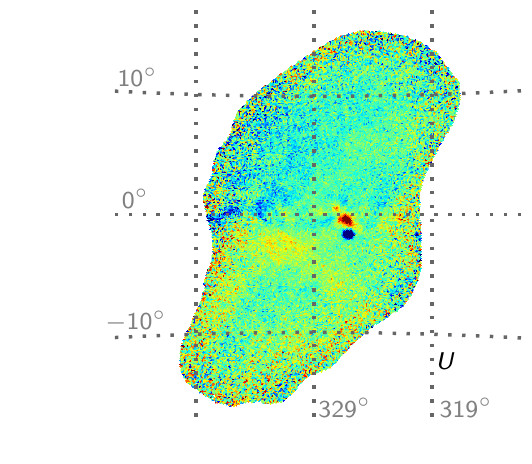}
}\vspace{-0.2in}
\mbox{
\includegraphics[width=0.36\linewidth, clip=true, trim=0.4in 0in 0.15in 0in]{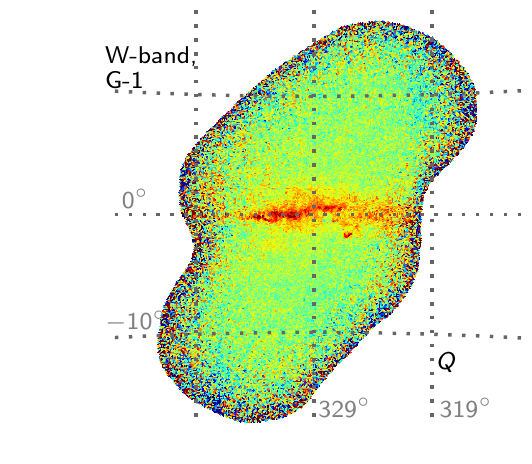}
\includegraphics[width=0.36\linewidth, clip=true, trim=0.45in 0in 0.1in 0in]{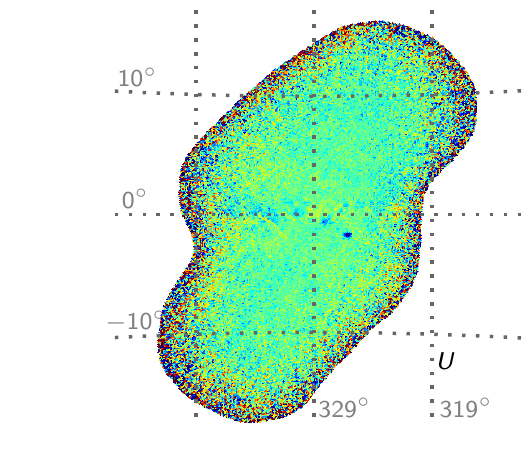}
}\vspace{-0.1in}
\mbox{
\includegraphics[width=0.36\linewidth, clip=true, trim=0.4in 0in 0.15in 0.1in]{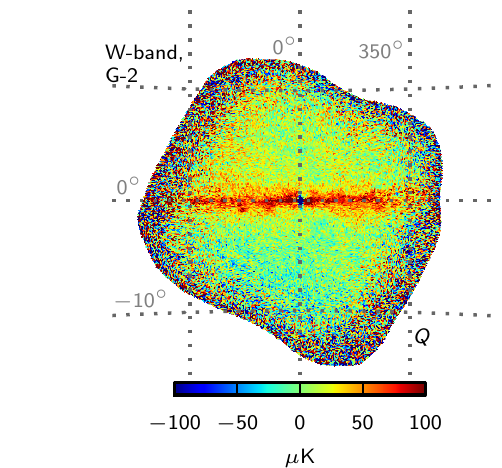}
\includegraphics[width=0.36\linewidth, clip=true, trim=0.45in 0in 0.1in 0.1in]{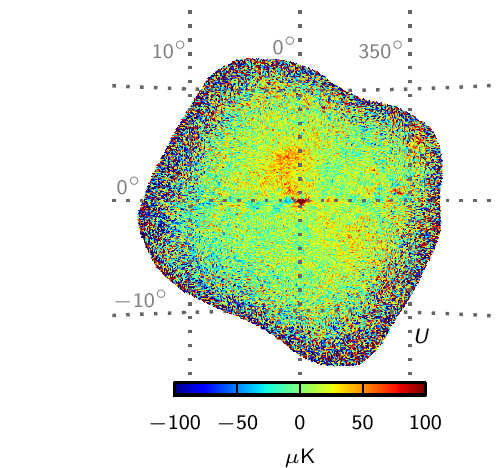}
}
\caption{Final co-added \QUIET$+$\WMAP\ maps. The top row shows the Q-band (43\,GHz) result for field G-1 (centered on Galactic coordinates $(l,b)=(329\deg,0\deg)$). Using the weight operators defined in \S~\ref{sec:coadd}, these maps are expressed as $\mathbf{F}_{\mathrm{Q}}\mathbf{m}_{\mathrm{Q}} + \mathbf{F}_{\mathrm{W}}\mathbf{m}_{\mathrm{W}}$. The middle and bottom rows show the equivalent maps for W-band (95\,GHz), both fields. The left and right columns show Stokes $Q$ and $U$, respectively. The grid cell width is $10\deg$.}
\label{fig:finalmaps_app}
\end{figure*}

\begin{figure*}[t]
\centering
\mbox{
\includegraphics[width=0.20\linewidth, clip=true, trim=0.3in 0in 0.1in 0.05in]{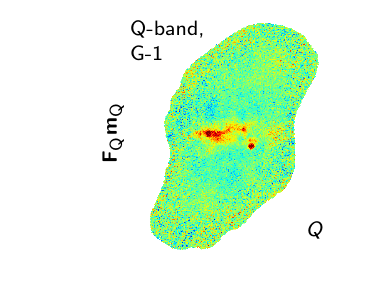}\hspace{-0.05in}
\includegraphics[width=0.20\linewidth, clip=true, trim=0.3in 0in 0.1in 0.05in]{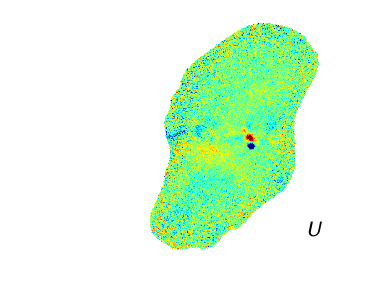}\hspace{-0.05in}
\includegraphics[width=0.20\linewidth, clip=true, trim=0.3in 0in 0.1in 0.05in]{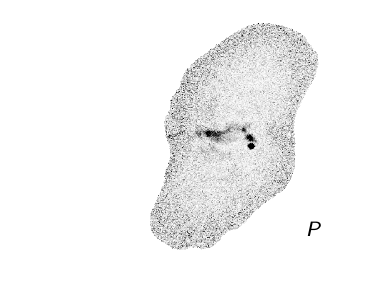}
}\vspace{-0.2in}
\mbox{
\includegraphics[width=0.20\linewidth, clip=true, trim=0.3in 0in 0.1in 0.05in]{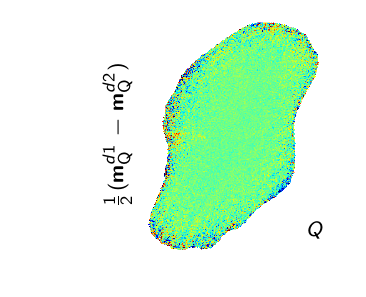}\hspace{-0.05in}
\includegraphics[width=0.20\linewidth, clip=true, trim=0.3in 0in 0.1in 0.05in]{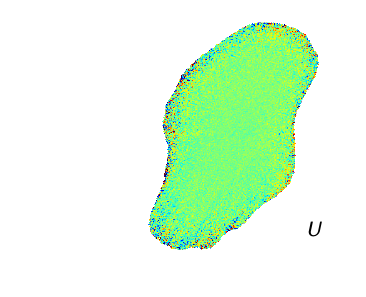}\hspace{-0.05in}
\includegraphics[width=0.20\linewidth, clip=true, trim=0.3in 0in 0.1in 0.05in]{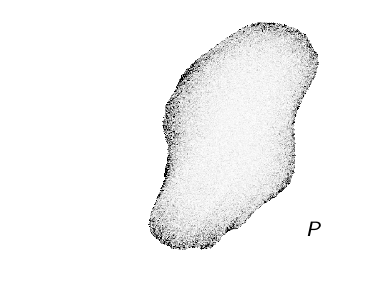}
}\vspace{-0.1in}
\mbox{
\includegraphics[width=0.20\linewidth, clip=true, trim=0.3in 0in 0.1in 0.05in]{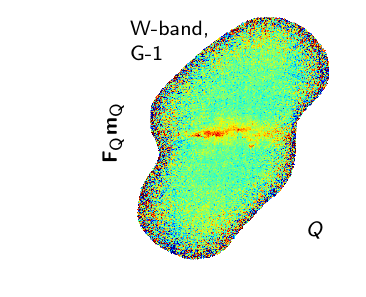}\hspace{-0.05in}
\includegraphics[width=0.20\linewidth, clip=true, trim=0.3in 0in 0.1in 0.05in]{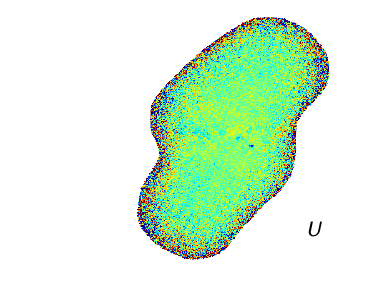}\hspace{-0.05in}
\includegraphics[width=0.20\linewidth, clip=true, trim=0.3in 0in 0.1in 0.05in]{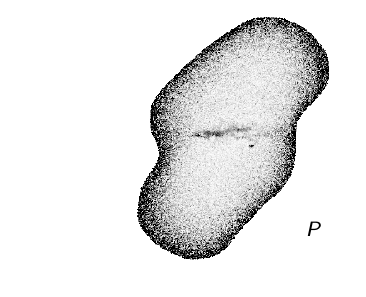}
}\vspace{-0.15in}
\mbox{
\includegraphics[width=0.20\linewidth, clip=true, trim=0.3in 0in 0.1in 0.05in]{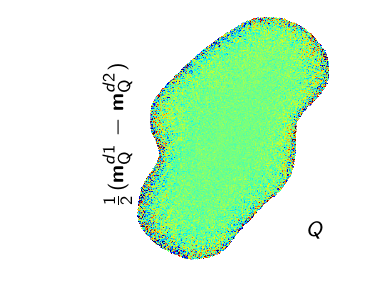}\hspace{-0.05in}
\includegraphics[width=0.20\linewidth, clip=true, trim=0.3in 0in 0.1in 0.05in]{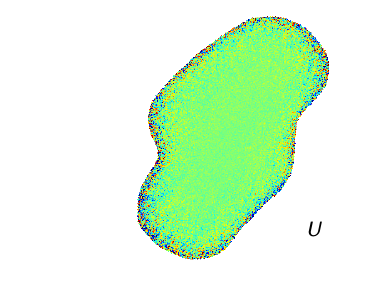}\hspace{-0.05in}
\includegraphics[width=0.20\linewidth, clip=true, trim=0.3in 0in 0.1in 0.05in]{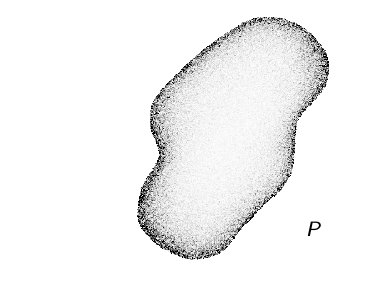}
}\vspace{-0.1in}
\mbox{
\includegraphics[width=0.20\linewidth, clip=true, trim=0.3in 0in 0.1in 0.05in]{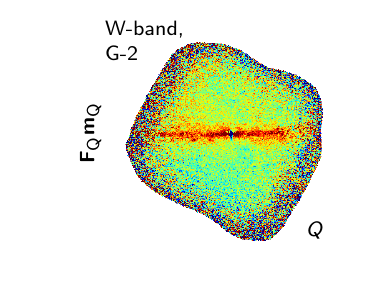}\hspace{-0.05in}
\includegraphics[width=0.20\linewidth, clip=true, trim=0.3in 0in 0.1in 0.05in]{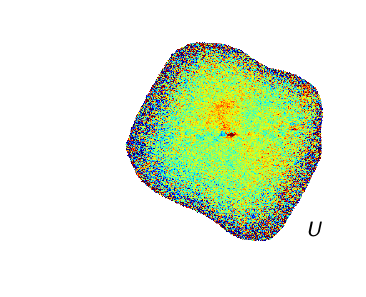}\hspace{-0.05in}
\includegraphics[width=0.20\linewidth, clip=true, trim=0.3in 0in 0.1in 0.05in]{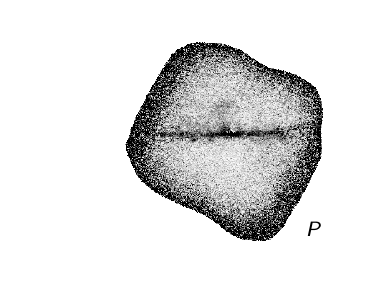}
}\vspace{-0.2in}
\mbox{
\includegraphics[width=0.20\linewidth, clip=true, trim=0.3in 0in 0.1in 0.1in]{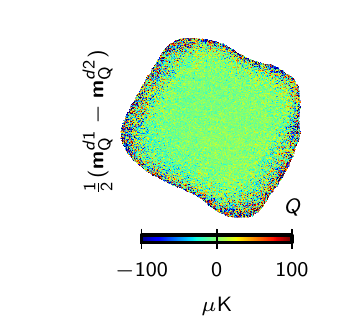}\hspace{-0.05in}
\includegraphics[width=0.20\linewidth, clip=true, trim=0.3in 0in 0.1in 0.1in]{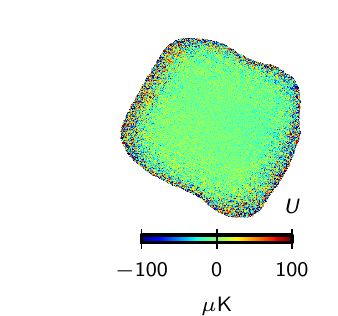}\hspace{-0.05in}
\includegraphics[width=0.20\linewidth, clip=true, trim=0.3in 0in 0.1in 0.1in]{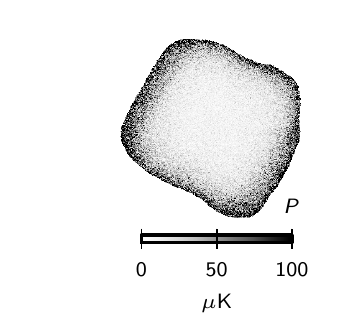}
}
\caption{Inverse-noise-variance-weighted \QUIET\ maps and deck-split half-difference maps. Columns show, from left to right, Stokes $Q$, Stokes $U$, and polarization amplitude $P$. The first row shows the \QUIET\ map $\mathbf{m}_{\mathrm{Q}}$, for the G-1 field (centered on Galactic coordinates $(l,b)=(329\deg,0\deg)$) in Q-band (43\,GHz), filtered to only contain the small-scale modes observable by \QUIET, as determined by the \QUIET\ weighting operator $\mathbf{F}_{\mathrm{Q}}$ defined in \S~\ref{sec:coadd}. The second row shows the corresponding deck-split half-difference map, $\frac{1}{2}(\mathbf{m}^{d1}_Q - \mathbf{m}^{d2}_Q)$. The following rows show the equivalent pairs of maps for  both fields in W-band (95\,GHz).}
\label{fig:filtered_app}
\end{figure*}

\begin{figure*}[t]
\centering
\mbox{
\includegraphics[width=0.28\linewidth, clip=true, trim=0.35in 0in 0.1in 0in]{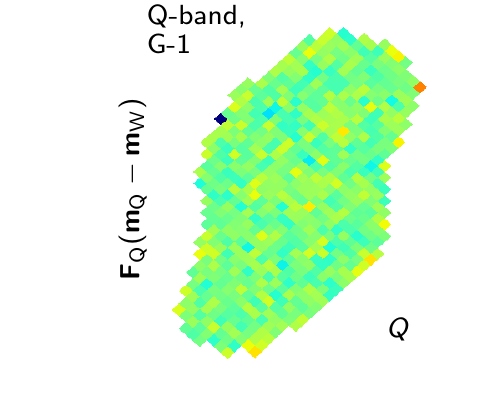}\hspace{-0.1in}
\includegraphics[width=0.28\linewidth, clip=true, trim=0.35in 0in 0.1in 0in]{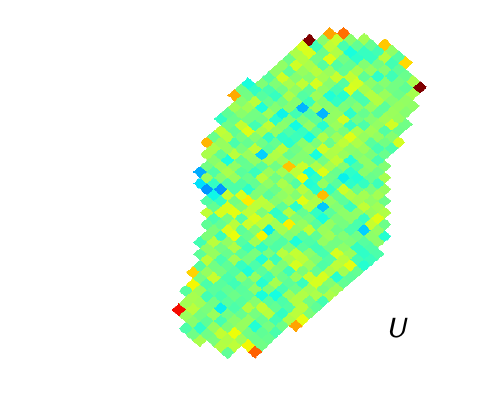}
}\vspace{-0.2in}
\mbox{
\includegraphics[width=0.28\linewidth, clip=true, trim=0.35in 0in 0.1in 0in]{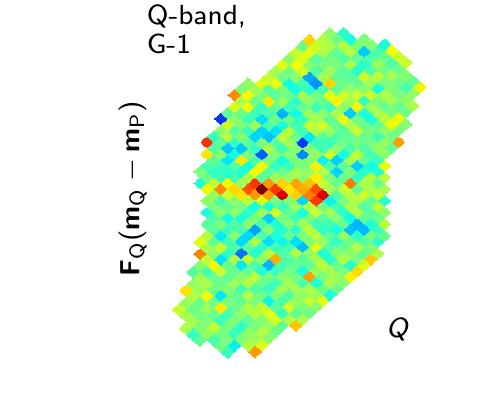}\hspace{-0.1in}
\includegraphics[width=0.28\linewidth, clip=true, trim=0.35in 0in 0.1in 0in]{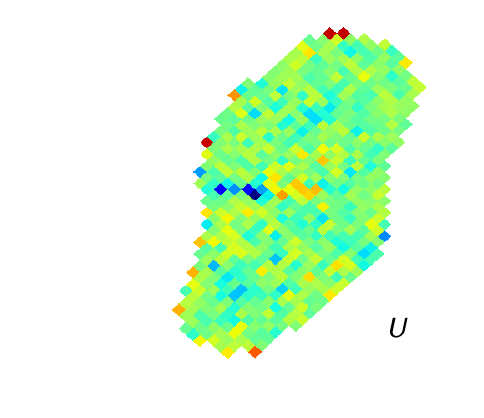}
}\vspace{-0.2in}
\mbox{
\includegraphics[width=0.28\linewidth, clip=true, trim=0.35in 0in 0.1in 0in]{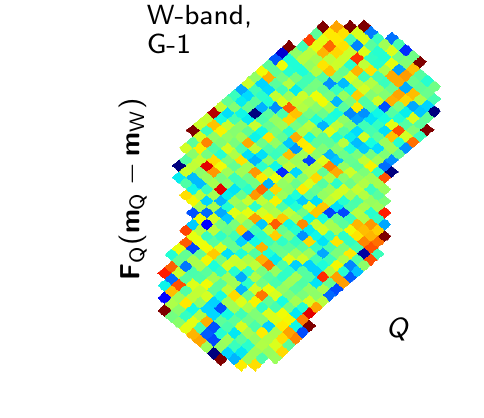}\hspace{-0.1in}
\includegraphics[width=0.28\linewidth, clip=true, trim=0.35in 0in 0.1in 0in]{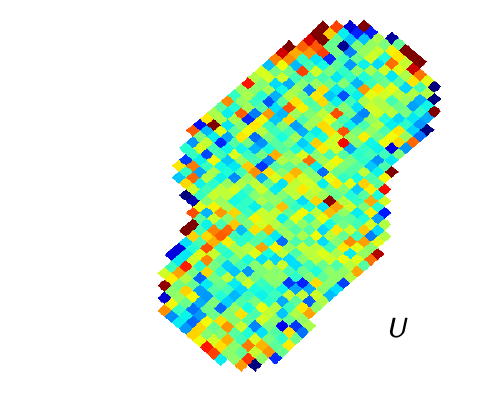}
}\vspace{-0.2in}
\mbox{
\includegraphics[width=0.28\linewidth, clip=true, trim=0.3in 0in 0.15in 0in]{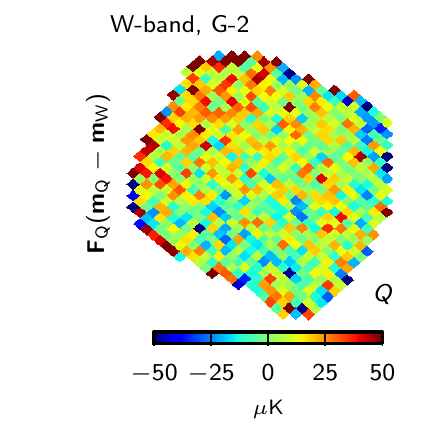}\hspace{-0.1in}
\includegraphics[width=0.28\linewidth, clip=true, trim=0.3in 0in 0.15in 0in]{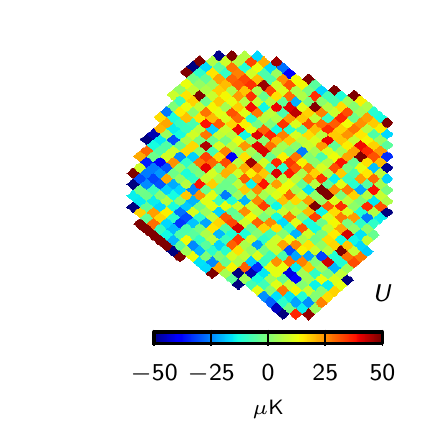}
}
\caption{Pairwise differences of \QUIET\, and \WMAP\ or \Planck\ maps, downgraded to HEALPix $N_{side}=64$ to suppress noise. All maps are weighted using the \QUIET\ weight operator $\mathbf{F}_{\mathrm{Q}}$, defined in \S~\ref{sec:coadd}, retaining only the small-scale modes observed by \QUIET\ in the differenced maps. The top row shows \QUIET$-$\WMAP\ for field G-1 (centered on Galactic coordinates $(l,b)=(329\deg,0\deg)$) in Q-band (43\,GHz). The second row shows the corresponding difference of \QUIET$-$\Planck, while the third and fourth rows show \QUIET$-$\WMAP\ for W-band (95\,GHz), both fields. Columns show, from left to right, Stokes $Q$ and Stokes $U$.}
\label{fig:diffmaps_app}
\end{figure*}

In \S~\ref{sec:maps} we presented both the raw \QUIET\ and the
co-added \QUIET$+$\WMAP\, Q-band sky maps as derived for the Galactic
center field, G-2. We also showed an internal consistency test for this field
between \QUIET, \WMAP, and \Planck, in the form of difference maps. In this
appendix we show corresponding plots for the remaining three data
combinations, namely the Q-band G-1 and W-band G-1 and G-2 fields. All
main conclusions remain unchanged compared to the original discussion,
and the following plots are reproduced for completeness and reference
purposes alone. Null-map statistics for all fields and and best-fit 
linear regression slopes for the Q-band data are listed in
Table~\ref{tab:patch_summary}.

Figure~\ref{fig:finalmaps_app} shows the final co-added
\QUIET+\WMAP\ Stokes $Q$ and $U$ parameter maps for Q-band G-1 (top
row), W-band G-1 (middle row), and W-band G-2 (bottom row),
corresponding to Figure~\ref{fig:finalmaps_gc_qband} for the Q-band
G-2 field in the main text. Computing the polarization amplitudes and
EVPAs from these leads to the maps shown in
Figures~\ref{fig:finalmaps_gcP} and \ref{fig:finalmaps_gbP}.

Figure~\ref{fig:filtered_app} shows the (filtered)
\QUIET\ contributions to the co-added sky maps and the deck-split
half-difference maps for each field combination, corresponding to
Figure~\ref{fig:compmaps_gcQ} in the main text. No significant
residuals are seen in any of these difference maps.  The Q-band G-2
case discussed in the main text therefore represents a conservative
worst-case scenario with respect to temperature-to-polarization
leakage.

Figure~\ref{fig:diffmaps_app} shows the difference maps
between \QUIET, \WMAP, and \Planck, all downgraded to
$N_{\textrm{side}}=64$ as in \S~\ref{sec:comparison}. Differences
with respect to \Planck\ are only evaluated for the Q-band G-1 field,
since \Planck\ does not provide a polarized W-band map at this time. As for the
G-2 field, we note a significant positive residual with respect to
\Planck\ in G-1, while no significant residuals are seen with respect
to \WMAP\ in either case. Finally, it could be noted that the 
W-band difference maps have a higher noise level. This is due 
to the high noise rms in the \WMAP\ W-band maps, as listed in 
Table~\ref{tab:patch_summary}.

\end{document}